\documentclass[11pt]{article}
\pdfoutput=1
\usepackage{amsmath}
\usepackage{amssymb}
\usepackage{graphicx,bbm,mathrsfs}
\usepackage{nicefrac}
\usepackage{slashed}
\usepackage{bbm}
\usepackage{geometry}
\usepackage{empheq}

\newcommand*\widefbox[1]{\fbox{\hspace{1.5em}#1\hspace{1.5em}}}

\usepackage{jheppub}

\usepackage{mathtools}

\usepackage{dcolumn}   
\usepackage{bm}        
\usepackage{graphicx,mathrsfs}
\usepackage{nicefrac}
\usepackage{multirow}
\usepackage{color}
\usepackage{mathtools}
\usepackage{makecell}

\usepackage{environ} 
\usepackage{lipsum} 
 \NewEnviron{Smaller11}{
           \scalebox{1.1}{$\BODY$} 
 } 
 \NewEnviron{Smaller08}{
           \scalebox{0.8}{$\BODY$} 
 }

\newcommand{\be}{\begin{equation}}
\newcommand{\ee}{\end{equation}}
\newcommand{\bea}{\begin{eqnarray}}
\newcommand{\eea}{\end{eqnarray}}

\newcommand{\tr}{\operatorname{tr}}

\newcommand{\PhiL}{\Phi^{(\ell)}}
\newcommand{\PhiH}{\Phi^{({h})}}
\newcommand{\n}{n}

\newcommand{\V}{\prod^\n_{i=1}\varphi_{i,0}}
\newcommand{\DeltaX}{\Delta_X}

\renewcommand{\c}{c}
\newcommand{\mur}{\mu_r}

\usepackage{wasysym}
\usepackage{hhline,colortbl}

\newcommand{\DeltaH}{\Delta_h}

\newcommand{\nn}{\nonumber}

\newcommand{\bb}{\bar{b} }

\newcommand{\p}{p}
\newcommand{\g}{g}
\newcommand{\f}{f}
\newcommand{\h}{h}

\newcommand{\tPhi}{\tilde \Phi}

\newcommand{\cB}{\star}
\newcommand{\cb}{\circ}

\usepackage{titlesec}

\titleformat*{\section}{\Large\bfseries}
\titleformat*{\subsection}{\large\bfseries}
\titleformat*{\subsubsection}{\large\bfseries}
\titleformat*{\paragraph}{\large\bfseries}
\titleformat*{\subparagraph}{\large\bfseries}

\makeatletter
\newcommand*{\prodsym}{%
  \DOTSB
  \mathop{
    \mathchoice
      {\rlap{\kern.3em\rotatebox[origin=c]{-90}{}}{\prod}}
      {\vcenter{\rlap{\kern.2em\rotatebox[origin=c]{-90}{}}}{\prod}}
      {\sum}{\sum}
  }\slimits@
}
\makeatother

\makeatletter
\DeclareFontFamily{OMX}{MnSymbolE}{}
\DeclareSymbolFont{MnLargeSymbols}{OMX}{MnSymbolE}{m}{n}
\SetSymbolFont{MnLargeSymbols}{bold}{OMX}{MnSymbolE}{b}{n}
\DeclareFontShape{OMX}{MnSymbolE}{m}{n}{
    <-6>  MnSymbolE5
   <6-7>  MnSymbolE6
   <7-8>  MnSymbolE7
   <8-9>  MnSymbolE8
   <9-10> MnSymbolE9
  <10-12> MnSymbolE10
  <12->   MnSymbolE12
}{}
\DeclareFontShape{OMX}{MnSymbolE}{b}{n}{
    <-6>  MnSymbolE-Bold5
   <6-7>  MnSymbolE-Bold6
   <7-8>  MnSymbolE-Bold7
   <8-9>  MnSymbolE-Bold8
   <9-10> MnSymbolE-Bold9
  <10-12> MnSymbolE-Bold10
  <12->   MnSymbolE-Bold12
}{}

\let\llangle\@undefined
\let\rrangle\@undefined
\DeclareMathDelimiter{\llangle}{\mathopen}%
                     {MnLargeSymbols}{'164}{MnLargeSymbols}{'164}
\DeclareMathDelimiter{\rrangle}{\mathclose}%
                     {MnLargeSymbols}{'171}{MnLargeSymbols}{'171}
\makeatother

\begin{document}

\vspace*{4mm}

\begin{center}

\thispagestyle{empty}
{\huge
On Holography in  General Background and the Boundary Effective Action from AdS to dS
 }\\[12mm]

\renewcommand{\thefootnote}{\fnsymbol{footnote}}

{\large  
Sylvain~Fichet \footnote{sfichet@caltech.edu }\,
}\\[8mm]

\end{center} 
\noindent
\quad\quad\quad \textit{ ICTP South American Institute for Fundamental Research  \& IFT-UNESP,}

\noindent \quad\quad\quad\quad \textit{R. Dr. Bento Teobaldo Ferraz 271, S\~ao Paulo, Brazil}
\\ 

\noindent \quad\quad\quad \textit{ Centro de Ciencias Naturais e Humanas, Universidade Federal do ABC,}

\noindent \quad\quad\quad\quad \textit{Santo Andre, 09210-580 SP, Brazil}

\addtocounter{footnote}{-1}

\vspace*{12mm}

\begin{center}
{  \bf  Abstract }
\end{center}

We study quantum fields on an arbitrary, rigid background with boundary. We derive the action for a scalar in the holographic basis that separates  the boundary and bulk degrees of freedom.
 A relation between Dirichlet and Neumann propagators valid for any background is obtained
  from this holographic action.
As a simple application, we derive an exact formula for the flux of bulk modes emitted from the boundary in a warped background. 
 We also derive a formula for the Casimir pressure on a $(d-1)$-brane depending  only on the boundary-to-bulk propagators, and apply it in AdS. 
Turning on couplings and using the holographic basis,  we evaluate the one-loop  boundary effective action in AdS  by means of the  heat kernel expansion.
We extract anomalous dimensions of single and double trace CFT operators generated by loops of heavy scalars and nonabelian vectors, up to third order in the large squared mass expansion. 
 From the boundary heat kernel coefficients we identify  CFT operator mixing and corrections to OPE data, in addition to the radiative generation of local operators. 
 We integrate out nonabelian vector fluctuations  in AdS$_{4,5,6}$ and obtain the associated  holographic Yang-Mills $\beta$ functions. 
 Turning to the expanding patch of dS, following  recent proposals, we  provide a boundary effective action generating the perturbative cosmological correlators using analytical continuation from dS to EAdS.  
We obtain the ``cosmological'' heat kernel coefficients in the scalar case and work out the  divergent part of the dS$_4$ effective action which renormalizes the cosmological correlators.  We find that bulk masses and wavefunction  can logarithmically run as a result of the dS$_4$ curvature, and that operators on the late time boundary are radiatively generated. More developments are needed to extract all one-loop information from the cosmological effective action.

\newpage
\tableofcontents

\newpage

\section{Introduction and Summary}
\label{se:intro}

Imagine a quantum field theory (QFT) supported on a background manifold with boundary. What can an observer standing on the boundary  learn about the QFT living in the interior of the manifold?
While this is a simple problem at the classical level, evaluating boundary observables at the quantum level is more challenging since it involves integrating over the quantum fluctuations occuring in the bulk of the system.    Integrating out the bulk quantum fluctuations can be done through the quantum effective action, the latter is then a functional of the boundary degrees of freedom, \textit{i.e.} a ``boundary effective action''.  

Such a setup --- a QFT seen from the boundary ---  is common in physics, and is typically referred to as ``holographic''. Holography actually refers to a variety of similar-but-not-equivalent concepts that we briefly review further below. 
Among all examples of QFTs on a background manifold with boundary, we can single out two cases of paramount importance: the  Anti-de Sitter (AdS)  and de Sitter (dS) spacetimes. These are the maximally symmetric curved spacetimes.

In AdS space, an observer standing on the  AdS  boundary sees a strongly-coupled CFT, with large number of colors $N$ if the bulk QFT is weakly coupled \cite{Aharony:1999ti,Zaffaroni:2000vh,Nastase:2007kj,Kap:lecture}.
Therefore holography in AdS leads to a profound connection between gravity and strongly-coupled gauge theories, opening new possibilities to better understand both. 
The holographic view of AdS was formulated two decades ago \cite{Witten:1998qj,Gubser:1998bc},   AdS/CFT  is now studied at loop level. In this work our application to AdS holography will be focused on systematic  one-loop  computations. 

The notion of holography in dS space is even more concrete: Cosmological observations suggest that the chronology of the Universe  has at least two phases with approximate dS geometry: the current expanding epoch  and the inflationary epoch. 
When we look at the sky and measure galactic redshifts or the CMB, we  are actually observers standing on the late time boundary of the expanding patch of dS, probing the dS interior with telescopes.
The inflationary phase is of great interest because it probes the highest accessible energies and the earliest period of the Universe. Remarkably, taking a glimpse into
the quantum fluctuations of the Early Universe is  possible by analyzing the cosmological correlators of the CMB. 
Classical and quantum calculations of cosmological correlators are much less advanced than those in AdS. In this work, we  focus on establishing a boundary quantum effective action that generates these correlators, 
with  the goal of extracting  loop-level information from it. 

In the litterature, depending on context, the term ``holography'' may either refer to the computation of boundary observables and related quantities, or to the identification of a dual $d$-dimensional theory reproducing these observables\,\cite{tHooft:1993dmi,Susskind:1994vu,Bousso:2002ju}.\,\footnote{
The latter usage is also sometimes referred to as  ``holographic principle''. One concrete incarnation is the emergence of dual $d-$dimensional theories from $d$+$1-$dimensional Chern-Simons theories, with for example the 3d CS/WZW correspondence \cite{CS_WZW}.
Another incarnation is the AdS/CFT correspondence. 
Holographic dualities can aim beyond weakly coupled QFT, exploring non-perturbative aspects of quantum gravity, black holes,  and entanglement entropy (see \textit{e.g.} \cite{Ryu:2006bv}). The present work is  about weakly coupled quantum fields.
In AdS/CFT a distinction is also  drawn between bulk theories with and without dynamical gravity --- in the latter case the conformal theory has no stress tensor. This distinction is unimportant for the present study which only involves interacting fields with spin-0 and 1 while interactions with the graviton sector are not considered. 
} 
Here we adopt the former usage: our focus is not on the existence and specification of dual boundary theories but simply on the evaluation and content of the boundary effective action itself. 
Only in the case of AdS background will we discuss data of the dual CFT.


The initial goal in this work is to derive the holographic action of a QFT on an arbitrary background.  
Apart from providing a  formalism computing holographic quantities beyond AdS, this approach  brings a somewhat different viewpoint on well-known AdS holography. 
In AdS holography, one may  wonder whether a given feature  is either a  manifestation of AdS/CFT or a more general property of the holographic formalism itself. If the latter is true, the feature under consideration is valid beyond AdS. Thus the boundary action on arbitrary background can, in this sense, be used to shed light on AdS/CFT features. 

Using the established holographic formalism, we then proceed with computing and investigating the  effects of bulk quantum  fluctuations on the physics seen from the boundary. Among these effects we can distinguish \textit{i)} the bulk vacuum bubbles (i.e. $0$pt diagrams) which cause a quantum pressure on the boundary and \textit{ii)} $(n>0)$pt connected bulk diagrams which are responsible for correcting/renormalizing the boundary theory. We investigate aspects from both \textit{i)} and \textit{ii)}.

From a technical viewpoint, the integration of quantum fluctuations at one-loop on mani\-folds with boundary is encoded into the heat kernel coefficients of the one-loop effective action \cite{DeWitt_original,DeWitt_original2,Gilkey_original,McAvity:1990we}.  The first heat kernel coefficients have been gradually computed along the past decades (see \cite{Vassilevich:2003xt} and references therein). To study the  boundary correlators, we introduce these  results into the framework of the holographic action. 
 An overarching theme of the second part of our study is therefore the
  encounter of the heat kernel with  holography, and especially with AdS/CFT. 


\paragraph{The holographic basis}

\begin{figure}[t]
\centering
	\includegraphics[width=1.0\linewidth,trim={0cm 6cm 0cm 8cm},clip]{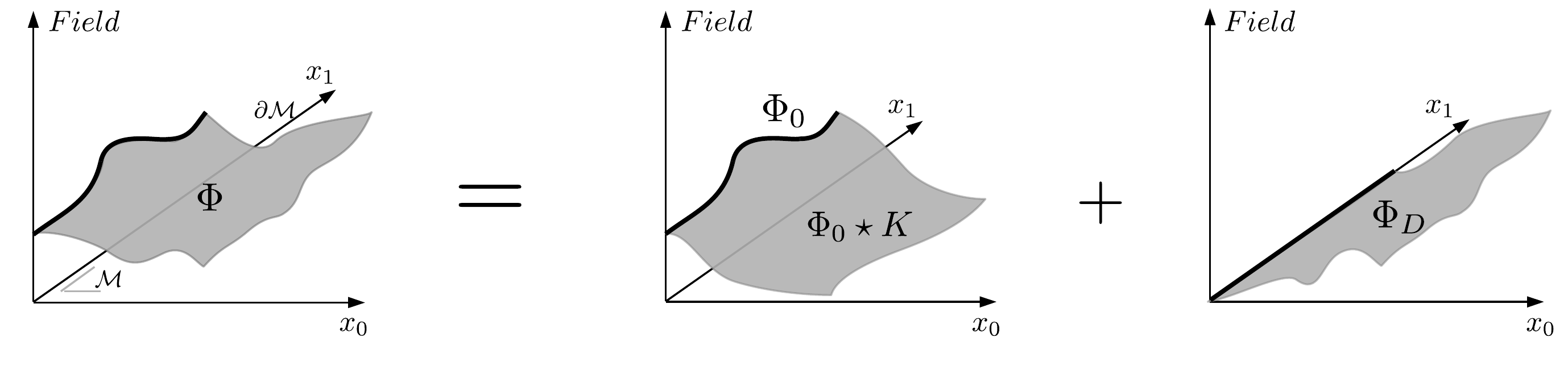}
\caption{
The holographic decomposition of a quantum field $\Phi$ where $\Phi_0$ is the boundary value of $\Phi$, $K$  is the boundary-to-bulk propagator, $\Phi_D$ is the Dirichlet component of $\Phi$. Here, the field is supported in  ${\cal M}=\mathbb{R}\times \mathbb{R}_+$ with  boundary at $x_0=0$.  
\label{fig:Hol_basis}}
\end{figure}

 A quantum field may or may not fluctuate on the boundary, \textit{i.e.} have respectively Neumann or Dirichlet boundary condition (BC).\,\footnote{
 Here and throughout this work, ``Neumann'' BC includes ``Robin'' BC. 
 For fields with spin, the field components consistently split into a subset with Neumann BC and a subset with Dirichlet BC.  } 
 In either case, the crux of the holographic approach is to separate the bulk and boundary degrees of freedom of the quantum field. Here we will isolate the pure bulk degrees of freedom by singling out the field component that vanishes on the boundary, that we refer to  as ``Dirichlet component''. 
 The  remaining degree of freedom on the boundary is then  encapsulated into a separate variable. 
 The value of  a  field at a given point $x^M$ in the bulk is  completely described by the Dirichlet  component  plus the (possibly fluctuating) boundary degree of freedom. The boundary degree of freedom  contributes remotely to  $\Phi(x)$, thus a propagator must be involved ---  we will see in Sec.~\ref{se:gen} that it is the so-called boundary-to-bulk propagator $K$.  Such a ``holographic'' decomposition of the quantum field is illustrated in Fig.\,\ref{fig:Hol_basis}.  Since a boundary observer  does not probe the Dirichlet modes, these can be completely integrated out to give rise to a boundary effective action. We will perform this operation at the loop level throughout this work.

\paragraph{Review}
The literature related to our study includes the following references. \\
\textit{Vacuum bubbles and quantum pressure}: 
We are unaware of a work  about evaluating Casimir pressure or energy in the presence of a $\mathbb{R}^d$ boundary beyond the much studied case of Minkowski background.
In our application we are interested in a difference of vacuum pressure between each side of the boundary, hence our setup has little connection to the formal calculations of  Casimir energy in AdS$_{d+1}$ from \cite{Beccaria:2014qea} and subsequent references. 
\\
\textit{Boundary correlators in AdS}:  There has been a lot of activity about computing and studying loop-level correlators\,\cite{ Cornalba:2007zb,
Penedones:2010ue, Fitzpatrick:2011hu,
Alday:2017xua,
Alday:2017vkk,
Alday:2018pdi,
Alday:2018kkw,
Meltzer:2018tnm,
Ponomarev:2019ltz,
Shyani:2019wed,
Alday:2019qrf,
Alday:2019nin,
Meltzer:2019pyl,
Aprile:2017bgs,
Aprile:2017xsp,
Aprile:2017qoy,
Giombi:2017hpr,
Cardona:2017tsw,
Aharony:2016dwx,
Yuan:2017vgp,
Yuan:2018qva,
Bertan:2018afl,
Bertan:2018khc,
Liu:2018jhs,
Carmi:2018qzm,
Aprile:2018efk,
Ghosh:2018bgd,
Mazac:2018ycv,
Beccaria:2019stp,
Chester:2019pvm,
Beccaria:2019dju,
Carmi:2019ocp,
Aprile:2019rep, 
Fichet:2019hkg,
Meltzer:2019nbs,
Drummond:2019hel, Albayrak:2020isk, 
Albayrak:2020bso,
Meltzer:2020qbr,
Costantino:2020vdu,
Carmi:2021dsn, Fitzpatrick:2011dm, Ponomarev:2019ofr, Antunes:2020pof,
Fichet:2021pbn}. 
 However to the best of our knowledge, such studies are always focused on specific diagrams, and not on the one-loop effective action. It seems that the AdS one-loop effective action has been used only in the very specific case of  the one-loop scalar potential (namely, for  constant scalar field with dimension $\Delta=d$)  \cite{Burgess:1984ti,
Inami:1985wu,
Camporesi:1993mz,
Gubser:2002zh,
Hartman:2006dy,
Giombi:2013fka,
Carmi:2018qzm}. 
The one-loop boundary effective action, through the heat kernel coefficients,   contains much more information on  the (bulk and boundary) divergences and on the long-distance EFT.  \\
\textit{Boundary  correlators in dS}: There has been a lot of activity about
computing cosmological correlators and understanding their  structure  in terms of conformal symmetry and singularities
\cite{Antoniadis:2011ib,Creminelli:2011mw,Maldacena:2011nz,Bzowski:2011ab,Kehagias:2012pd,Mata:2012bx,Kundu:2014gxa,Kundu:2015xta,Ghosh:2014kba,Pajer:2016ieg,Arkani-Hamed:2015bza,Arkani-Hamed:2018kmz,Farrow:2018yni,Baumann:2019oyu,Green:2020ebl,Sengor:2021zlc,Wang:2021qez}.\footnote{See also \cite{Green:2020ebl,Goodhew:2020hob,Pajer:2020wxk,Jazayeri:2021fvk,Melville:2021lst,Goodhew:2021oqg,Baumann:2021fxj} for extension to models without invariance under special conformal transformations.} 
A bootstrap program analogous to flat space amplitudes techniques has also been developed, often at the level of the dS wavefunction coefficients, with \textit{e.g.} cutting rules, dispersion relations and positivity bounds,
see for example \cite{Maldacena:2011nz,Raju:2012zr,Arkani-Hamed:2017fdk, Arkani-Hamed:2018bjr,Benincasa:2018ssx,Cespedes:2020xqq,Sleight:2020obc,Meltzer:2020qbr, Baumann:2020dch, Jazayeri:2021fvk,Melville:2021lst,Goodhew:2021oqg,Meltzer:2021bmb,Meltzer:2021zin, Baumann:2021fxj,Gomez:2021qfd,Bonifacio:2021azc}.
In this work we build on recent developments for computing the cosmological correlators via analytical continuation from dS to EAdS \cite{Sleight:2020obc,Sleight:2019mgd, Sleight:2019hfp}, see also \cite{Balasubramanian:2002zh, Maldacena:2002vr, Harlow:2011ke, Anninos:2014lwa} for earlier works. We build  on a proposal from \cite{DiPietro:2021sjt} to define an EAdS effective action that generates the perturbative cosmological correlators.
\textit{Added in v2}: Along similar lines, the recent work Ref.\,\cite{Heckelbacher:2022hbq}  computed loop corrections to cosmological correlators using the EAdS formulation.

\subsection{Outline}

Since our study intertwins  a number of themes and results, we end this introduction with a guide to the sections and their relationships. 

Essential conventions can be found in section \ref{se:conv} (see also \ref{app:HK}). We work with both Euclidian and Lorentzian metric depending on the section. We work with scalar and vector fields, restricting mostly to scalars for conceptual discussions (vectors are introduced in section \ref{se:HOLEA}). 

In section~\ref{se:gen}, we introduce the holographic basis for a scalar field in an arbitrary background with boundary, and compute the action in this basis. The holographic basis will later be an important ingredient to derive the boundary effective action, and it also brings some insights on the behaviour of the fields in the presence of the boundary. The boundary component ``$K \Phi_0$'' of the holographic basis is on-shell in the bulk while off-shell on the boundary, in which case AdS/CFT can apply. This direction is pursued in sections~\ref{se:AdSLoop},~\ref{se:YM}.


In section~\ref{se:warped_holography} we apply our general formulation to a more specific class of Lorentzian warped background, obtaining holographic action and propagators.   This section  is essentially a review accompanied with some  scattered new results and observations. For example we find a simple formula for the flux of modes emitted from the boundary. 

 In section~\ref{se:Pressure}  we consider a $(d-1)$-brane (a.k.a. interface or domain wall) in the warped background. We show how   to compute the one-loop quantum pressure on the brane using our formalism. 
 This  section is about integrating the bulk modes at one-loop at the level of 0pt diagrams, \textit{i.e.} vacuum bubbles. In contrast,   the following sections are about connected correlators.

In section~\ref{se:interactions} we introduce interactions in the holographic basis and describe in details the general structure of the boundary action. The structure of the long-distance EFT, which is then used in sections \ref{se:HOLEA},~\ref{se:AdSLoop},~\ref{se:YM}, is discussed in details. This section also points out a connection between a certain type of AdS Witten diagram and a large $N$ diagrammatic expansion in the CFT.

In section~\ref{se:HOLEA} we introduce the one-loop boundary effective action, which is computed by the heat kernel coefficients in the holographic basis. This section  essentially contains a rewriting of known heat kernel results, with a discussion on the one-loop effective potential as an aside. The background and fluctuations for both scalar and vector fields are considered.

In sections~\ref{se:AdSLoop} and \ref{se:YM} the overall goal is to extract AdS and CFT results from the heat kernel coefficients.  We evaluate  the one-loop boundary effective action in  AdS$_{d+1}$ background. In section \ref{se:AdSLoop} we integrate out a scalar fluctuation interacting with  scalars, while  in section \ref{se:YM} we integrate out a nonabelian vector, interacting with either vectors or charged scalars. 
Since AdS/CFT applies when using the holographic basis, we  can extract CFT data  from the heat kernel coefficients. 
This is an algebraic  (as opposed to diagrammatic) computation from  the AdS side of leading non-planar effects in the CFT.

In section~\ref{se:dS} the overall goal is to extract one-loop corrections to cosmological correlators from the heat kernel coefficients. To put the ``cosmological'' effective action in a convenient form  we use  analytical continuation from dS to Euclidian AdS (EAdS) space. 
The boundary effective action expressed in EAdS connects to the AdS results from sections~\ref{se:AdSLoop}, \ref{se:YM}, (upon proper translation of AdS/EAdS conventions). 
We give some concrete results  on renormalization in dS$_4$ for inflaton-like scalar fields, extracting information from the EAdS heat kernel coefficients. Finally we  discuss propagation in dS in the presence of boundary operators.

The appendices contain a proof of the discontinuity equation (\ref{app:der}), details and checks on  propagators in warped background (\ref{app:prop}), a derivation of the double holographic action (\ref{app:double}), basics of the heat kernel   (\ref{app:HK}) and some explicit expressions for cosmological correlators (\ref{app:dS_corr}).

\subsection{Summary of Results}

\subsubsection*{General background }

\begin{itemize}

\item We derived the action of a scalar field  in the holographic basis in a general background. 
The action is  diagonal and the Neumann-Dirichlet identity ``$G_N=G_D+K  G_0  K$" immediately follows, proving that such a relation is independent of the background geometry.  This relation provides  a trivial understanding of the effect of  boundary-localized bilinear operators on the propagator. 

 \item We evaluate the holographic action in a generic warped background, including also a dilaton background, and find that it is essentially as simple as in AdS. 
 We obtained a general formula for the flux of bulk modes emitted from the boundary. 
  In the AdS limit we find $G_D=G_{\Delta=\Delta_+}$ and $G_N=G_{\Delta=\Delta_-}$ for any $\Delta$, which, together with $G_N=G_D+K  G_0  K$,  reproduces a known identity for AdS propagators. 
   We derive the ``double'' holographic action in the presence of two boundaries, as a functional of the two boundary variables (App.~\ref{app:double}).

\item Integrating out the free bulk modes, we obtain a simple formula for the quantum pressure induced by a bulk scalar fluctuation on a  $(d-1)$-brane , expressed only  as a function of boundary-to-bulk propagators. We recover the scalar Casimir pressure of  $(d+1)$-dimensional flat space then study the quantum pressure on a brane in AdS$_{d+1}$ (or a point particle if $d=1$). For AdS$_2$ we find that the point particle is attracted towards the AdS$_2$ boundary. For a conformally massless scalar, the field  effectively sees a flat half-plane and the known 2d  Casimir force is recovered.  For higher dimensions the pressure can  be understood as a  contribution to the brane tension. For example, if the brane is static at the  tree level, at quantum level it acquires a velocity driven by the quantum pressure.  Logarithmic divergences renormalize the brane tension for even $d$.


\end{itemize}

\subsubsection*{Anti-de Sitter background  }

\begin{itemize}

\item

The AdS boundary effective action for fields fluctuating on the boundary generates boundary diagrams
with $\Delta_-$ internal lines. Upon using the Neumann-Dirichlet identity, the diagrams are decomposed into diagrams whose internal lines are either $\Delta_+$ or boundary propagators. 
We point that diagrams with only internal boundary lines can be computed by using a   perturbative  expansion scheme in $\frac{1}{N}$ directly in the holographic CFT.
 The resulting  large $N$ CFT diagrams  are built from ``vertices'' made from the  amputated   $n>2$pt CFT correlators at leading order in $\frac{1}{N}$, connected to each other by the mean field  (\textit{i.e.} $N=\infty$) $2$pt CFT correlator. An analogous diagrammatic approach to  large $N$  CFT was introduced in \cite{Petkou:1994ad}, here  we found out how it arises from the AdS side.

\item 
In AdS$_{d+1}$ space, we evaluate the  boundary effective action arising from integrating out a heavy scalar fluctuation at one-loop via the heat kernel formalism, assuming scalar interactions.
AdS/CFT applies since the holographic basis has boundary components which are on-shell in the bulk, therefore we can  extract non-planar CFT data from the heat kernel coefficients. 
 We obtain anomalous dimensions of scalar single trace opera\-tors up to fourth order in the large $\DeltaH$ expansion. We compare a 3d result with the corresponding anomalous dimension from an exact  AdS$_3$ bubble  and obtain perfect agreement at all  available orders.
We also derive contributions to the  anomalous dimensions of double-trace operators ${\cal O} \square^n {\cal O}$.

\item From boundary heat kernel coefficients, we obtain corrections to the OPE data in the form of a ``wavefunction renormalization'' of the CFT operator. We also obtain that a mixing  between the $\cal O$ and ${\cal O}^2 $ operators arises at one-loop.
Finally we obtain the one-loop correction/renormalization    to local operators on the boundary.  Such local operators can, under certain conditions, be interpreted as multitrace deformations of the CFT. 
More generally,  the local operators generated on the boundary include  mass, kinetic and interaction terms, whose existence is  known from the extradimension literature but which had, to the best our knowledge, never been computed exactly. 

\item
 In AdS$_{4,5,6}$, we integrate out  nonabelian vector fluctuations and deduce the  holographic  $\beta$ function of the gauge coupling. For any of these dimensions the boundary gauge coupling has a logarithmic running, which  comes  from either bulk or boundary heat kernel coefficients depending on spacetime dimension.  For $d=6$ the logarithmic running of the dimensionful gauge coupling is a consequence of the nonzero curvature of spacetime. We also give the 6d result in general background. 
 In the case of heavy nonabelian vectors (\textit{i.e.} highly nonconserved CFT currents), we compute the anomalous dimensions of the operators associated with light scalars in arbitrary representation of the gauge group. We also obtain the anomalous dimensions of scalar double trace operators generated by the heavy vectors.  
 
 \end{itemize}
 
 \subsubsection*{de Sitter background}

\begin{itemize}

\item
Turning to the expanding patch of dS, we evaluate the dS 2pt functions directly in momentum space and  perform the analytical continuation from dS to EAdS also used in \cite{Sleight:2019hfp,Sleight:2019mgd,Sleight:2020obc,DiPietro:2021sjt}. 
Following a proposal from \cite{DiPietro:2021sjt} we establish a boundary effective action which is the generating functional of  amputated cosmological correlators. 
Working in a simple scalar case,  we obtain the ``cosmological'' one-loop effective action.  More developments are however needed to extract all one-loop information from the cosmological heat kernel coefficients. 

\item 
Focussing on  scalar fields in dS$_4$ we extract the  divergences from the cosmological one-loop effective action. In the case  of a massless field with nonderivative quartic interactions, we find that the bulk mass (and thus the associated scaling dimension) runs logarithmically as a result of the finite Hubble scale. Along the same line a boundary-localized mass term is also radiatively generated. The normalization of the corresponding boundary CFT operator also  runs. 
In the case  of a massless field with derivative quartic interactions (\textit{e.g.} the inflaton)  we obtain wavefunction renormalization in the bulk controlled by the Hubble scale, and the radiative generation of a boundary-localized kinetic term. 
The cosmological effective action provides the beta functions for all of these operators. 
Finally we point that such boundary-localized operators should be included in the dS action from the start, and that dS propagators may be substantially deformed in their presence,  as dictated by the Neumann-Dirichlet identity. 

\end{itemize}

\subsection{Definitions and Conventions}

\label{se:conv}

We consider a  $d+1$-dimensional manifold ${\cal M}$ with boundary $\partial {\cal M}$.  The bulk and boundary are taken to be sufficiently smooth such that Green's identities apply. The metric is either Euclidian or Lorentzian with $(-++\ldots)$ signature. 
Latin indices $M,N,\ldots =\{0,1,\dots d\}$  index the bulk coordinates,  denoted by $x^M$. The bulk metric $g_{MN}$ is defined by $ds^2=g_{MN}dx^M dx^N$. 
A point belonging to the boundary is labelled as $x^M_0\equiv x^M|_{\partial \cal M}$ in the bulk coordinates. 
Greek indices  $\mu,\nu,\ldots=\{0,1\,\dots d-1\}$ index boundary coordinates, denoted by $y^\mu$. The boundary is described by the embedding 
$x^M=x^M_0(y^\mu)$. 
Defining $E^M_\mu=\frac{\partial x_0^M}{\partial y^\mu}$, the induced metric is defined  by $\bar g_{\mu\nu}=E^M_\mu E^N_\nu g_{MN}$. 
Let $e_\perp^M(y^\mu)$ be the unit  vector  normal to the boundary at the point $y^\mu$ and outward-pointing.
We refer to the contraction $\partial_\perp\equiv e_\perp^M\partial_M $ as the normal derivative. 
   The $\partial_\mu\equiv E_\mu^M\partial_M $ derivatives are referred to as transverse.\,\footnote{ The unit normal vector satisfies $e^M_\perp e^N_\perp g_{MN}  =1$ and $e^M_\perp E^\mu_M =0  $  by definition.
     Refs.\,\cite{Branson:1999jz,Vassilevich:2003xt} used orthonormal vielbeins.
   In our notations, the orthonormal vielbein in $\partial \cal M$ is defined by $\bar g_{\mu\nu} = e_\mu^a e_\nu^a \eta_{ab} $. The orthonormal vielbein in $\cal M$ is $(e^M_\perp , E^M_\mu e^\mu_a )$ ---   the $(e^M_\perp , E^M_\mu)$ basis  forms a non-orthonormal vielbein in   ${\cal M}$. 
     Here we do not need to use these vielbeins explicitly  apart from the normal vector. See \textit{e.g.} \cite{Blau_GR} for  details on hypersurface geometry. 
}

We consider  a  QFT  on the ${\cal M}$ spacetime background. The boundary value of  a field $\Phi$ is denoted by $\Phi_0=\Phi|_{\partial {\cal M}}$. 
The ``Dirichlet''  component of  $\Phi$ that vanishes on the boundary   is denoted as $\Phi_D$.
We will routinely switch between Euclidian and Lorentzian metric via Wick rotation $x^0_E=i x_L^0 $. We remind the convention for the actions,  $S_E=-i S_L$. Propagators are related by $G_E=i G_L$.

We define the bulk inner product as
\be
p \cB q = \int_{\cal M} d^{d+1}x\sqrt{g} \,p(x) q(x) \,. \label{eq:cB}
\ee
We define the boundary inner product as 
\be
p \cb q = \int_{\partial \cal M} d^{d}y\sqrt{\bar g} \,p(x_0) q(x_0) \,. \label{eq:cb}
\ee
\footnote{
The identity elements for the bulk and boundary products are respectively $\frac{1}{\sqrt{g}_x}\delta^{d+1}(x-x')$ and $\frac{1}{\sqrt{\bar g}_{y}}\delta^{d}(y-y')$.
}\,This is useful to manipulate Green's identities. For example  Green's first identity is 
 $\partial_M\Psi \cB \partial^M \Phi = -\Psi \cB \square \Phi + \Psi\cb \partial_\perp \Phi  $ in this notation.

\section{The Holographic Action  in a General Background}
\label{se:gen}

We derive   the action in the holographic basis for  an arbitrary, smooth manifold with boundary. 
We assume Euclidian signature. The  analogous result in Lorentzian signature  can be obtained via analytical calculation of the Euclidian result.\,\footnote{In Lorentzian signature the derivation of the holographic action for a timelike boundary is identical to the Euclidian case. The case of a spacelike boundary is possibly more subtle because time-ordering in the 2pt functions matters. It is not treated in this section. Holographic calculations in dS space (which has spacelike boundary) are done in section~\ref{se:dS} relying on analytical continuation to Euclidian space.  }
 It is sufficient to focus on a scalar field $\Phi$ to avoid  spin-related technicalities.
This section is somewhat technical, the reader willing to skip the details can go to the main result Eq.\,\eqref{eq:free_action_final}.

\subsubsection*{Action and Green's functions}

 The fundamental action is denoted $S[\Phi]$. 
 The general partition function with a generic bulk source $J$ is  $Z[J] = \int {\cal D} \Phi \exp[{- S[\Phi]-\int  dx^{d+1}\sqrt{g} J\Phi}] $. 
 The quadratic part of the action takes the form
\be
S[\Phi]=\frac{1}{2}\int_{\cal M} d^{d+1}x\sqrt{g} \left( \partial_M \Phi \partial^M \Phi +m^2\Phi^2 \right)+ \frac{1}{2}\iint_{\partial {\cal M}} d^d yd^d y'\sqrt{\bar g}_y\sqrt{|\tilde g|}_{y'} \Phi_0{\cal B}\Phi_0 \,+\, {\rm int.} \,
\label{eq:free_action}
\ee
In the boundary term we have included the most general form of the bilinear operator ${\cal B}$. It can contain transverse derivatives and can be non-local.
Using the product notations of Eqs.\,\eqref{eq:cB}\,,\eqref{eq:cb}, we have 
\be
S[\Phi]=\frac{1}{2}\left(\partial_M \Phi \cB \partial^M\Phi +m^2\Phi\cB \Phi + \Phi_0\cb {\cal B} \cb \Phi_0 \right)+ {\rm int.}
\ee

 We can identify the wave operator by applying Green's first identity to the bulk term, giving $\partial_M \Phi \cB \partial^M\Phi +m^2\Phi\cB\Phi = \Phi \cB (-\square +m^2)\Phi +$boundary terms, with
  $\square$ the  Laplacian on $\cal M$.
  
  Finally, the inverse of the  $-\square+m^2$  operator is the  propagator $G$,  \be(-\square+m^2)G(x,x')=\frac{1}{\sqrt{g}_x}\delta^{d+1}(x-x')
  \label{eq:EOM_G}
  \,.\ee 
We denote by $G_D$ the propagator with  boundary condition
  \be
 \quad\quad\quad\quad\quad\quad G_D(x_0,x)=0\,\quad\quad \forall \, x_0\in {\partial\cal M},\, x\in { \cal M} \quad\quad\quad {\rm (Dirichlet)} \label{eq:Dirichlet}
  \ee
  The Neumann boundary condition is a bit more subtle. 
  We denote by $G_N$ the propagator with the boundary condition
  \be
  \quad\quad\quad\quad\quad\quad
\partial^\perp_{x_0}  G_N(x_0,x)= c  \,\quad\quad \forall \, x_0\in {\partial\cal M},\, x\in { \cal M} \quad\quad\quad {\rm (Neumann)}
\label{eq:Neumann}
  \ee
  where 
  \be
  c=\begin{cases}
  0 \quad {\rm if} \quad m\neq0
  \\
 -  \frac{1}{{\rm Vol}(\partial \cal M)} \quad {\rm if} \quad m=0 \,.
  \end{cases}
  \ee
Such a distinction is required for consistency of the boundary condition with the bulk equation of motion. For $m\neq 0 $, 
the volume integral of Eq.\,\eqref{eq:EOM_G} gives, upon use of the divergence theorem, $\int_{\partial \cal M} d^dy \sqrt{\bar g}_y \partial^\perp_{x_0}  G(x_0,x)=0$.\,\footnote{
One uses $\int_{\cal M} d^{d+1}x\sqrt{g}_x G(x,x')= G\cB {\bm 1}_{\cal M}  =\langle\Phi\rangle _{{\bm 1}_{\cal M}} $. Applying the EOM to this equation gives $m^2 \langle\Phi\rangle_{{\bm 1}_{\cal M}} = {\bm 1}_{\cal M} $, thus $\int_{\partial \cal M} d^{d+1}x\sqrt{g}_x G(x,x')=\frac{1}{m^2}{\bm 1}_{\cal M}$. This identity is then used in the integral of Eq.\,\eqref{eq:EOM_G}, resulting in $c=0$ for $m\neq 0$. 
} 
For $m=0$, one has instead $\int_{\partial \cal M} d^dy \sqrt{\bar g}_y \partial^\perp_{x_0} G(x_0,x)=-1$. This is Gauss's law on $\cal M$ \cite{Jackson:100964}.  We will show that the subtlety about the Neumann BC has no consequence for physical results.\,\footnote{The value of $c$ for $m=0$ could more generally be any unit-normalized distribution on $\partial \cal M$. The subsequent results hold in the general case,  here we use  constant $c$ for simplicity. } 

Finally, we compute the discontinuity in the normal derivative of the propagator with endpoints on the boundary. A derivation is given in App.~\ref{app:der}, the result is
\be
\partial^\perp_{x^-_0}  G(x^-_0,x'_0) -
\partial^\perp_{x_0}  G(x_0,x^{'-}_0) 
=  \frac{1}{\sqrt{\bar g}_{y}}\delta^d(y-y') 
\quad\quad\quad {\rm (Discontinuity)}
\label{eq:disc}
\ee
where the minus superscript denotes bulk points in the vicinity of the boundary. 
In the first term of the l.h.s  of Eq.\,\eqref{eq:disc}, the $x'_0$ point is exactly on the boundary while $x^-_0 $ is in the vicinity of the boundary. In the second term $x_0$  is exactly on the boundary while $x^{'-}_0 $ is in the vicinity of the boundary. 
  For a Neumann propagator the second term of the l.h.s of Eq.\,\eqref{eq:disc} is  set by the Neumann boundary condition, giving
\be\partial^\perp_{x^-_0}  G_N(x^-_0,x'_0)=   \frac{1}{\sqrt{\bar g}_{y}}\delta^d(y-y') +c 
\label{eq:DiscN}
\,.\ee 
  This discontinuity equation is  needed for holographic calculations.

 \subsubsection*{Holographic Basis}

We  work out the  holographic decomposition of a scalar field $\Phi$ with Neumann BC \textit{i.e.} that fluctuates on the boundary.
The starting point is to split the field variable into boundary and bulk degrees of freedom,
\be
\int {\cal D}\Phi= \int {\cal D}\Phi_0 ~ \int_{\Phi|_{ {\partial {\cal M}}}=\Phi_0} {\cal D}\Phi_{\rm bulk}\,. \label{eq:measure}
\ee
The  bulk degrees of freedom on the r.h.s  describe the set of fluctuations leaving the boundary value unchanged. These bulk  modes  satisfy thus  Dirichlet condition on the boundary. 
We introduce the ``Dirichlet'' component that satisfies  
\be  {\cal D}\Phi_{\rm bulk}= {\cal D}\Phi_{D}\,\quad\quad {\rm and }\quad\quad \Phi_D|_{\partial \cal M}=0\,.\ee


How is $\Phi$ decomposed into the ``holographic basis'' $(\Phi_0,\Phi_D)$ ? 
Let us  write $\Phi$ in a general form $\Phi(x)=\Phi_0 \cb K(x) +a \Phi_D(x)$ and determine  the $a$  and $K$ functions. 
To proceed we consider the classical value of $\Phi$, $\langle\Phi\rangle_J(x)=\langle\Phi_0\rangle_J \cb K(x) +a \langle\Phi_D\rangle_J(x) $ sourced by a generic source $J$, and  satisfying $\langle\Phi\rangle_J(x)=G_N\cB J(x)$.

If we choose a source $J_D$ that vanishes on the boundary, there is no  contribution from the boundary, $\langle\Phi_0\rangle_{J_D}=0$. We get $\langle\Phi\rangle_{J_D}=\langle\Phi_D\rangle_{J_D}$ and thus $a=1$. If, conversely, we choose a boundary-localized current $J_0(x)|_{\partial {\cal M}}=J_0(x_0)$, the Dirichlet component does not contribute, $\langle\Phi_D\rangle_{J_0}=0$. In that case the field is purely sourced from the boundary,   $\langle\Phi\rangle_{J_0}(x)=\langle\Phi_0 \rangle_{J_0} \cb K(x)$.

To obtain the $K$ function, we need to use  Green's third identity \footnote{ Here Green's third identity takes the form 
 $\langle\Phi\rangle(x)= -  \langle\Phi_0\rangle \cb \partial_\perp  G (x) +  \partial_\perp \langle\Phi_0\rangle \cb G (x) $.}
in the case of a Neumann problem. Using the Neumann boundary condition
Eq.\,\eqref{eq:Neumann}, we get
\be
\langle\Phi\rangle_{J_0}(x)= c' \bar \Phi_0 + \partial_\perp \langle\Phi_0\rangle_{J_0} \cb G_N (x) \label{eq:Grren3N}
\ee
with $c'=-{\rm Vol}({\partial \cal M}) c $. The $ \bar \Phi_0 $ constant is the average value of the field over the boundary \cite{Jackson:100964} and will drop from the calculations. 
To proceed, we define the boundary-to-boundary propagator \be G_0(x_0,x_0')=G_N(x,x')|_{x,x'\in\partial {\cal M}}\,. \ee
We also introduce its inverse $G^{-1}_0$ as 
\be
G^{-1}_0 \cb G_0(x_0,x_0') = \frac{1}{\sqrt{\bar g}}_{y} \delta^{d}(y-y')\,. 
\ee
We evaluate Eq.\,\eqref{eq:Grren3N}  on the boundary, which gives the value $\langle\Phi\rangle_{J_0}(x_0)=\langle\Phi_0\rangle_{J_0}$ we are interested in. 
Using the expression of $\langle\Phi_0\rangle_{J_0}$ we substitute the quantity $\partial_\perp \langle\Phi_0\rangle$ in Eq.\,\eqref{eq:Grren3N}  by making use of the inverse boundary propagator. 
The  $\bar \Phi_0$ constant cancel.\,\footnote{
To see this, first note that $\bar \Phi_0$ can be written as $\bar \Phi_0=G_N\cb \bar J_0$ (where $\bar J_0$ is an averaged boundary source), then  use that $\bar\Phi_0$ is constant in the bulk, which implies 
$\bar \Phi_0=G_N\cb \bar J_0=G_0\cb \bar J_0$. 
 It follows that $\bar\Phi_0 \cb G_0^{-1}\cb G_N= \bar\Phi_0 $, such that $\bar\Phi_0 $ cancels throughout the evaluation and does not appear in Eq.\,\eqref{eq:GJ0}. 
} The result is 
\be
\langle\Phi\rangle_{J_0}(x)= \langle\Phi_0\rangle_{J_0} \cb G^{-1}_0 \cb G_N(x)\,. \label{eq:GJ0}
\ee
It follows that the $K$ function is
\be
K(x_0,x)= G^{-1}_0\cb G_N(x_0,x) \label{eq:Bo2B_def}
\ee
That is, $K$ is the bulk propagator with an endpoint on the boundary, and \textit{amputated} by a boundary-to-boundary propagator. It satisfies $\Phi_0\cb K|_{\partial {\cal M}}=\Phi_0$. The quantity
$K$  is itself typically called the ``boundary-to-bulk propagator''. 

Summarizing, we have determined $K$ and $a$ by considering the expectation value of $\Phi$.  The expression of the bulk field $\Phi$  in the holographic basis is  found to be \be
\Phi(x)=\Phi_0 \cb K(x) +\Phi_D(x) \, \label{eq:hol_basis}
\ee
with $K=G_0^{-1}\cb G_N$. 

In the Neumann case considered here, $\Phi_0$ fluctuates. If instead the field has Dirichlet BC, the decomposition is the same but $\Phi_0$ is not dynamical and corresponds to the boundary  data of the Dirichlet problem.

\subsubsection*{Action in the Holographic Basis}

We plug the obtained expression of $\Phi$ in the $(\Phi_0,\Phi_D)$ basis
into the partition function. This is 
\begin{align}
Z[J] 
  = \int {\cal D} \Phi_0{\cal D} \Phi_D \exp\left[
 - S[\Phi_0\cb K+\Phi_D] -  ( \Phi_0 \cb K+\Phi_D) \cB J  \label{eq:Z_hol}
  \right]\,.
\end{align}
The  action Eq.\,\eqref{eq:free_action} takes the form
\be
S[\Phi]=\frac{1}{2}\left( \partial_M( \Phi_0\cb K+\Phi_D) \cB \partial_M( K \cb \Phi_0+\Phi_D) + \Phi_0 \cb {\cal B} \cb \Phi_0
\right) \,+\,{\rm interactions}\,. \label{eq:free_action1}
\ee
We apply Green's first identity to each of the three bulk terms
\be
\partial_M( \Phi_0\cb K)\cB \partial^M (K \cb \Phi_0)+
2\partial^M \Phi_D\cB \partial_M (K \cb \Phi_0 ) +
\partial_M \Phi_D\cB \partial^M \Phi_D \label{eq:free_action2}
\ee
For the middle one, one applies the identity such that the obtained Laplacian acts on $K\cb\Phi_0$. 
The action becomes
\begin{align}
S[\Phi]=& \frac{1}{2}\left(  \Phi_0\cb K \cB (-\square+m^2) K \cb \Phi_0 + 2 \Phi_D \cB (-\square+m^2) K\cb \Phi_0+  \Phi_D \cB (-\square+m^2) \Phi_D
\right) \nn \\ &
+ \frac{1}{2}\left(
\Phi_0 \cb K \cb \partial_\perp K \cb \Phi_0
+ 2 \Phi_D \cb  \partial_\perp K \cb\Phi_0  + 
\Phi_D \cb \partial_\perp \Phi_D \right)
\,+\,{\rm int.}\, \label{eq:free_action3}
\end{align} 
The first two  terms of the first line vanish because  $(-\square+m^2)K=0$ in the bulk. The last two terms of the second line vanish because of the Dirichlet condition, $\Phi_D|_{\partial{\cal M}}=0$. We also used that $\partial_\perp \Phi_0=0$ and $\Phi_0\cb K|_{\partial M}=\Phi_0$. 

The remaining terms are $\Phi_D\cB(-\square +m^2)\Phi_D+ \Phi_0 \cb  \partial_\perp K \cb \Phi_0$. To evaluate the term with $\partial^{\perp} K=\partial_\perp G \cb G^{-1}_0 $ we note that
the point of $G$ on which the derivative does not act belongs to  the boundary. Therefore this is a derivative of the $\partial^\perp_{x^-_0} G(x^-_0,x_0)$ form, which requires the use of the discontinuity equation Eq.\,\eqref{eq:DiscN}. We obtain
\be
\Phi_0 \cb  \partial_\perp K \cb \Phi_0 = \Phi_0 \cb  G^{-1}_0 \cb \Phi_0 + c'\bar J_0 \cb \Phi_0 \,.\ee 
The extra term $\bar J_0\cb \Phi_0$ (with $\bar\Phi_0=G_0 \cb \bar J_0$) present in the massless case amounts to a shift of the  boundary value of the generic $J$ source and can thus be absorbed by a redefinition of $J$.
 This  explicitly shows that this term has no physical relevance.

Combining all the pieces, we find that the partition function  of a scalar field supported on an arbitrary background  with boundary and  fluctuating on the   boundary (\textit{i.e.} Neumann BC) is
\begin{empheq}[box=\widefbox]{align}
Z[J] 
   & =  \nn  \\    \int {\cal D}  &   \Phi_0{\cal D}    \Phi_D   \exp \Bigg[
   -\frac{1}{2} \Big(\Phi_D\cB (-\square +m^2)\Phi_D 
+  \Phi_0\cb\left(G^{-1}_0+{\cal B}\right)\cb \Phi_0
\Big) \nn 
\\ \label{eq:free_action_final}
 &\quad\quad\quad\quad\quad\quad  -  (\Phi_0 \cb K+\Phi_D)\cB J
 +{\rm interactions}\Bigg] 
\end{empheq}
The holographic action in Eq.\eqref{eq:free_action_final} is  diagonal. 
The Dirichlet modes have canonical kinetic term. We often  equivalently refer to them as ``bulk modes''. 
On the other hand, the boundary degree of freedom has a nontrivial, generally nonlocal self-energy. 
 While the ${\cal B}$ piece is a generic surface term, the  $G^{-1}_0$ piece of this holographic self-energy reflects the fact that the boundary degree of freedom knows about the bulk modes.

\subsubsection*{Propagator}

Perturbative amplitudes are  obtained by taking $J$ derivatives of Eq.\,\eqref{eq:free_action_final} or of related quantities such as the generator of connected correlators. 
In particular, we can compute the propagator
\be
\langle\Phi(x)\Phi(x')\rangle=\frac{1}{\sqrt{g}_x\sqrt{g}_x'}\frac{\delta^2}{\delta J(x)\delta J(x')}\log Z[J]\,.
\ee
We denote it as $\langle\Phi(x)\Phi(x')\rangle=G^{\cal B}_{N}(x,x')$ since, as verified below, it is the Neumann propagator in the presence of the boundary term ${\cal B}$. 
Defining the inverse of the boundary operator as 
\be\left[G^{-1}_0+{\cal B}\right]^{-1}\left(G^{-1}_0+{\cal B}\right)(x_0,x_0') = \frac{1}{\sqrt{\bar g}}_{x_0} \delta^{d}(y-y')\,,
\ee
we obtain
\begin{empheq}[box=\widefbox]{equation}
G^{\cal B}_N(x,x')= G_D(x,x') +K \cb \left[G^{-1}_0+{\cal B}\right]^{-1} \cb K(x,x') 
\, \label{eq:prop_hol_gen} 
\end{empheq}

Let us verify that $G^{\cal B}_N$ satisfies a Neumann boundary condition. To do so we act with $\partial_\perp$ and convolute with a boundary field $\hat\Phi_0$ \textit{i.e.} we act on Eq.\,\eqref{eq:prop_hol_gen} with $\hat\Phi_0 \cdot \partial_\perp $. The action of $\partial_\perp$ on the first term is evaluated using the discontinuity equation Eq.\,\eqref{eq:DiscN}. The second term is evaluated using 
 Green's third identity, giving $\hat\Phi_0 \cb \partial_\perp G_D(x)=-\hat\Phi (x)$. One gets that $ \hat \Phi_0 \cb (\partial_\perp + {\cal B} ) G^{\cal B}_N(x_0,x)=0$. This is true for any $\hat \Phi_0$, therefore 
\be
 (\partial_\perp + {\cal B} ) G_N(x_0,x) =0 \quad\quad \forall \, x_0\in {\partial\cal M},\, x\in { \cal M}
\ee
which is the Neumann boundary condition in the presence of the boundary term. 

We have obtained a ``holographic'' representation of the bulk Neumann propagator in terms of Dirichlet and boundary-to-bulk propagators, valid in any background. We sometimes refer to it as the ``Neumann-Dirichlet'' identity. The two terms on the r.h.s of Eq.\,\eqref{eq:prop_hol_gen} correspond to the propagation of Dirichlet modes and to the propagation of the boundary degree of freedom --- connected to the bulk endpoints by a boundary-to-bulk propagator. 
We can notice that the boundary term takes the form of a Dyson resummation of the boundary operator ${\cal B}$. The effect of the boundary action in $G^{\cal B}_N$ can be recovered by dressing  $G_N$ with ${\cal B}$  boundary insertions. We can also see that whenever ${\cal B}$ becomes infinite, the $G_N^{\cal B}$ propagator reduces to the Dirichlet one, \textit{i.e.}
\be
G_N^{\cal B} \Big|_{{\cal B}\to\infty} = G_D \,. 
\ee

\section{Review: Holography in a  Warped Background}
\label{se:warped_holography}

We discuss aspects of  the holographic formalism for a scalar field in the case of a generic warped   background with flat boundary. Since this setup has been studied to death for two decades,  this section can be considered as mostly review with bits of less known results. This section also provides sanity checks of the general results from  Sec.~\ref{se:gen}. 

\subsection{Warped Background}
\label{se:warped_formalism}

We consider a $d+1$-dimensional Lorentzian  conformally-flat background with  $d$-dimensional Poincaré symmetry along the constant-$z$ slices. 
The background metric takes  the  form
\begin{align}	ds_{\rm warped}^2	=\g_{MN} \, dx^{M}dx^{N}=\frac{1}{\rho^2(z)} \left( \eta_{\mu\nu}dy^\mu dy^\nu+dz^2 \right) \,\label{eq:metric:general}\end{align}
 with a boundary at $z=z_0$ constraining $ z\geq z_0 \,. $ 
 \footnote{
 The normal, outward pointing component of the vielbein obtained from the metric is
 $e^M_\perp=-\rho(z) \delta^M_z\hat u_z $. Hence the normal derivative is $\partial_\perp =-\rho_0 \partial_z $. 
 }

The quadratic action in Lorentzian signature  is
\begin{align}	S &= - \int_{\cal M} d^{d+1}x \sqrt{|\g|}e^{-\varphi}	\left(		\frac{1}{2} \partial_M\Phi \partial^M\Phi+
\frac{1}{2} m^2 \Phi^2		\right)	+ \frac{1}{2}\iint_{\partial {\cal M}} d^d x \sqrt{\bar\g}e^{-\varphi_0}
\Phi_0 B\left[\partial_\mu\partial^\mu\right] \Phi_0+
\cdots \, \label{eq:SPhi}\end{align}
where $\varphi(z)$ is a dilaton background. 
This action covers many cases considered in the literature (for a sample, see \textit{e.g.} \cite{Karch:2006pv,Gursoy:2007cb,Gursoy:2007er, Batell:2008zm, Cabrer:2009we,vonGersdorff:2010ht}). 
For  $\rho(z)=kz$, $z_0=0$, $\varphi=0$, the  background reduces to the AdS$_{d+1}$ Poincaré patch with curvature $k$.  

We Fourier transform along the $z$ slices, using $\Phi(x^M)=\int \frac{d^{d}p}{(2\pi)^d} e^{ip^\mu y_\mu }\Phi_p(z)$.
The wave operator takes the form 
\be
{\cal D} = -\square+m^2= -\rho^{d+1}e^\varphi \partial_z\left(\rho^{1-d} e^{-\varphi}\partial_z\right)+\rho^2 p^2+m^2\,
\ee
with $p = \sqrt{\eta_{\mu\nu}p^\mu p^\nu}$.

 The physical $p^2$ takes both signs in Lorentzian signature. With the mostly plus metric, we have $p^2>0$ for spacelike momentum, $p^2<0$ for timelike momentum.
In the free theory, $p^2$ is made slightly complex to resolve the non-analyticities arising  for timelike momentum. This corresponds to the inclusion of an infinitesimal imaginary shift $p^2+i\epsilon$, $\epsilon\rightarrow 0$. 
$\epsilon>0$ is consistent with causality and  defines  the Feynman propagator.
The $i\epsilon$ shift will often be left implicit in our notations.

The homogeneous solutions of ${\cal D}$ are denoted $\f$, $\h$ with ${\cal D} \f=0$, ${\cal D} \h=0$. The Wronskian of these solutions, $W=\f \h'-\f'\h$, satisfies
\be
W(z)=C \rho^{d-1}e^{\varphi}\, \label{eq:W_gen}
\ee
where the only unknown is the overall  constant $C$, which depends on the choice of $\f,\h$ solutions.

\subsubsection*{Regularity condition}

We specify a  regularity condition  on a hypersurface  away from the boundary, at $z=z_1>z_0$.  On this surface we assume that the $\h$ solution blows up while $\f$ is the regular solution. The precise condition at $z_1$ is 
\be
\frac{\f(z)}{\h(z)}\xrightarrow{z\to z_1} 0\,,\quad \frac{\f'(z)}{\h'(z)} \xrightarrow{z\to z_1} 0 \, . \label{eq:RC}
\ee
For timelike momentum,  taking  $\f$ as an outgoing wave and $\h$ as an ingoing wave, the condition Eq.\,\eqref{eq:RC} amounts to a outgoing wave condition. This condition is assumed throughout this section.

\subsubsection*{Asymptotics}

In any region where the condition
\be
|p^2| \gg m^2 + \left(\partial_z\log W\right)^2 + \partial^2_z\log W \label{eq:asymptotic_cond}
\ee
is verified, the solutions of the EOM admit the asymptotic behaviour
\be
\sqrt{W(z)} e^{\sqrt{p^2}z}\,,\quad \sqrt{W(z)} e^{-\sqrt{p^2}z}\,. \label{eq:asymptotic_fsW}
\ee
This can be shown by performing a field redefinition $\Phi=\tilde \Phi \rho^{\frac{d-1}{2}e^{\varphi/2}} $ in the action.

\subsubsection{Propagators}

\label{se:props}

All the propagators satisfy the bulk equation of motion
\be
{\cal D} G(x,x')=-i \rho^{d+1}e^\varphi\delta^{d+1}(x-x')\,.  \label{eq:eom_G}
\ee
We define $\f(z_0)=\f_0$, $\partial_z \f|_{z_0}=\f'_0 $ and similarly for $\h$, $\rho$ and $\varphi$. 
The  propagators take the following form.

\noindent\textit{Neumann propagator}:
\be
G^{B}_N(p;z,z')=\frac{i}{C}\left(  
\frac{\rho_0 \h'_0+ {B} \h_0}{\rho_0f'_0+{B} \f_0} \f(z_<)  -\h(z_<) 
 \right)  \f(z_>)\,
 \label{eq:propa_B2B_N}
\ee
\textit{Boundary-to-boundary propagator}:
\be
G_0^B(p)\equiv G^{B}_N(p;z_0,z_0)=i\rho^d_0e^{\varphi_0} \frac{\f_0}{\rho_0\f'_0+{B} \f_0}
 \label{eq:propa_b2b}
\ee
\textit{(Amputated) Boundary-to-bulk propagator}:
\be
K(p,z)= \rho^{d}_0e^{\varphi_0} \frac{\f(z)}{\f_0} \,
\ee
\textit{Dirichlet propagator}:
\be
G_D(p;z,z')=\frac{i}{C}\left( 
\frac{\h_0}{\f_0} \f(z_<)  - \h(z_<) 
 \right)  \f(z_>)\,
 \label{eq:propa_B2B_ND}
\ee
We also have  $G_0=G_0^{B=0}$. 
In some of the above expressions, the Wronskian at $z_0$ (Eq.\,\eqref{eq:W_gen}) has been used.

In App.~\ref{app:warped_prop} we show that these propagators satisfy the general relations  derived in Sec.~\ref{se:gen}. 
For example one gets by direct calculation the relation
\begin{align}
G_N^B(p;z,z')-G_D(p;z,z') & =K(p,z)K(p,z') \left(G^{-1}_0-i \rho_0^{-d}e^{-\varphi_0}B\right)^{-1}
\end{align}
which verifies Eq.\,\eqref{eq:prop_hol_gen} upon translating to Euclidian conventions.

\subsubsection{Holographic Action}

\label{se:action_warped}

The holographic basis is \be \Phi(p,z)=\frac{f(p,z)}{f_0(p,z_0)}\Phi_0(p,z_0)+\Phi_D(p,z)
\label{eq:holbasis_AdS}
\,.\ee
We remind that $f$ is the regular solution of the EOM  away from the boundary, see Eq.\,\eqref{eq:RC}. 
We   then derive the holographic action. 
We  introduce the spectral representation of the Dirichlet component,
\be
\Phi_D(p;z)=\sum_\lambda \phi^\lambda_D(p) f^\lambda_D(z)
\ee
where each Dirichlet mode $f^\lambda_D$ satisfies the EOM ${\cal D} f^\lambda_D(z) =0$ with $p^2=-m^2_\lambda$. The mode distribution may  be either discrete or continuous depending on the background, we denote the summation  by  $\sum_\lambda$ either way.\,\footnote{See \textit{e.g.}
\cite{Karch:2006pv,Gursoy:2007cb,Gursoy:2007er, Batell:2008zm, Cabrer:2009we,vonGersdorff:2010ht} for some examples of warped backgrounds  featuring either discrete or continuum spectra beyond pure AdS.} 
The Dirichlet modes  satisfy the orthogonality and completeness relations 
\be
\int dz \rho^{1-d}e^{-\varphi}  f_D^\lambda(z) f_D^{\lambda'}(z) = \delta_{\lambda\lambda'}\,,\quad 
\sum_\lambda  f_D^\lambda(z) f_D^\lambda(z') = \rho^{d-1}e^{\varphi}  \delta(z-z')\,.
\ee

Using orthonogonality of the Dirichlet modes and the other properties of the  holographic variables previously derived, we get the partition function for a scalar fluctuating on both bulk and boundary of the warped background,
\begin{align}
Z[J] 
   & =  \nn  \\    \int {\cal D}  &   \Phi_0{\cal D}    \phi_D   \exp \Bigg[
   \frac{i}{2}\int\frac{d^dp}{(2\pi)^{d}}\left( -\sum_\lambda\,\phi^\lambda_D ( p^2 +m_\lambda^2)\phi^\lambda_D 
+  \sqrt{\bar\g}e^{-\varphi_0}\Phi_0\left(\rho_0\frac{f_0'}{f_0}+B\right) \Phi_0 \right)
 \nn 
\\ \label{eq:free_action_final_warped}
 &\quad\quad\quad\quad\quad  -i \int \frac{d^dp}{(2\pi)^{d}} \int dz \sqrt{|\g|}e^\varphi  \left( \frac{f(z)}{f_0} \Phi_0   +   \Phi_D(z)  \right)J(z)
 +{\rm interactions}\Bigg]
\end{align}

This is a consistent with  the main formula Eq.\,\eqref{eq:free_action_final} upon translating into Euclidian conventions. 
The $p$-dependence of the fields is left implicit, it is $\Phi \varphi=\Phi(p)\varphi(-p)$ for each  monomial of the quadratic action.

\subsubsection{Holographic Mixing}

\label{se:mixing}

In the class of warped backgrounds considered here, the holographic self-energy may feature a pole indicating the existence of a $d$-dimensional free field,   here denoted $\tilde \Phi_0$. This is the mode satisfying the bulk  EOM with $p^2=m_0^2\ll m^2$, which can always exist   upon appropriate tuning of the boundary terms in $B$ (see \textit{e.g.} discussion in \cite{Fichet:2019owx}). 
In the presence of this mode, a variant of the holographic basis is to let $\Phi(z;p)=\tilde \Phi_0(p) K(z,p^2=m^2_0)+\Phi_D(z;p)$ (see Refs.\,\cite{Batell:2007jv,Batell:2007ez} for the original proposal in the case of a slice of AdS).  In this basis all the degrees of freedom are free fields but the resulting  holographic action is nondiagonal. The action contains a cross term between
$\tilde \Phi_0$ and $\Phi_D$, taking the form
\be
\int dz \rho^{1-d}e^{-\varphi} \,\tilde \Phi_0(p) (p^2-m^2_0) K(p;z) \sum_\lambda f(z)_D^\lambda\phi^\lambda_D(-p)
\ee
upon  integration by parts. This term induces both 
  kinetic and mass mixing between the $\tilde \Phi_0$ and $\phi^\lambda_D$ fields --- this is how the boundary degree of freedom knows about the bulk modes in this basis.   The interpretation of this ``holographic mixing'' is discussed in details in \cite{Batell:2007jv,Batell:2007ez}.
  In our present work the holographic action is instead exactly diagonal, to the price of having a nontrivial self-energy for the boundary degree of freedom.

\subsection{Properties of AdS Propagators}

\label{se:AdS_properties}

We review the propagators in  AdS background (with regularized boundary)  from the viewpoint of our formalism.

In AdS we have
$\rho(z)=kz $
with  $k$ the AdS curvature. We take $g_{MN}\equiv g^{\rm AdS}_{MN}$ throughout this subsection. 
The scalar bulk mass is written as \be m^2=\Delta(\Delta-d)k^2\,. 
\label{eq:AdS_bulkmass}
\ee 
Defining $\Delta_+=d-\Delta_- > \Delta_- $,  the solutions to the wave equation ${\cal D}\Phi=0$ near the AdS boundary ($z=0$) are\,\footnote{In the notation of Sec.~\ref{se:props}, the solutions satisfying the regularity condition Eq.\,\eqref{eq:RC} are
$\f(z)=z^{d/2}K_{\Delta_+-d/2}(\sqrt{p^2}z)$, $\h(z)=z^{d/2}I_{\Delta_+-d/2}(\sqrt{p^2}z)$.  $\f$ and $\h$ match respectively onto the $-$ and $+$ asymptotics  discussed here.
}
\be
\Phi(x,z)=z^{\Delta_+}(A_+(x)+ O(z^2))+z^{\Delta_-}(A_-(x)+ O(z^2))
\label{eq:Phi_exp_standard}
\ee
For $\frac{m^2}{k^2}>-\frac{d^2}{4}+1$, the $-$ modes are non-normalizable. For $-\frac{d^2}{4}<\frac{m^2}{k^2}<-\frac{d^2}{4}+1$, both $+$ and $-$ modes are normalizable.
The modes can be selected by imposing appropriate condition on the AdS boundary, \be (z\partial_z-\Delta_\pm) \Phi|_{z=0}=0 
\label{eq:BC_st}
\ee

However in AdS calculations (\textit{e.g.} for Witten diagrams and  AdS/CFT), it is often necessary to regulate the AdS boundary, which amounts to truncate spacetime such that $z>z_0$ with $z_0\neq 0 $. Even if $z_0$ is infinitesimal, the propagators reflect on the boundary and take the form given in section\,\eqref{se:props} instead of being simply $-iC^{-1}g(z_<)f(z_>)$. 
Moreover, in our formalism  we have standard Neumann and Dirichlet BC on the regulated boundary. How does this language compare to the BC Eq.\,\eqref{eq:BC_st} ?
To understand how our regulated formalism matches onto the unregulated one, we start from our holographic basis in AdS (given in Eq.\,\eqref{eq:holbasis_AdS}) and take the near-boundary asymptotic values for $K\propto \frac{f}{f_0}$ and $f^\lambda_D$.  The result is
\be
\Phi=\Phi_0 \left(\frac{z}{z_0}\right)^{\Delta_-}\left(1+ O(z^2)\right) +z^{\Delta_+}\left(A_+(x)+ O(z^2)\right) 
\label{eq:holbasis_AdS_as}
\ee
for any $\Delta$. 
In case of Neumann BC, the  modes that dominate near the boundary are from the first term \textit{i.e.} scale as $z^{\Delta_-}$. In case of a Dirichlet BC, $\Phi_0$ does not fluctuate, and the remaining modes from the second term \textit{i.e.} scale  as $z^{\Delta_+}$.  Hence the $-$ modes correspond to Neumann BC and the $+$ modes correspond to Dirichlet BC.

\subsubsection{AdS Bulk Propagators}

The Neumann/Dirichlet$\equiv \mp$ identification implies that the regulated bulk propagators satisfy
\be
 G_{\Delta=\Delta_-}=G_N \,,\quad \quad G_{\Delta=\Delta_+}=G_D\,
\ee
when $z_0\to 0$. 
Only the second propagator exists if $\frac{m^2}{k^2}>-\frac{d^2}{4}+1$ because $-$ modes are not normalizable in that case.  

We can readily apply the general results of section \ref{se:gen}. 
The general relation  between $G_D$ and $G_N$ obtained in Eq.\,\eqref{eq:prop_hol_gen}  becomes here
\be
G_{\Delta=\Delta_-}(p;z,z')=G_{\Delta=\Delta_+}(p;z,z')+K(p,z)K(p,z')G_{\Delta=\Delta_-}(p;z_0,z_0)\Big|_{z_0\to 0}\,
\label{eq:prop_holo_AdS}
\ee
in Fourier space. The $K$ are computed from $G_N$, and thus here from the $G_{\Delta=\Delta_-}$ propagator. 
The Neumann-Dirichlet identity Eq.\,\eqref{eq:prop_holo_AdS} has been known in the AdS literature see \textit{e.g.} \cite{ Hartman:2006dy,Falkowski:2008yr,Giombi:2018vtc}, but was not given a deeper explanation. Our approach shows that such a relation is intrinsic to the holographic formalism and exists for any background geometry.

We can also comment on the conformal spectral representation of AdS propagators (see \cite{Leonhardt:2003qu, Cornalba:2007fs, Paulos:2011ie}, and \cite{Costa:2014kfa, Liu:2018jhs,Carmi:2018qzm, Meltzer:2019nbs, Costantino:2020vdu,Fichet:2021pbn} for some applications).
The $\Delta_+$ propagator can be written as
\be
G_{\Delta=\Delta_+} = k^{d-1}\int_{\mathbb{R}} d\nu P(\nu,\Delta) \Omega_\nu
\ee
where $\Omega_\nu$ is a known harmonic kernel and $P(\nu,\Delta)=\frac{1}{\nu^2+(\Delta-d/2)^2 }$.
In contrast \cite{Carmi:2018qzm}, the $\Delta_-$ propagator takes the form
\be
G_{\Delta=\Delta_-} = k^{d-1}\int_{\mathbb{R}+C_u+C_d} d\nu P(\nu,\Delta) \Omega_\nu
\label{eq:prop_spec_M}
\ee
where $C_{u,d}$ are contours wrapping the $\nu=\pm i(\Delta-d/2)$ clockwise and counterclockwise respectively. 
We verified explicitly using Ref.\,\cite{Costantino:2020vdu} that the harmonic kernel satisfies
\be
i\frac{2\pi}{\nu k} \Omega(p;z,z')  = K(p,z)K(p,z')G_{\Delta=\Delta_-}(p;z_0,z_0)\Big|_{z_0\to 0}\,.
\ee
This implies that Eq.\,\eqref{eq:prop_spec_M} is equivalent to Eq.\,\eqref{eq:prop_holo_AdS}. The $\mathbb{R}$ integral gives $G_D$, while the $C_{u,d}$ integrals give the $KKG$ term.

\subsubsection{AdS Boundary-to-Bulk Propagators}

\label{se:AdSB2b}

The AdS propagators need to properly encode the regulated AdS boundary
in order to obtain the correct  CFT correlators, see \textit{e.g.} \cite{Freedman:1998bj, Klebanov:1999tb}.  Such a regularization is built-in in our approach, since all propagators are evaluated in the presence of the regulated boundary without  any approximation. 

Let us work out the CFT 2pt function in our formalism. We start from our general definition $K(x) = G_N \cb [ G_0 ]^{-1}$. One can easily evaluate $\partial_\perp K$ using  the discontinuity equation (as in section \ref{se:gen}) giving  $\partial_\perp K= [ G_0 ]^{-1} $. The subsequent evaluations are best done in momentum space because the final result may have contact terms, that we will need to take into account in subsequent calculations. 
 We have
\be
\partial_\perp K =- \frac{1}{\sqrt{\bar \g}} \int \frac{d^dp}{(2\pi)^d}\frac{kz \partial_z\left(z^{d/2}K_{\Delta-\frac{d}{2}}(pz_0)\right)_{z=z_0} }{z_0^{d/2}K_{\Delta-\frac{d}{2} }(pz_0) }
\ee
which matches the usual  result \cite{Freedman:1998bj}. This expression is exact and is invariant under $\Delta_+\leftrightarrow \Delta_-$. 

We then focus, as customary, on the large chordal distance limit $\zeta(x_0,x_0')\gg\frac{1}{k^2}$ (see App.~\ref{app:warped_prop} for position space expressions), which amounts to the
$(y-y')^2\gg z^2_0$ limit, and thus to $p^2 z z_0 \ll 1 $ in momentum space. For any value of $\Delta$, the leading term scales as $p^{2\Delta_+-d}$, where $\Delta_+>\Delta_-$.
Evaluating the Fourier transform and plugging the result into the Euclidian boundary action $S^E=\frac{1}{2}\Phi_0 \cb \partial_\perp K \cb \Phi_0+\ldots$
gives the  asymptotic result 
\begin{align}
& S^E_{\rm 2pt}= 
\frac{1}{2}\int d^dy d^dy' \sqrt{\bar \g}  \left[
 \frac{{\cal C}_{\Delta_+}  \eta_{\Delta_+}}{(y-y')^{2\Delta_+}}+
 \Delta_- k \delta^{d}(y-y')
 + O\left(\zeta^{-\Delta_+-1}\right)\right]\Phi_0(y)\Phi_0(y') 
  \label{eq:S2pt}
\end{align}
  with
 \be  {\cal C}_\Delta=  \frac{(2\Delta-d)\Gamma(\Delta)}{\pi^{d/2}\Gamma(\Delta-\frac{d}{2})}\,,\quad\quad\quad  \eta_\Delta = k z^{2\Delta-d}_0
\ee
The nonlocal part of $S^E_{2{\rm pt}}$ gives rise to the CFT 2pt function with the correct normalization factor ${\cal C}_\Delta$. The remaining dimensionful factor $\eta_\Delta$ (or $\sqrt{\bar \g} \eta_\Delta$ ) can be absorbed into the normalization of the CFT operators. 
We  see that, in addition to the standard nonlocal part, there is  a contact term. 

 If one tries instead to evaluate $[G_0]^{-1}$ in position space 
 by taking the $\zeta\gg \frac{1}{k^2}$ limit   and using a conformal integral to perform the inversion, one gets precisely the nonlocal part of Eq.\,\eqref{eq:S2pt} but not the contact term. One could also try to evaluate the whole $G_N \cb [ G_0 ]^{-1}$ expression in position space, by taking the $\zeta\gg \frac{1}{k^2}$ limit for $G_N$ and $G_0$. Such an approximation gives a divergent result for $z\to z_0$. We trace back these discrepancies to the fact that the $\zeta\gg \frac{1}{k^2}$ limit, which requires large $|y-y'|$,  does not commute in general with the convolutions in $y$-space---which involve integrals over arbitrary values of $y$. 
 We conclude that the complete result is Eq.\,\eqref{eq:S2pt}. In the following we will see that the contact term in Eq.\,\eqref{eq:S2pt} is needed to ensure consistency of results with  conformal symmetry.

\subsection{On Boundary Flux and Unitarity}
\label{se:flux}

Here we present an elementary application of the  formalism of Sec.~\ref{se:warped_formalism} in the case where $\Phi$ fluctuates on the boundary (\textit{i.e.} Neumann BC).
We consider the generic warped background of Eq.\,\eqref{eq:metric:general} and assume that
some isolated states localized on the boundary  collide to form $\Phi$ states.  What is the flux of $\Phi$ modes emitted from the boundary?  To obtain the answer we will use a unitarity cut. 

\begin{figure}[t]
\centering
	\includegraphics[width=0.7\linewidth,trim={1cm 8cm 0cm 3cm},clip]{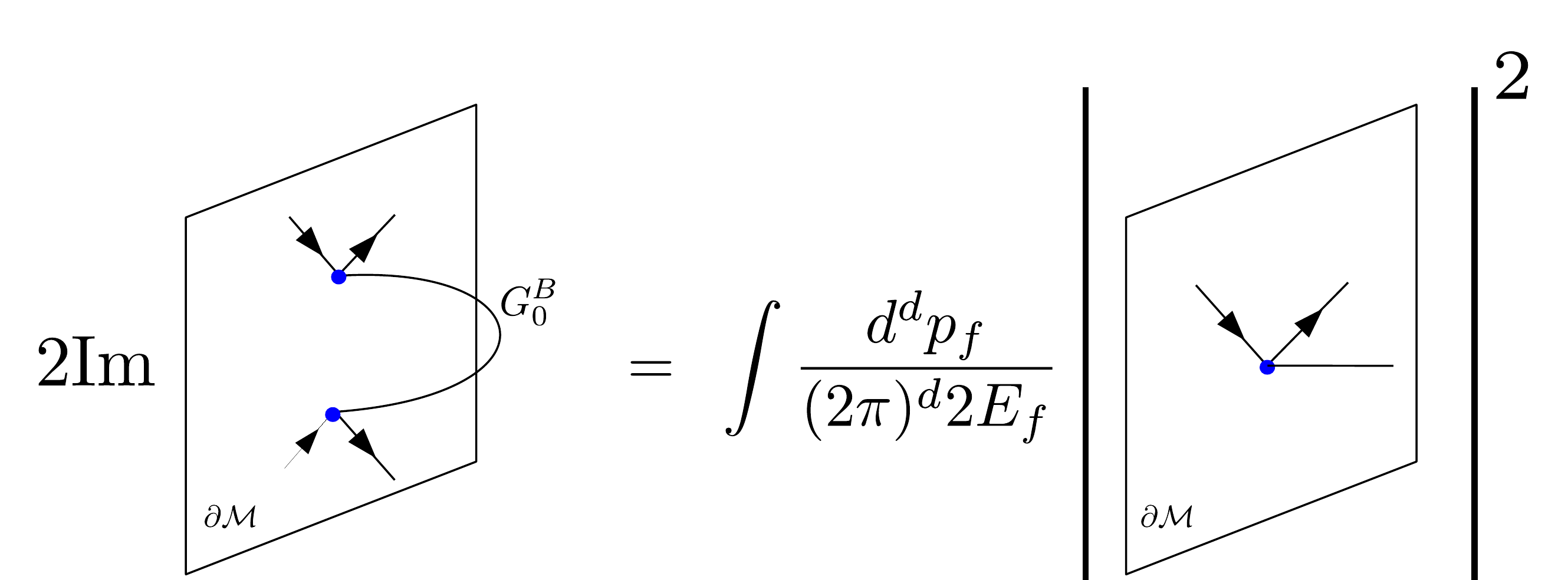}
\caption{ The $i{\cal A}_{\bar\Psi\Psi\to \bar\Psi\Psi}$ amplitude and its unitarity cut. 
\label{fig:Flux}}
\end{figure}

We denote the boundary-localized states by $\Psi, \bar\Psi $ and consider the $2\to 2$ timelike process where  $\Phi$ is produced from a boundary-localized interaction, \textit{e.g.} a $\bar\Psi\Psi \Phi|_{\partial \cal M}$ coupling. The corresponding scattering amplitude is denoted by $i{\cal A}_{\bar\Psi\Psi\to \bar\Psi\Psi}$ and is represented in Fig.\,\ref{fig:Flux}.  
This amplitude is proportional to the boundary-to-boundary propagator given in Eq.\,\eqref{eq:propa_b2b},
\be
i{\cal A}_{\bar\Psi\Psi\to \bar\Psi\Psi} =-\alpha G^B_0(p)
\ee
where $\alpha$ is a positive coefficient encoding couplings and other overall constants.

The production rate of bulk modes, $\sigma_{\bar\Psi\Psi\to \Phi}$, can be obtained from a unitary cut of the $i{\cal A}$ amplitude. 
The unitarity cut amounts to take the imaginary part of ${\cal A}$, one gets 
\be
\sigma_{\bar\Psi\Psi\to \Phi}= 2{\rm Im}{\cal A}   = 2\alpha\, {\rm Im}\left[
i G^B_0(p)
\right] =2 \alpha \rho^d_0e^{\varphi_0}
\frac{\rho_0}{|\rho_0\f'_0+{B} \f_0|^2 }{\rm Im} \left[ \f_0^\dagger \f'_0\right]
\ee
We then evaluate ${\rm Im} \left[ \f_0^\dagger \f'_0\right]$.
By definition, $\f$ is the  outgoing wave solution. The conjugate $\f^\dagger$ must contain the ingoing wave solution, $\f^\dagger=\h+\ldots$ where we chose a unit coefficient for $\h$ without loss of generality. 
It follows that  the imaginary part takes the form ${\rm Im} \left[ \f_0^\dagger \f'_0\right]=\frac{1}{2i}(\f'_0 \h_0-\f_0 \h'_0)$ and is thus computed by the Wronskian Eq.\,\eqref{eq:W_gen}.  Because of $\f^\dagger=\h+\ldots$ one has  $W=-W^\dagger$, thus the Wronskian is imaginary. Thus the Wronskian can be written as 
$W=-i c \rho^{d-1} e^{\varphi}$ with $c\in {\mathbb R}$.

What is the sign of $c$ ? Assuming that the EOM is regular everywhere, the Wronskian cannot be zero. Thus ${\rm Im} W$ has a definite sign  everywhere, encoded into $c$. To determine the sign of $c $ we consider the large $p$ asymptotic limit obtained in Eq.\,\eqref{eq:asymptotic_fsW}, with $f\sim  e^{ip z}$, $g\sim  e^{-ip z}$. The corresponding asymptotic Wronskian is $-2i p$, therefore $c>0$.

Putting the pieces together we  obtain that the imaginary part is given by 
${\rm Im} \left[ \f_0 \f'^\dagger_0\right]=\frac{1}{2}  c \rho^{d-1}_0 e^{\varphi_0} $. 
As a result the production rate of  bulk modes is given by
\be
\sigma_{\bar\Psi\Psi\to \Phi}=
\alpha c \frac{\rho^{2d}_0e^{2\varphi_0}}{|\rho_0\f'_0+{B} \f_0|^2 } \,.
\ee
This is a very simple, exact formula for the flux  emitted from the boundary.
We  see that only the regular solution near the boundary is needed to determine it. 
We can also notice that the $c>0$ condition derived above ensures  that unitarity is respected.

\subsection{Summary}

While this section is mostly a review, there are lessons to take away. 
We have shown that for the warped background Eq.\,\eqref{eq:metric:general}, 
even in the presence of a dilaton background, the holographic action and the  propagators take simple, explicit expressions, making holographic calculations in this class of background essentially as simple as in the Poincar\'e patch of AdS. We used this background to perform checks of the general formalism of Sec.~\ref{se:gen}.  

Going back to AdS, we have cast  a new light on an elementary aspect of the  propagators, the Neumann-Dirichlet identity Eq.\,\eqref{eq:prop_holo_AdS}, which is understood to be an intrinsic property of the holographic formalism, and not as  a specificity of AdS.
We have also verified that our approach gives the CFT 2pt function with correct normalization and we have  taken into account the 2pt contact term. 

Finally we have derived a simple general formula for the flux of bulk modes emitted from the boundary by using a unitarity cut on a boundary $2\to2$ exchange diagram.

\section{Quantum Pressure on a Brane in Warped Background}

\label{se:Pressure}

Throughout the rest of the paper,  our focus is essentially on integrating out  bulk modes at the quantum level.
In the present section we consider free bulk modes and their effect on the boundary. More precisely, we assume that space is subdivided by a codimension-$1$ interface or domain wall, here referred to as brane. Our aim is to derive  the pressure on the brane induced by the quantum fluctuations of the field.
Apart from formal interest this setup is also relevant because it serves in braneworld  models, where the motion of the brane in the bulk is relevant for cosmology (see for example \cite{Kraus:1999it,Hebecker:2001nv,Langlois:2002ke}).

\subsection{Preliminary Observations}

\begin{figure}[t]
\centering
	\includegraphics[width=0.7\linewidth,trim={5cm 5cm 5cm 4.5cm},clip]{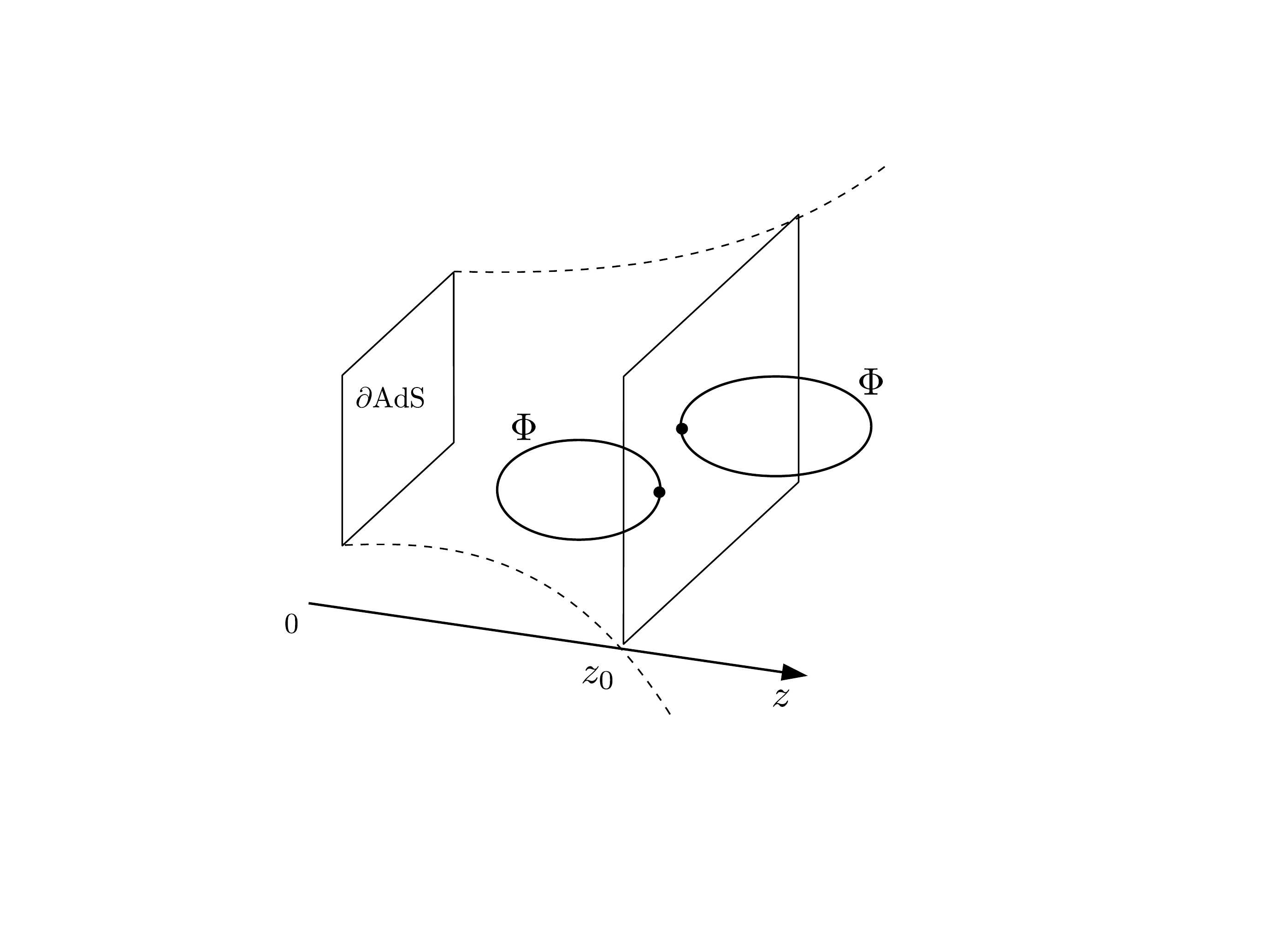}
\caption{ A brane (\textit{i.e.} an interface or domain wall) along the $z=z_0$ slice in the Poincaré patch of AdS$_{d+1}$. Loops of the scalar field $\Phi$ in each region induce a finite pressure on the brane. 
\label{fig:AdS_Brane}}
\end{figure}

We consider a free scalar $\Phi$  supported on the class of warped background discussed in section~\ref{se:warped_holography}. The brane is along the  $z=z_0$ slice as shown in Fig.\,\ref{fig:AdS_Brane}.  
Our aim is to investigate  the pressure on the brane induced by fluctuations of $\Phi$. The field may have either boundary condition on the brane,  we focus here on integrating the Dirichlet component $\Phi_D$ on either side of the brane. 
We refer to the  $z<z_0$ and $z>z_0$ regions as ``left'' and ``right''. 
Our method to obtain  the quantum pressure is to vary the energy of the quantum vacuum with respect to the position of the brane, $z_0$ (this method was used in \cite{Brax:2018grq} in flat space).\,\footnote{The   energy of the QFT vacuum is a divergent quantity. However, if one varies it with respect to a physical parameter (such as the distance between two plates), the resulting variation is a physical observable and thus must be  finite. If divergences still appear in the expression of the vacuum pressure, they have a physical meaning and have to be treated in the framework of renormalization.}




In the presence of  curvature and/or  dilaton background, the  variation in $z_0$ can be tricky to evaluate. It is useful to  rescale the fields in the action as follows 
\be
\Phi_D=\rho^{d-1}e^\varphi \hat\Phi_D\, \label{eq:Phi_redef}
\ee
to absorb the geometric prefactors into the fields. With this field redefinition,  the  prefactors are moved into an effective position-dependent mass term for $\hat\Phi_D$, that leads to no difficulties in the calculation of the variation. 
This trick  simplifies the calculations and, as we will see, renders manifest the small-distance limit for which the flat metric is asymptotically recovered.  

Starting from the partition function with no sources, $Z(z_0)$, the vacuum energy is given by
\be
E_{\rm vac}(z_0)=\frac{W(z_0)}{T}\,,\quad\quad\quad W(z_0)=i\log Z(z_0)\,
\ee
with  the overall time factor $T=\int dt$. This factor is irrelevant for the static quantities we compute. 

The leading order contribution from the bulk modes is the one-loop determinant,  $E_{\rm vac}(z_0)=-\frac{i}{2}{\rm Tr}\log\left({\cal D}\right)=-\frac{i}{2}\sum_\lambda \int A_{d-1}\frac{d^{d}p}{(2\pi)^d} \log\left(p^2+m^2_\lambda\right)$ where $A_{d-1}$ is the brane spatial volume. Taking the $z_0$ variation leads to the definition of the vacuum pressure on the $d-1$-brane:
\be
-\frac{F}{A_{d-1}}=\frac{\delta E_{\rm vac}(z_0)  }{\delta z_0 }=\frac{1}{2}\sum_\lambda \int \frac{d^{d}p_E}{(2\pi)^d} \frac{1}{p_{E}^2+m^2_\lambda}\frac{\delta m_\lambda^2}{\delta z_0}\,. \label{eq:P_1}
\ee
We have performed a Wick rotation to integrate in the Euclidian momentum $p_E$.
A positive (negative) value of the pressure means that the boundary is pushed towards positive (negative) $z$. 

We could  have started from a theory in Euclidian space --- the metric  signature is irrelevant for the quantity of interest. We  work in Lorentzian signature for which the brane has $d-1$ spatial dimensions.

For $d=1$ the $0$-brane is a point particle. 
For $d>1$  the brane is an extended object and can have a localized energy density (\textit{i.e.} brane tension)
\be
S\supset \int dx^d \sqrt{\bar \g} \sigma\,\,. 
\ee
Depending on the value of $\sigma$ the brane may be static or have  velocity in the bulk --- the solutions of Einstein's equation for this system  have been classified in \cite{Kraus:1999it}. Here we assume that $\sigma$ is tuned such that the brane is static, as in the RS2 model.

\subsection{Vacuum Pressure from Bulk Modes}

What is the mass variation  $ \delta m^2_\lambda /\delta z_0 $ appearing in Eq.\,\eqref{eq:P_1}? To determine it, we consider  a piece of the action for one given Dirichlet mode,\,\footnote{
In this section  we use the rescaled field $\hat \Phi$  everywhere (see Eq.\,\eqref{eq:Phi_redef}), thus  the profiles and propagators are understood as $\hat f_\lambda$, $\hat G$. The hats will be omitted throughout the section. 
}
\be
S_\lambda=-\frac{1}{2}\int^\infty_{z_0} dz  \left[
\partial_M (f^\lambda(z) \phi^\lambda_D )
\partial^M (f^\lambda(z)\phi^\lambda_D )  +m^2 (f^\lambda(z)\phi^\lambda_D )^2 \right]
\label{eq:S_lam}
\ee
We can evaluate the variation $ \delta S_\lambda /\delta z_0$ in two ways.

First, one can integrate by part and use  the orthogonality relation of the modes, which gives 
\be
S_\lambda = -\frac{1}{2} (p^2+m_\lambda^2) (\phi^\lambda_D )^2
\ee
The variation in $z_0$ is then found to be
\be
\frac{\delta S_\lambda }{\delta z_0} = -\frac{1}{2} \frac{\delta m^2_\lambda }{\delta z_0} (\phi^\lambda_D )^2
\ee

Second, we can instead apply the variation directly to Eq.\,\eqref{eq:S_lam}.  The  only contribution is from the change of integration domain. This gives
\be
\frac{\delta S_\lambda }{\delta z_0} =  \frac{1}{2} 
 (\phi^\lambda_D )^2 \partial_z f^\lambda(z)\partial_{z'} f^\lambda(z')\bigg|_{z,z'=z_0} 
\ee
We  thus obtain the mass variation\,\footnote{
I am grateful to  E.\,Ponton  for providing insights on this trick in an early unpublished work  \cite{Loop_Quiros}.}
\be
 \frac{\delta m^2_\lambda }{\delta z_0} =-
 \partial_z f^\lambda(z)\partial_{z'} f^\lambda(z')\bigg|_{z,z'=z_0} \,.
 \label{eq:dm}
\ee

Plugging Eq.\,\eqref{eq:dm} into Eq.\,\eqref{eq:P_1}, we recognize the spectral representation of the Dirichlet propagator with Euclidian momentum. We obtain the contribution to the pressure from the right region ($z>z_0$)
 \be
 \frac{F_R}{A_{d-1}}=\frac{i}{2} \int\frac{d^dp_E}{(2\pi)^d}  
 \partial_{z}\partial_{z'}G_D(p_E;z,z')\bigg|_{z,z'=z_0}
 \label{eq:P_2}
 \ee

We proceed similarly with the contribution from the left region ($z<z_0$).  We obtain  the total pressure
 \be
 \frac{F}{A_{d-1}}=\frac{i}{2} \int\frac{d^dp_E}{(2\pi)^d}  
\left[ \partial_{z}\partial_{z'}G^R_D(p_E;z,z')-\partial_{z}\partial_{z'}G^L_D(p_E;z,z')\right]_{z,z'=z_0}
 \ee
 where $G^L_D$, $G^R_D$ are the Dirichlet propagators  in the left and right regions respectively. 
 Moreover we can use the explicit expression for the Dirichlet propagator given in Eq.\,\eqref{eq:propa_B2B_ND} ---applied to the rescaled field $\hat \Phi_D$. 
The regular solution in the left and right regions are denoted by $f_L$, $f_R$. 
Using spherical coordinates and defining $\p=|p_E|$, 
the final result for the quantum vacuum  pressure is found to be\,\footnote{
We notice a resemblance with the result from the $D$-sphere presented in \cite{Milton_Sphere}.
Our result, however, does not have any unwanted divergences apart from those renormalizing the brane tension. 
}
\begin{align}
\frac{F}{A_{d-1}}&=-\frac{1}{2^{d}\pi^{d/2}\Gamma(d/2)} \int d\p \p^{d-1} \left(
\frac{f_L'(z_0)}{f_L(z_0)}
+\frac{f_R'(z_0)}{f_R(z_0)}
\right) \nn
\\
& = -\frac{1}{2^{d}\pi^{d/2}\Gamma(d/2)} \int d\p \p^{d-1}
 \partial_z\log \left(K_L(\p,z)K_R(\p,z)\right)\Big|_{z=z_0}\,
\label{eq:P_gen}
\end{align}
where $K_{L,R}$ are the brane-to-bulk propagators in the left and right regions. Moreover, comparing to Eq.\,\eqref{eq:free_action_final_warped}, we see that the quantum pressure is proportional to the sum of the holographic  self-energies from each side of the brane.

Some sanity checks can be done. 
In the limit of large Euclidian momentum, when spacetime curvature and mass are negligible, the regular solutions are asymptotically exponentials as dictated by Eq.\,\eqref{eq:asymptotic_fsW}. In this limit we have $f_R(z)\sim e^{\p z}$, $f_L(z)\sim e^{-\p z}$. As a result the integrand of Eq.\,\eqref{eq:P_gen} vanishes asymptotically in this limit. This is the expected flat space behaviour.
Note  that this does not imply that the integral is automatically finite. Depending on how fast the metric becomes flat at small distance, \textit{i.e.} depending on the large $\p$ behaviour, there can be divergences from the $\p$ integral, that we will treat using standard dimensional regularization.

Second, we recover the Casimir pressure in a Minkowski interval \cite{Ambjorn:1981xw}. We assume the presence of a second brane in the left region, at $z=z_0-{\ell}<z_0$. We have $f_R = e^{-\p z}$, while $f_L=\sin (\p(z-z_0+\ell))$. 
We obtain $\partial_z\log K_L(z)=\p\coth(\p \ell)$, $\log \partial_z K_R(\p,z)=-\p$, giving exactly the scalar Casimir pressure between  two plates of spatial dimension $d-1$,
\be
\frac{F_{\rm flat}(\ell)}{A_{d-1}}= - \frac{1}{2^{d-1}\pi^{d/2}\Gamma(d/2)} \int d\p  \frac{\p^d}{e^{2\p\ell}-1}=-
\frac{d\,\Gamma\left(\frac{d+1}{2}\right)\zeta(d+1)}{{2^{d+1}\pi^{(d+1)/2}}\ell^{d+1}}
\,. \label{eq:Casimir_Flat}
\ee
In particular for $d=3$ we recover the well-known result $F/A_{2}=-\pi^2/480 \ell^4$.

The fact that the pressure on the $z_0$ brane is negative means that it is attracted towards negative $z$, consistent with the fact that the second brane is placed  at the left of $z_0$.

\subsection{Vacuum  Pressure in AdS}

We turn to a single brane in AdS metric.  
We consider the fluctuations from a bulk scalar field with mass $m^2=\Delta(\Delta-d)k^2=(\alpha^2-\frac{d^2}{4})k^2$ (with  $\alpha\geq 0$), existing in both  left ($z<z_0$) and right ($z>z_0$) regions. 
The solutions to the bulk EOM for the rescaled field are $\sqrt{z}I_\alpha(\sqrt{p^2}z)$, $\sqrt{z}K_\alpha(\sqrt{p^2}z)$ in any dimension.

Using the general formula Eq.\,\eqref{eq:P_gen},  the pressure induced by the scalar fluctuations is
\be
\frac{F_{{\rm AdS}_{d+1}}}{A_{d-1}} = -\frac{1}{2^{d}\pi^{d/2}\Gamma(d/2)} \int d\p \p^{d-1}
 \partial_z\log \left(z I_\alpha(\p z) K_\alpha(\p z) \right)\Big|_{z=z_0}\,
\label{eq:P_AdS}
\ee
This expression can be evaluated analytically only in particular cases and limits. However, general features can already be deduced. By dimensional analysis the pressure must scale as 
\be
\frac{F_{{\rm AdS}_{d+1}}}{A_{d-1}} \propto \frac{1}{z^{d+1}}\,.
\ee

Regarding the sign of the pressure, a qualitative statement can be done when the integral is finite such that no regularization is needed. We notice that the integrand of Eq.\,\eqref{eq:P_AdS} is positive  for $\alpha\geq \frac{1}{2}$. 
Under these restrictions we know that the pressure is negative, \textit{i.e.} the quantum pressure pushes the brane towards the AdS boundary.

\subsubsection{Conformally Massless Scalar}
For $\alpha=\frac{1}{2}$ in any $d$, the scalar is conformally massless upon a Weyl transformation to flat space. For this value of $\alpha$, the field in AdS effectively behaves just as in flat half-space with boundary at $z=0$. Accordingly, when setting $\alpha=\frac{1}{2}$ in Eq.\,\eqref{eq:P_AdS},  the vacuum pressure reduces to the flat space result with separation  $z_0$ corresponding to the distance between  the   $z=0$ boundary and the brane ($z=z_0$),
\be
\frac{F_{{\rm AdS}_{d+1}}(z_0)}{A_{d-1}}\bigg|_{\alpha=\frac{1}{2}}=\frac{F_{{\rm flat}_{d+1}}(z_0)}{A_{d-1}} \,.
\ee

\subsubsection{Uniform Expansion and  Flat Space Limit}

In the large $\p z$ limit and for $\alpha \neq \frac{1}{2}$  the  asymptotic behaviour of the Bessel functions leads to an asymptotically vanishing integrand in Eq.\,\eqref{eq:P_AdS}. However, if one uses this expansion to  approximate the $\p z >1$ region of the integral, 
highly inexact results occur.   This can be traced back to the fact that the bulk mass is  neglected in the usual large $\p z$ asymptotics. 
Instead  we need a limit of the Bessel functions where  the mass term is kept in the large $pz$ limit so that massive (as opposed to massless) flat space QFT  appears.
This  limit is realized by the ``uniform expansion'' of Bessel functions, \cite{abramowitz+stegun}
giving
\be
I_\alpha(\alpha y )K_\alpha(\alpha y ) = \frac{ u }{2\alpha}\left(1+\frac{
u^2-6u^4-5u^6
}{8\alpha^2}+O\left(\frac{1}{\alpha^4}\right)
\right) \,\quad \quad {\rm for} \quad \alpha\to \infty 
\label{eq:Unif_Exp}
\ee
with $u=(1+y^2)^{-\frac{1}{2}}$.
This limit amounts to taking the flat space limit in the sense of taking zero AdS curvature $k\to 0 $ at fixed bulk mass and using coordinates $z=\frac{1}{k}e^{k \eta}$ with fixed $\eta$, such that both $\alpha$ and $z$ simultaneously tend to infinity as $\propto \frac{1}{k}$.

\subsubsection{AdS$_2$}

  \begin{figure}
\centering
	\includegraphics[width=0.6\linewidth,trim={0cm 0cm 0cm 0cm},clip]{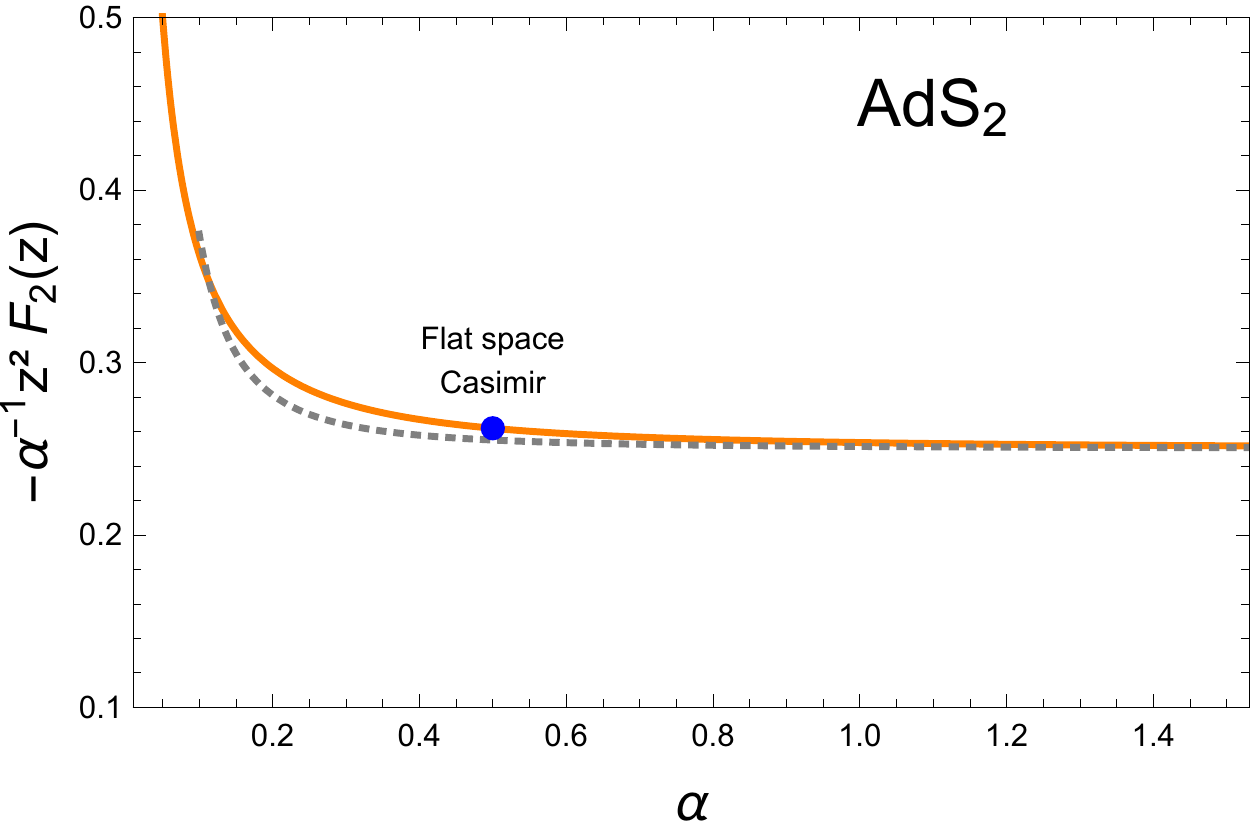}
\caption{ Force from the quantum vacuum felt by a point particle in $AdS_2$ in the presence of a scalar field with mass $m^2=(\alpha^2-\frac{1}{4})k^2$. For a conformally massless scalar $\alpha=\frac{1}{2}$, the flat space Casimir force between the particle and the boundary is recovered. For any $\alpha$ the point particle is attracted towards the AdS boundary.  The dotted line shows the
result from large mass expansion, Eq.\,\eqref{eq:FAdS2}.
\label{fig:Casimir_AdS2}}
\end{figure}

For $d=1$,  the ``brane'' is a point particle, thus we talk about force instead of pressure.  
The force computed by Eq.\,\eqref{eq:P_AdS} is finite. 
 We evaluate the force  at large $\alpha$  using the expansion Eq.\,\eqref{eq:Unif_Exp}. 
\be
F_{\rm AdS_{2}}\bigg|_{\alpha\gg 1} =-\left(\frac{\alpha}{4}- \frac{1}{256\pi \alpha} +O\left(\alpha^{-3}\right)\right)\frac{1}{z^2_0}
\label{eq:FAdS2}
\ee
We compute numerically the pressure and find excellent agreement down to $\alpha\sim \frac{1}{2}$. 
The force is shown in Fig.\,\ref{fig:Casimir_AdS2}.
At $\alpha=\frac{1}{2}$ the flat space Casimir pressure $F_2=-\frac{\pi}{24 z^2_0}$ is recovered. 
The force is found to be negative for any value of the bulk mass: The point particle  is attracted towards the AdS boundary.

\subsubsection{AdS$_{>2}$}

For $\alpha\neq \frac{1}{2}$ and $d\geq 2$, divergences appear in the integral Eq.\,\eqref{eq:P_AdS}. 
These divergences can be seen using the uniform expansion Eq.\,\eqref{eq:Unif_Exp}, which gives
\be
 \partial_z\log \left(z I_\alpha(\p z) K_\alpha(\p z) \right)\Big|_{z=z_0}=
\frac{1}{z} \frac{\alpha^2}{\alpha^2  +z^2\p^2}
-\frac{
4 \alpha^4 z\p^2-10 \alpha^2 z^3\p^4 +z^5\p^6
}{4(\alpha^2+z^2\p^2)^4}+\ldots
 \, \label{eq:int_AdS}
\ee
The degree of divergence for each term under the $\int d\p \p^{d-1}$ integral is then obvious. 
Through Eq.\,\eqref{eq:int_AdS} we also see that the integrand takes the same form as in massive flat space, with terms of the form $1/(\p^2+\Delta)^n$. The integrals can thus be treated using textbook dimensional regularization. That is, power-law divergences are automatically removed and only logarithmic divergences remain. 
There are log divergences  for even $d\geq 2$ only.  

The result for general $d$ is
\be
\frac{F_{\rm AdS_{d+1}}}{A_{d-1}}\bigg|_{\alpha\gg 1} =
-\left(\frac{\alpha^d}{2^{d+1}\pi^{d/2}}\Gamma\left(1-\frac{d}{2}\right) +O\left(\alpha^{d-2}\right)\right)
\frac{1}{z^{d+1}_0}
\ee
The divergences at even $d\geq 2$ appear via the Gamma function, as customary from dimensional continuation. 

Since the Gamma function  with negative argument can take both signs, the expression can take either sign depending on the dimension.
For odd $d$, keeping the leading term, the expression gives
\be
\frac{F_{\rm AdS_{4}}}{A_{2}}\bigg|_{\alpha\gg 1} \sim \frac{\alpha^3}{8\pi} \frac{1}{z^{4}_0} \,,\quad\quad \quad 
\frac{F_{\rm AdS_{6}}}{A_{4}}\bigg|_{\alpha\gg 1} \sim -\frac{\alpha^5}{48\pi^2} \frac{1}{z^{6}_0}\,\quad\ldots
\ee

What quantity is being renormalized for even $d$? As can be seen by comparing with AdS$_2$, the divergences are tied to the brane being an extended object, hence we can expect that the quantity being renormalized is brane-localized, and the only candidate is the brane tension. The brane tension takes the form 
\be
S\supset \int dx^d z^{-d}_0 (\sigma+\delta \sigma)\,
\ee
where $\delta \sigma$ is the $z_0$-independent counterterm. 
We introduce $d=n-\epsilon$ with $n$ a positive even integer.
Varying in $z_0$,  the $\delta\sigma$ counterterm cancels the divergence for
\be
d\delta\sigma =\frac{(-1)^{d/2}\alpha^d}{2^{d}\pi^{d/2}\Gamma(d/2)\epsilon}    +{\rm finite} \label{eq:CT}
\ee
The finite quantum contribution to the pressure is then proportional to $\log(z_0 \mu) z^{-d-1}_0$, where $\mu$ is the renormalization scale. The finite part in the counterterm Eq.\,\eqref{eq:CT} is absorbed in the definition of the renormalization scale. 
The final result for the quantum pressure felt by the brane is 
\be
\frac{F_{\rm AdS_{d+1}}}{A_{d-1}}\bigg|_{\alpha\gg 1} \sim
\frac{(-1)^{d/2}\alpha^d}{2^{d+1}\pi^{d/2}\Gamma(d/2)} \frac{\log(z_0 \mu)}{z^{d+1}_0} \,,\quad {\rm for~even}~ d\geq 2 
\ee
at leading order in the $1/\alpha$ expansion. 

The quantum contributions arising  in AdS$_{>2}$  may dominate the motion of the brane if the bare tension is tuned such that the brane is static at classical level (see \cite{Kraus:1999it} for brane motion as a function of  brane tension.) It could be interesting to investigate the consequences of this detuning in \textit{e.g.} braneworld models.

\section{Interactions and Boundary Correlators: Overall Picture }
\label{se:interactions}

 In this section and the following ones we turn on the  interactions.  This section is a prelude where  we discuss the overall structure of the boundary effective action and make some general observations.

 We consider, as in Secs.\,\ref{se:gen},\,\ref{se:warped_holography}, a field that can fluctuate on the boundary \textit{i.e.} satisfies Neumann BC. We will discuss the boundary effective action for such field and how the generated correlators relates to those with Dirichlet BC. 
 We will then notice an interesting implication in the AdS case. Finally we discuss some general features  of the  long-distance holographic EFT  obtained by integrating out heavy degrees of freedom.

Interactions are easily written in terms of holographic variables in the general action  Eq.\,\eqref{eq:free_action_final}. For example the coupling  $\Phi^n(x)$ takes the form 
\be
(\Phi_0 \cb K+\Phi_D)^n(x) \,. \label{eq:hol_int}
\ee
This leads to monomials of the form $(\Phi_0 \cb K)^k\Phi_D^{n-k}$. Each of these vertices contribute to the  correlators. 
In this work we are interested in  boundary observables, which are defined by insertions of boundary-localized sources $J_0=J|_{\partial \cal M}$ in the partition function $Z[J_0]$.

\begin{figure}
\centering
	\includegraphics[width=0.32\linewidth,trim={0cm 8cm 17cm 0cm},clip]{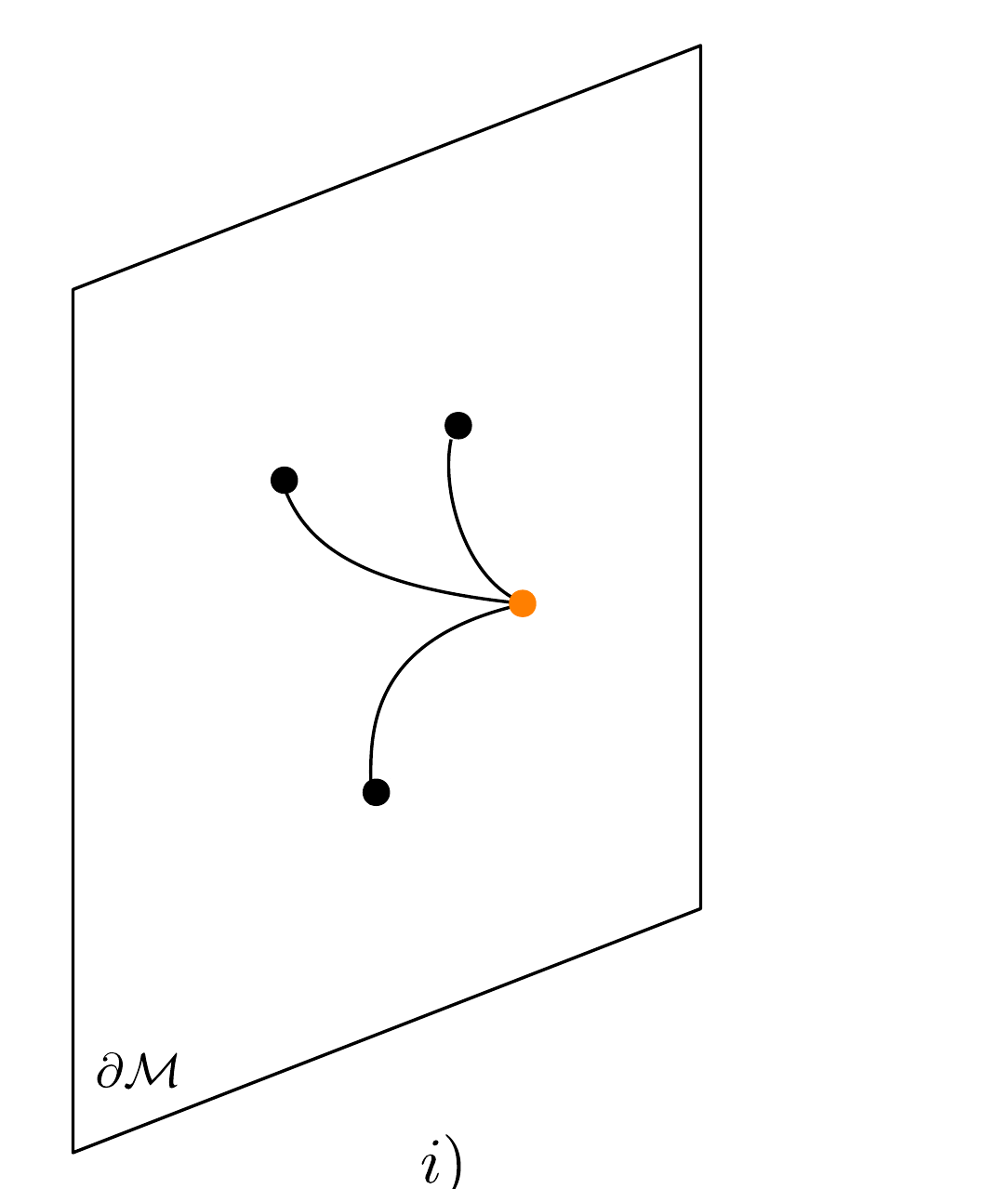}
	\includegraphics[width=0.32\linewidth,trim={0cm 8cm 17cm 0cm},clip]{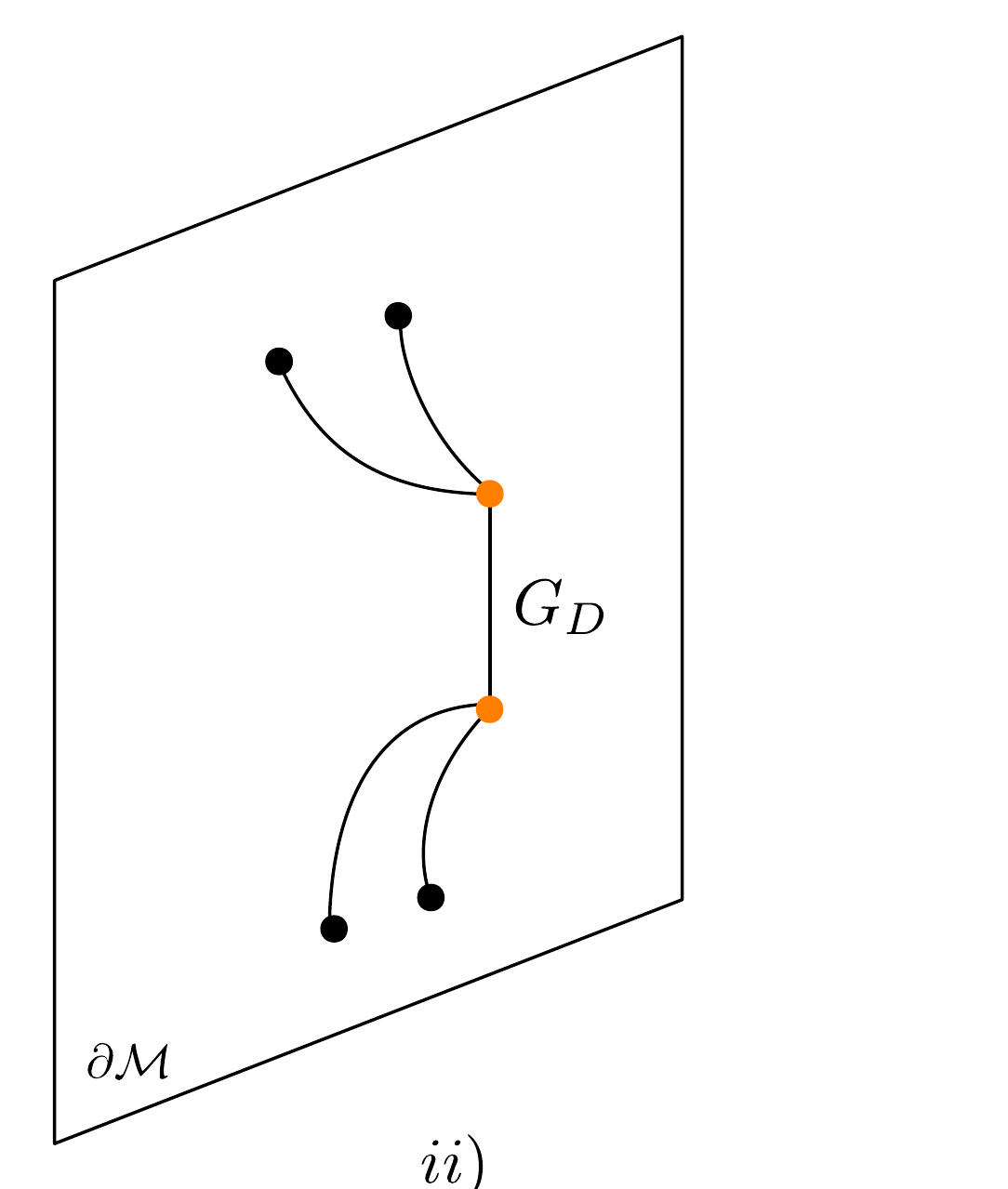}
	\includegraphics[width=0.32\linewidth,trim={0cm 8cm 17cm 0cm},clip]{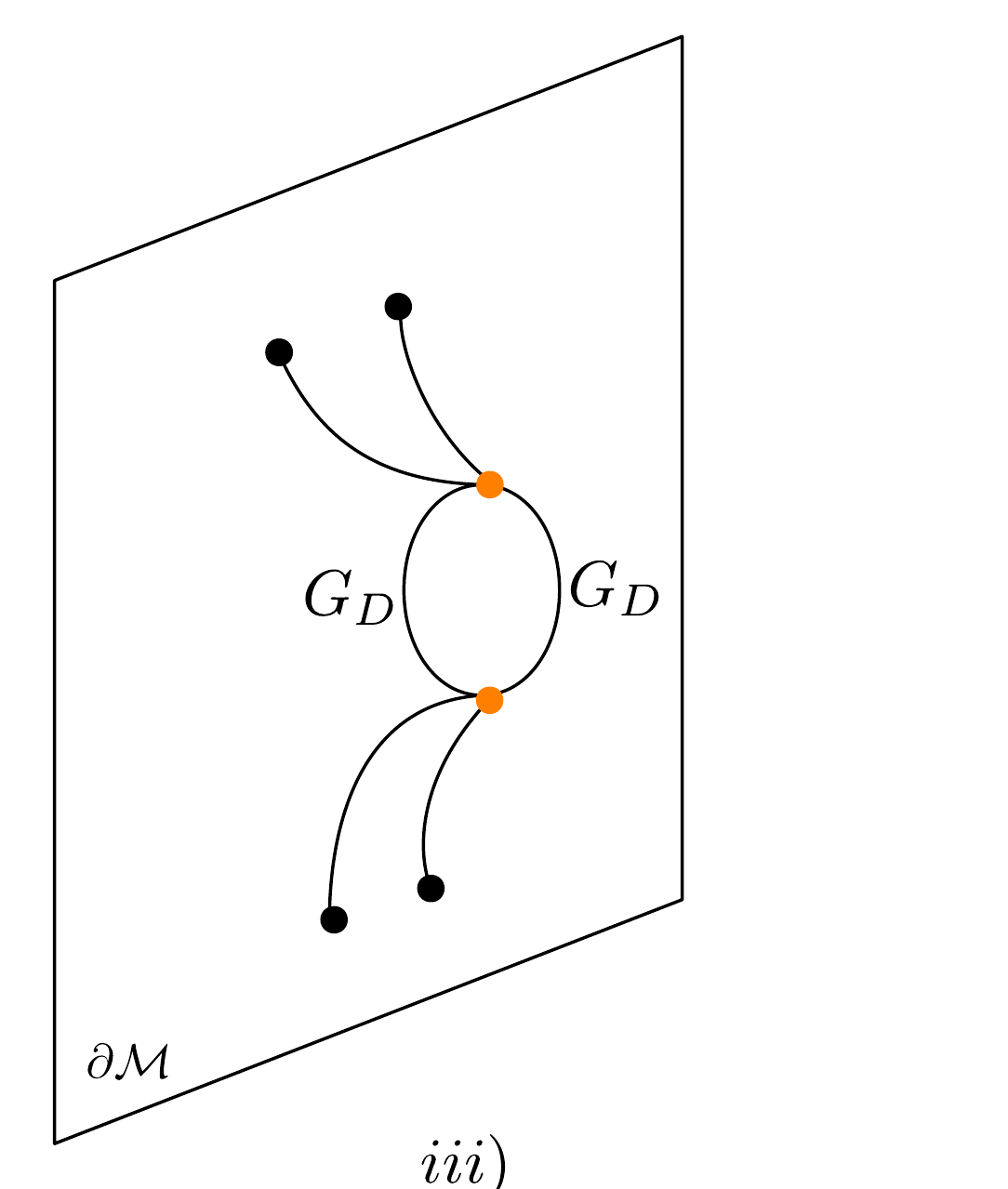} \\
	\includegraphics[width=0.32\linewidth,trim={0cm 8cm 17cm 0cm},clip]{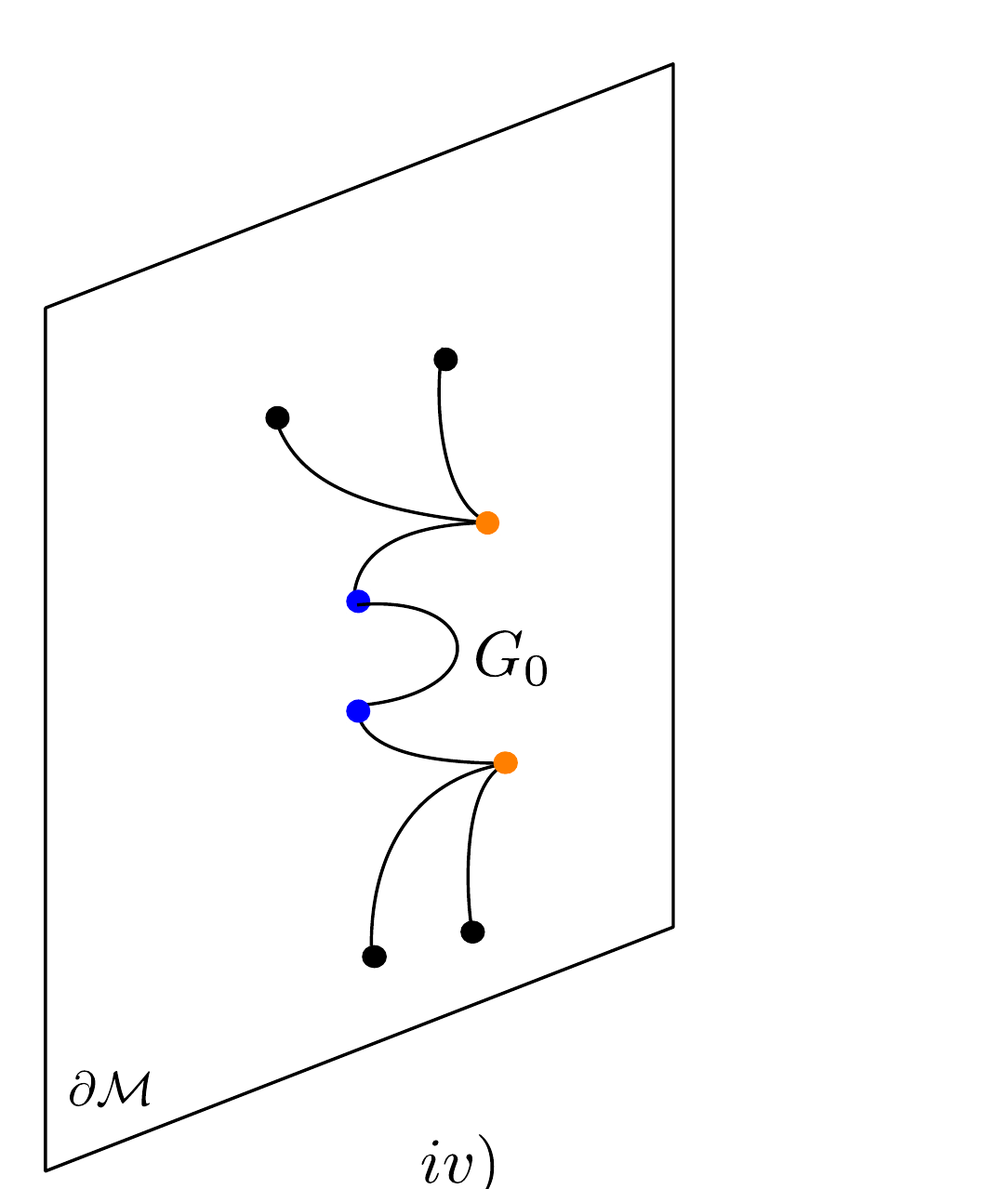}
	\includegraphics[width=0.32\linewidth,trim={0cm 8cm 17cm 0cm},clip]{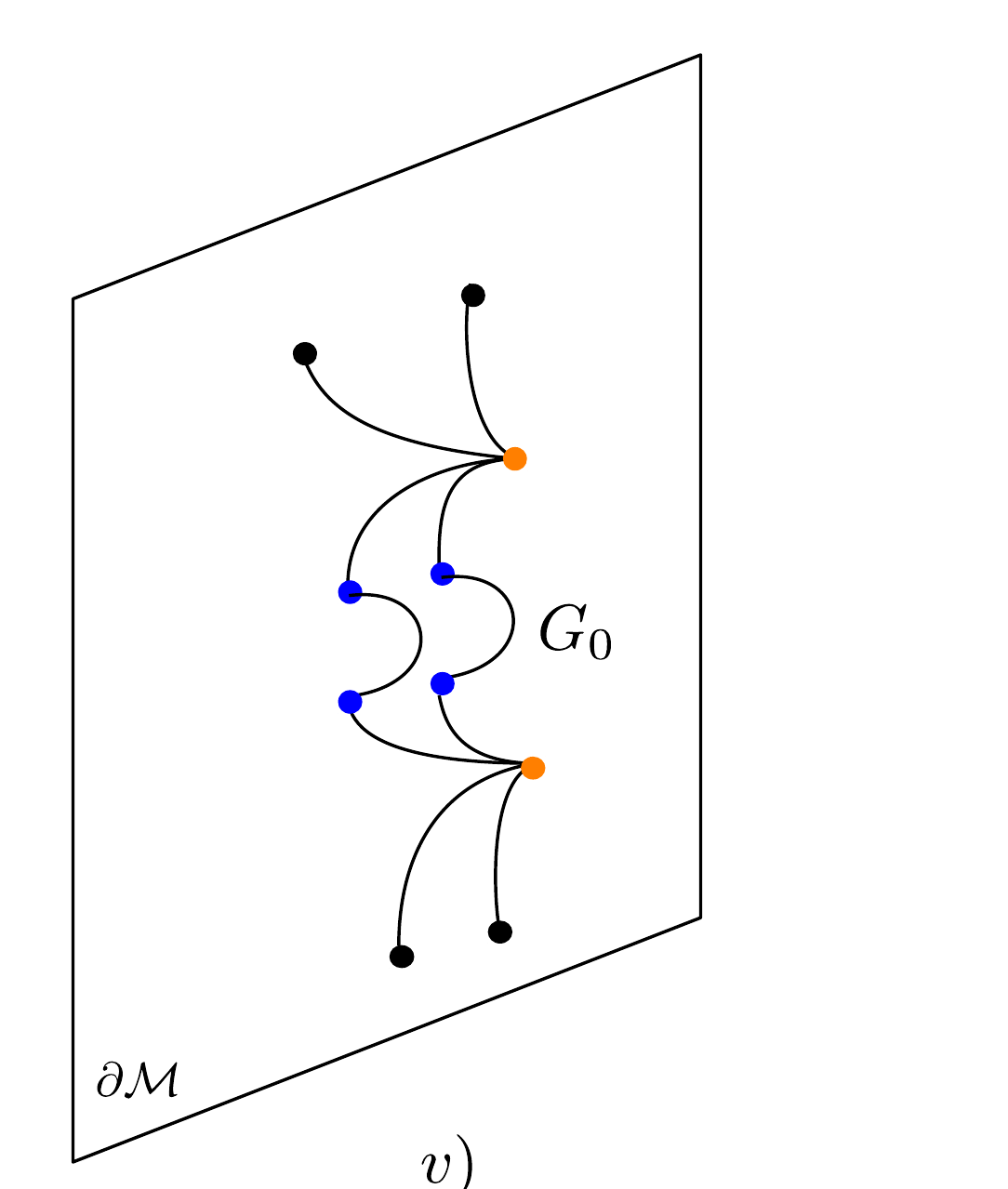}
		\includegraphics[width=0.32\linewidth,trim={0cm 8cm 17cm 0cm},clip]{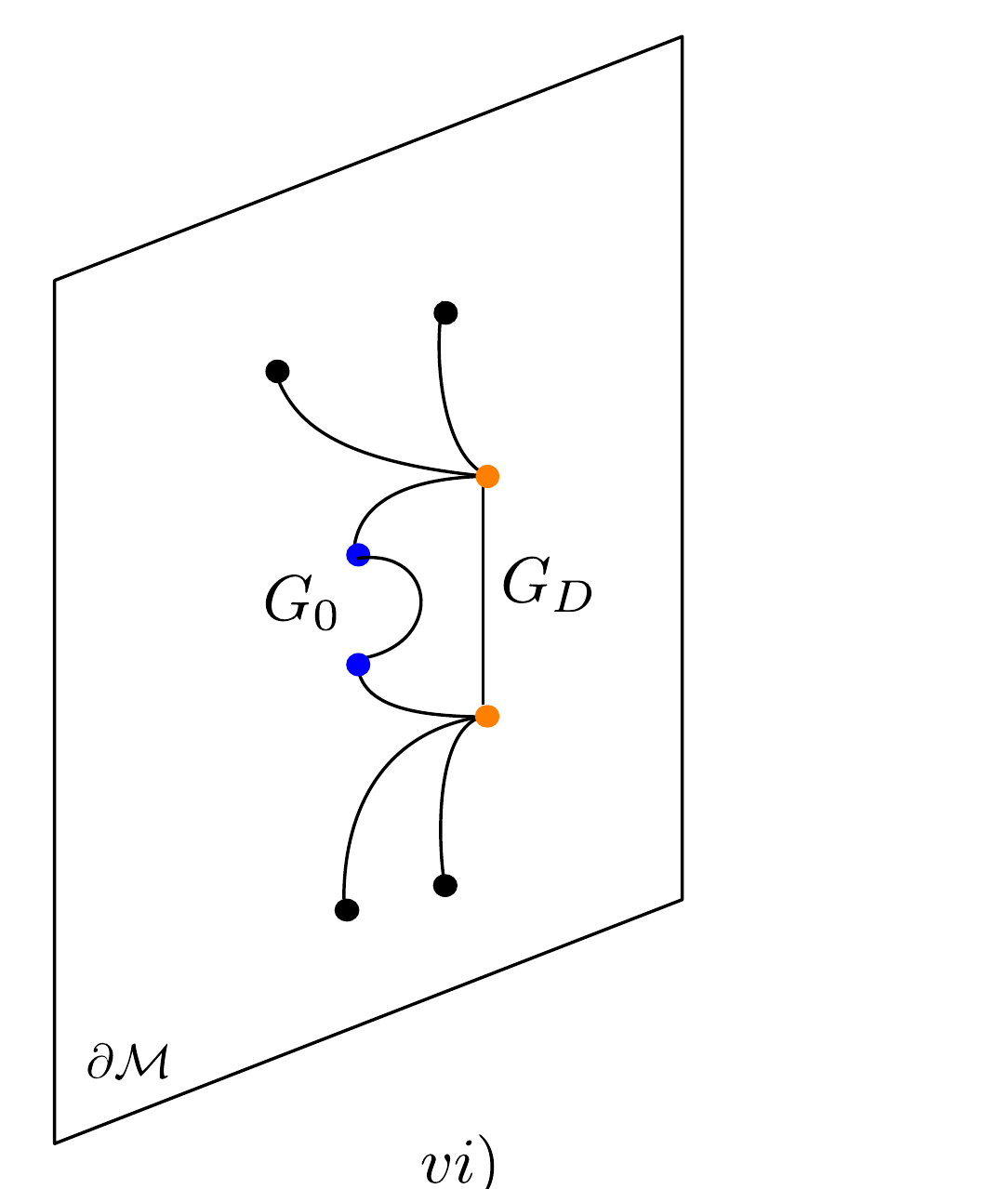}
	\caption{
Sample of boundary connected correlators. Orange and blue points are respectively integrated over ${\cal M}$ and ${\partial \cal M}$.
\label{fig:Holo_Diags_ex}}
\end{figure}

A $J_0$ source probes only the boundary degrees of freedom and not the Dirichlet modes.  Let us first  integrate over the Dirichlet modes in $Z[J_0]$. This gives 
\be
Z[J_0]= \int {{\cal D}  \Phi_0} e^{-S_D[\Phi_0]-\Phi_0 \cb J_0}\, \label{eq:SD_def}
\ee
where we introduced a nonlocal boundary action $S_D$ that we refer to as the ``Dirichlet action''.  It is a functional of the boundary degree of freedom $\Phi_0$.
At perturbative level, the Dirichlet action encodes the diagrams with only $G_D$ internal lines , that we refer to as Dirichlet diagrams. 
These are for example diagrams \textit{i)} to \textit{iii)} in Fig.\,\ref{fig:Holo_Diags_ex}. Since $\Phi_0$ is a dynamical field, $S_D$ can be seen as a fundamental action  in $\Phi_0$ --- encoding very nonlocal operators.
In AdS the Dirichlet action  encodes the  Witten diagrams with $\Delta_+$ propagators in internal lines.

Among all the operators in $S_D$ it is useful to single out the class of operators 
\be
\int_{{\cal M}} d^{d+1}x \sqrt{| \g|} \prod^\n_{i=1} K_i  \cb \Phi_{i,0} (x)  
\ee
These are the operators generated by a single bulk vertex, they are 
in this sense the most local operators. We define the associated ``holographic vertices''
\be\Lambda_n(x_{0,i})=\int_{{\cal M}} d^{d+1}x \sqrt{| \g|} \prod^n_{i=1} K(x_{0,i},x)
\label{eq:Lambdan}
\ee 
A 3pt holographic vertex is shown as diagram \textit{i)}  in Fig.\,\ref{fig:Holo_Diags_ex}. 
The $\Lambda_n$ are in one-to-one correspondence with bulk vertices and in AdS are proportional to contact Witten diagrams.

 Let us turn to the definition of boundary  correlators generated by $Z[J_0]$.  They are  given by\,\footnote{We remind that in our notation, the boundary coordinates $y^\mu $   appear through the  $x^M_0(y^\mu)$ embedding.} 
\be
\langle \Phi_0(x_{0,a}),\Phi_0(x_{0,b} )\ldots\rangle \,. 
\ee
In particular, the connected boundary correlators are generated by taking the derivatives of $W[J_0]=\log Z[J_0]$ (this is the convention for Euclidian signature, there is an extra $i$ for Lorentzian signature). Given the definition of the Dirichlet action Eq.\,\eqref{eq:SD_def}, the boundary connected diagrams are built from   boundary subdiagrams with only Dirichlet lines (\textit{i.e.} Dirichlet subdiagrams) connected to each other by boundary lines. Examples of such connected correlators are shown as diagrams \textit{iv)} to \textit{vi)} in Fig.\,\ref{fig:Holo_Diags_ex}. 

We can finally define the boundary effective action, \textit{i.e.}
the generating functional of 1PI  boundary connected correlators via the Legendre transform
\be
\Gamma[\Phi_{0,{\rm cl }}]=-W[J_0]+ J_0 \cb \Phi_{0,{\rm cl}} \,.   
\ee
Since the Legendre transform is done with respect to the boundary source, it amputates boundary-to-boundary propagators on each leg of the correlators. Likewise,  the notion of one-particle irreducibility (1PI) is here meant with respect to boundary-to-boundary lines only.
Thus in Fig.\,\ref{fig:Holo_Diags_ex} only diagram \textit{iv)} is 1PR, all the other ones are 1PI since they cannot be split by cutting a single line on the boundary. 
Since boundary-to-boundary correlators are amputated on each external leg, the 1PI correlators obtained by taking derivatives of $\Gamma[\Phi_{0,{\rm cl }}]$ have amputated boundary-to-bulk propagators as external legs (see definition Eq.\,\eqref{eq:Bo2B_def}). 

\subsubsection*{Relation to Unitarity Cuts and Transition Amplitudes}

Unitarity cuts on Dirichlet subdiagrams act on internal $G_D$ propagators. We have shown in section \ref{se:AdS_properties} that in AdS we have $G_D=G_{\Delta=\Delta_+}$, which admits a momentum spectral representation (see \textit{e.g.} \cite{Costantino:2020vdu,Meltzer:2020qbr}).  A unitarity cut acting on such a propagator gives a Wightman propagator, which takes a split form \cite{Meltzer:2020qbr}.  It follows that  cutting on a AdS Dirichlet subdiagram gives rise to a squared AdS ``transition amplitude''. 
We checked that this picture of unitarity cuts similarly holds for an arbitrary warped background (see also  \cite{Sivaramakrishnan:2021srm} for developments on transition amplitudes in general background).   
This is due to the fact that  the momentum-space Wightman propagator on an arbitrary warped background must take a split form, as can be deduced by generalizing the calculations in \cite{Costantino:2020vdu}.

\subsection{On Holographic CFT Correlators}

We have seen that a generic contribution to a given $n$pt boundary correlator is made of Dirichlet subdiagrams (\textit{i.e.}  with only  $G_D$ in internal lines) connected to each other by boundary lines. 
Formally, this substructure of the  diagrams appears when singling out the Dirichlet action in the partition function, Eq.\,\eqref{eq:SD_def}.  
We also know that when the background is AdS, the boundary correlators are identified as correlators of a strongly coupled large $N$  CFT. We can thus wonder: How does the substructure of a diagram in terms of Dirichlet subdiagrams  translate into the holographic CFT?

The answer will  certainly involve the notion of large $N$ corrections. First we stress  that a Dirichlet subdiagrams can be of arbitrary loop  order, hence the  Dirichlet subdiagrams  encode plenty of $\frac{1}{N}$ corrections. Our focus here is rather on the structure of the entire diagram. In order to figure out the CFT meaning of an entire diagram,
let us take an example. 

We consider  \textit{v)} in Fig.\,\ref{fig:Holo_Diags_ex} which is induced by quartic couplings and amounts to
$\Lambda_4\cb G^2_0 \cb \Lambda_4$.\,\footnote{The simplest example would be $\Lambda_3\cb G_0 \cb \Lambda_3$ but it is 1PR,  we rather focus on a 1PI diagram.  }
In the CFT language this bubble diagram corresponds to\,\footnote{Strictly speaking 
the 4pt CFT correlator  also contains contribution from $\Lambda_3\cb G_0 \cb \Lambda_3$ at leading order in large $N$, corresponding to  including box and triangle diagrams in the example.  For our discussion we can safely focus on the bubble diagram only.  }
\be
\langle{\cal O}(y_a){\cal O}(y_b){\cal O}(u){\cal O}(v)\rangle \left[ \langle {\cal O}(u) {\cal O}(u') \rangle\right ]^{-1} \left[ \langle {\cal O}(v) {\cal O}(v') \rangle\right ]^{-1} \langle{\cal O}(u'){\cal O}(v'){\cal O}(y_c){\cal O}(y_d)\rangle
\label{eq:4ptCFT1}
\ee
Here the inverse and the $\cb$ convolution in spacetime coordinates are written in matrix notation. 
We can see that Eq.\eqref{eq:4ptCFT1} is built solely from other CFT correlators, evaluated at leading order in the large $N$ expansion. The large $N$ scaling is $\sim \frac{1}{N^2}$ for the 4pt correlators and $\sim 1 $ for the 2pt correlator. 

  We can  recognize the structure of a perturbative expansion scheme at the level of the large $N$ CFT, where the expansion is in  the small parameter $\frac{1}{N}$. 
To better recognize the diagrammatic structure, we define the amputated $4$pt correlator,
\begin{align} \langle{\cal O}(y_a) & {\cal O}(y_b){\cal O}(y_c){\cal O}(y_d)\rangle_{\rm amp}=
\\
& \left[ \langle {\cal O}(y_a) {\cal O}(y_a') \rangle\right ]^{-1} 
\dots 
\left[ \langle {\cal O}(y_d) {\cal O}(y_d') \rangle\right ]^{-1}
\langle{\cal O}(y'_a){\cal O}(y'_b){\cal O}(y'_c){\cal O}(y'_d)\rangle  \nn
\end{align}
This amounts to a 4pt ``vertex''  in the diagrammatic language.  This vertex scales as $\sim \frac{1}{N^2}$. 
The expression Eq.\,\eqref{eq:4ptCFT1} then becomes
\begin{align}
\langle {\cal O}(y_a) {\cal O}(y_a') & \rangle  \langle {\cal O}(y_a) {\cal O}(y_a') \rangle 
\langle{\cal O}(y'_a){\cal O}(y'_b){\cal O}(u){\cal O}(v)\rangle_{\rm amp} \label{eq:4ptCFT2} \\ \nn & \times\langle   {\cal O}(u) {\cal O}(u') \rangle \langle   {\cal O}(v) {\cal O}(v') \rangle   \\ \nn & 
\quad\quad\quad\quad\times \langle{\cal O}(u'){\cal O}(v'){\cal O}(y'_c){\cal O}(y'_d)\rangle_{\rm amp}
 \langle {\cal O}(y'_c) {\cal O}(y_c) \rangle  \langle {\cal O}(y'_d) {\cal O}(y_d) \rangle 
\end{align}
We recognize the structure of a (non-amputated) bubble diagram --- the second line contains the two internal lines of the bubble.

We can reproduce the same steps for more complicated topologies built from Dirichlet subdiagrams, with same outcome.  One can also prove the perturbative structure at all order by working at the level of the path integral, using the Dirichlet action as a generator of the Dirichlet subdiagrams.

The conclusion is  that AdS diagrams involving internal boundary lines (\textit{e.g.} \textit{iv}-\textit{vi)} in Fig.\,\ref{fig:Holo_Diags_ex}) correspond in the holograhic CFT  to diagrams made of amputated $n>2$pt  CFT correlators  connected by mean field 2pt correlators. Such expressions can, at least in principle, be directly computed in the  CFT.
Interestingly, an analogous diagrammatic approach to  large $N$  CFT has been introduced in \cite{Petkou:1994ad} in the context of evaluating a 4pt CFT correlator. 
Here we have found how it appears from the AdS side.

An implication of the above observations is that, when calculating an AdS Witten diagrams with $\Delta_+$ propagators in internal lines, one actually computes operators of the Dirichlet action $S_D$. If one wants to obtain the complete boundary correlator generated by the boundary action $\Gamma[\Phi_0]$ at a given order in $\frac{1}{N}$, it is necessary to include 
the  extra contributions taking the form of large $N$  CFT diagrams described above. Examples from Fig.\,\ref{fig:Holo_Diags_ex} are $\textit{ii)}+\textit{iv)}$, $\textit{iii)}+\textit{iv)}+\textit{v)}$. 
Equivalently, we can say that $\Delta_-$ propagators have to be used in internal lines instead of  $\Delta_+$ propagators.

\subsection{Long-distance  Holographic EFT}
\label{se:EFT_discussion}

A field propagating in internal lines  may have mass $m$  much higher than the inverse distance scale $\zeta^{-1}$ involved in the correlators, $m\gg \zeta^{-1}$. In that case one can integrate the heavy field out using a large-mass expansion. This produces a series of effective operators depending only on the light degrees of freedom that  encodes the effect of the heavy field in the long-distance regime.

 In the large $m$ expansion, bulk diagrams are expanded as a series of local bulk interactions suppressed by powers of $m$. 
From the viewpoint of the boundary effective action, these bulk vertices map one-to-one onto contributions to the holographic vertices $\Lambda_n$. In analogy with flat space low-energy EFT, the contributions from the heavy field to the boundary effective action can be cast into a long-distance effective Lagrangian. 
Of course, unlike the familiar EFT Lagrangians from flat space, the holographic long-distance EFT is nonlocal in $\partial \cal M$ --- even though the bulk vertices are local, nonlocality  arises from    convolution with  the boundary-to-bulk propagators. 

Denoting  the heavy field by $\PhiH$ and the light field by $\PhiL$, the long-distance Dirichlet action
$S_{D,{\rm eff}}+ S^\partial_{D,{\rm eff}} $
generated by integrating $\PhiH_D$  
takes the schematic structure
\begin{align}
S_{D,{\rm eff}}[\PhiL_0,\PhiH_0] & = S_D[\PhiL_0,\PhiH_0] +\sum_{n,\alpha } \frac{a_{n,\alpha}}{m^{c_{n,\alpha}}}\Big(\Lambda_n\Big)_{y_1,y_2,\ldots y_n} \prod^n_{i=1}\Phi^{\ell,  h}_0(y_i) +\ldots
\label{eq:LDeff} \\
S^\partial_{D,{\rm eff}}[\PhiL_0,\PhiH_0] & = S^\partial_D[\PhiL_0,\PhiH_0] +\sum_{n,\alpha } \frac{b_{n,\alpha}}{m^{d_{n,\alpha}}}\Big({\cal B}_n\Big)_{y_1,y_2,\ldots y_n} \prod^n_{i=1}\Phi^{\ell,  h}_0(y_i) \label{eq:LDeffbound}
\end{align}
where  $a_{n,\alpha}$, $b_{n,\alpha}$ encode products of bulk couplings and $c_{n,\alpha},d_{n,\alpha}>0$. The boundary degree of freedom of the heavy field is not integrated out and thus remains in the Dirichlet action. 
The $S^{(\partial)}_D$ are the fundamental bulk and boundary actions  with renormalized constants. The ellipses in $S_{D,{\rm eff}}$ denote Dirichlet subdiagrams other than the holographic vertices $\Lambda_n$. The ${\cal B}_n$ are ---possibly nonlocal--- boundary terms generated when integrating out $\PhiH_D$.

Integrating out the $\PhiH_0$ degree of freedom is more subtle because the holographic self-energy can in principle contain a light degree of freedom (see also discussion in section \ref{se:mixing}). If present, the light degree of freedom has to be appropriately singled out, therefore this operation  requires some care. 
When integrating out $\PhiH_0$ except for a possible light mode, additional contributions to the boundary operators ${\cal B}_n$ in $S^\partial_{D,{\rm eff}}$ are generated, while the bulk terms for $\PhiL_0$ in $S_{D,{\rm eff}}$ remain  unchanged. This is because the boundary-to-bulk propagators of $\PhiH$ get shrunk to the boundary and expands into a series of boundary-localized local terms.   As a result, the heavy $\PhiH_0$ only contributes  to the boundary effective operators. This feature here explained conceptually  will show up in the explicit calculation of the one-loop boundary effective action.

Computing the long-distance EFT arising from integrating out a heavy  bulk field is fairly simple at tree-level. The key ingredient is the large-mass expansion of the propagator. This expansion is obtained by inverting the bulk EOM, for example by multiplying by  $\sum^\infty_{k=0} \frac{\square^{k}}{m^{2{k}}}$ and solving order by order. 
The propagator in the large mass limit is expressed as a sum of derivatives of the delta function 
 \be
 G_D(x,x')=  \frac{1}{\sqrt{|g|}}\sum^\infty_{k=0} \frac{\square^k}{m^{2(k+1)}} \delta^{d+1}(x,x')\,
 \label{eq:AdS_prop_series}
 \ee
  giving rise to local bulk vertices and thus to the holographic vertices.\,\footnote{In AdS$_{d+1}$, Eq.\,\eqref{eq:AdS_prop_series}  proves that the harmonic kernel satisfies
  $  \int_{\mathbb{R}} d\nu (\nu^{2}+\frac{d^2}{4})^r\Omega_\nu(x,x') =  \frac{1}{\sqrt{|g|}} k^{1-d}\left(\frac{-\square}{k^2}\right)^r \delta^{d+1}(x,x') $.  }
Integrating out a bulk field at loop level is much more technical, this is where the holographic effective action  becomes powerful.

\section{The Boundary One-loop Effective Action }

\label{se:HOLEA}

In this section we integrate out the bulk modes at one-loop in the presence of interactions. 
From the technical viewpoint this section is mostly a review in the sense that it consistently gathers existing heat kernel results from the literature. 

\subsection{Preliminary Observations}

In order to compute the effective action we are going to use a background field method. That is, one  separates the fields into background and fluctuation,
$\Phi=\Phi|_{\rm bg}+\Phi|_{\rm fl}$.
If we identify 
$\Phi|_{\rm bg} \equiv \Phi_0 \cb K $, $\Phi|_{\rm fl} \equiv \Phi_D$, the fluctuation  has Dirichlet BC, while if we also let the boundary degree of freedom fluctuate, \textit{i.e.} $\Phi_0=\Phi_{0,{\rm bg}}+\Phi_{0,{\rm fl}}$,
the fluctuation has Neumann BC, \textit{i.e.} $\Phi|_{\rm fl} \equiv \Phi_{0,{\rm fl}} \cb K + \Phi_{D} $.
Integrating out  quantum fluctuations at one-loop gives rise to the one-loop effective action, 
\be
\Gamma=\Gamma_{\rm cl} +\Gamma_{1-{\rm loop }} +\ldots
\ee
$\Gamma_{1-{\rm loop }}$ in the presence of background fields is best evaluated via the heat kernel method, which gives rise to the Gilkey-de Witt ``heat kernel coefficients'' \cite{DeWitt_original,DeWitt_original2,Gilkey_original,McAvity:1990we}. The heat kernel coefficients are local covariant quantities built from  geometric invariants of the background. Here these  invariants  will be expressed in terms of  holographic variables.

The heat kernel coefficients induced from a light field provide one-loop divergences while the heat kernel coefficients from a heavy field provide both one-loop divergences and a long distance EFT.
The Dirichlet component of the fluctuation contributes to the bulk  and boundary heat kernel coefficients while the boundary component  contributes solely to the boundary heat kernel coefficients, in accordance with the observations in Sec.~\ref{se:EFT_discussion}.
 In all cases, the geometric invariants are  built from the light field background and thus depend on $\PhiL_0,\PhiL_D$. 
 Upon integration in $\PhiL_D$ the resulting boundary effective action takes the general form shown in Eqs.\,\eqref{eq:LDeff},\,\eqref{eq:LDeffbound}.
In this work, for the bulk piece   Eq.\,\eqref{eq:LDeff} we will only focus on the holographic vertices $\Lambda_n$.

 There has been a plethora of studies of perturbative amplitudes in AdS background. AdS loop diagrams and their properties have been investigated in
Refs.\,\cite{ Cornalba:2007zb,
Penedones:2010ue, Fitzpatrick:2011hu,
Alday:2017xua,
Alday:2017vkk,
Alday:2018pdi,
Alday:2018kkw,
Meltzer:2018tnm,
Ponomarev:2019ltz,
Shyani:2019wed,
Alday:2019qrf,
Alday:2019nin,
Meltzer:2019pyl,
Aprile:2017bgs,
Aprile:2017xsp,
Aprile:2017qoy,
Giombi:2017hpr,
Cardona:2017tsw,
Aharony:2016dwx,
Yuan:2017vgp,
Yuan:2018qva,
Bertan:2018afl,
Bertan:2018khc,
Liu:2018jhs,
Carmi:2018qzm,
Aprile:2018efk,
Ghosh:2018bgd,
Mazac:2018ycv,
Beccaria:2019stp,
Chester:2019pvm,
Beccaria:2019dju,
Carmi:2019ocp,
Aprile:2019rep, 
Fichet:2019hkg,
Meltzer:2019nbs,
Drummond:2019hel, Albayrak:2020isk, 
Albayrak:2020bso,
Meltzer:2020qbr,
Costantino:2020vdu,
Carmi:2021dsn, Fitzpatrick:2011dm, Ponomarev:2019ofr, Antunes:2020pof,
Fichet:2021pbn}. 
 The one-loop effective action and its applications, however, seem fairly underrepresented in the literature. 
To the best of our knowledge, it has  mostly been used to compute the one-loop effective potential, see \cite{Burgess:1984ti,
Inami:1985wu,
Camporesi:1993mz,
Gubser:2002zh,
Hartman:2006dy,
Giombi:2013fka,
Carmi:2018qzm}. 
The one-loop potential is only a particular case that we briefly discuss below. The full one-loop effective action contains much more information on the long-distance limit of amplitudes and on their divergences.

\subsection{The One-loop Effective Potential}
\label{se:Veff}

There is a special case for which the one-loop effective action can be computed exactly. 
For  general spacetime, this special case is defined as follows. 
It is the case when the one-loop effective action encodes a scalar background with constant value on the boundary (\textit{i.e.} constant in the $y^\mu$ coordinates) and with appropriate scaling in the bulk  such that the background-dependent mass  amounts exactly to a constant bulk mass for the fluctuation,
\be
 m^2_{\rm fluctuation }[\Phi_{\rm background}](z)= {\rm cste}
\ee
or equivalently $
 X[\PhiL_0](z)= {\rm cste} \,
 $ in the notation of section \ref{se:HK}. 
  Under this condition, if the free propagator is known, then the one-loop effective potential is automatically known since its derivative is given by the propagator evaluated at coincident endpoints. 

For example, in the case of a scalar fluctuation $\PhiH_D$ in AdS, and assuming that the action  contains 
\be   \int d^{d+1} x \sqrt{|\g|} \frac{1}{2}(\PhiH_D)^2 U[\PhiL_0]\,, \ee
the effective potential can be computed if the $U[\PhiL_0] $ background is constant in $z$.

Here $U[\PhiL_0]$ is in general a composite operator, for concreteness we choose a monomial $U[\PhiL_0]=\prod^r_{i=1}\PhiL_{0,i} K_i(z)$. Requiring constant $U[\PhiL_0]$ does not impose that the individual $ K_i(z)$ are each constant in $z$, only  their product has to be.   In  AdS, the condition for having constant $U[\PhiL_0]$ then becomes  a condition on the $z$ scaling of its elements. Since we have $K_i(z)\propto z^{a_i}$, $U[\PhiL_0]$ is constant if the powers  satisfy
\be
\sum_{i=1}^r a_i =0 \label{eq:suma}
\ee
This condition seems to be usually left implicit in the literature.

Along these lines we can ask: What is the CFT dual of the $U[\PhiL_0]$ background in AdS? 
A possible answer is a follows.  
Provided we can apply the $\Delta_+$ branch of AdS/CFT \cite{Klebanov:1999tb} to  $U[\PhiL_0]$ itself, the resulting source term scales as $z^0$ and the associated fluctuation  scales as $z^d$ near the boundary. This corresponds to a marginal operator ${\cal O}_U$ (\textit{i.e.} $\Delta_U=d$) in the CFT, appearing as $S_{\rm CFT}+  \int d^d x  J_U {\cal O}_U$ with $J_U\equiv U_0=\prod^r_{i=1}\PhiL_{0,i}  $. When $U_0$ takes its $x$-independent background value, $J_U$ is a mere constant sourcing the VEV of ${\cal O}_U$. The nonzero VEV of ${\cal O}_U$ does not break conformal symmetry since it is marginal. Instead, the values of $J_U$ define a continuous family of CFT in which $\langle {\cal O}_U \rangle $ can be nonzero. According to this picture it follows  that the AdS one-loop potential encodes the leading non-planar effects of   $\langle {\cal O}_U \rangle $  on the CFT data. This may deserve further investigation.

When departing from the specific case of constant background field discussed here,  a derivative expansion in the slowly-varying  background can be used. This line of thinking ultimately leads to  the background field method and to the heat kernel expansion, which is our next topic.

\subsection{Review of the Heat Kernel Coefficients}
\label{se:HK}

We use Lorentzian signature. 
The one-loop effective action takes the form 
\be
\Gamma_{1-{\rm loop}} = (-)^F \frac{i}{2}{\rm Tr}\log\left[\left( -\square +m^2 +X(\PhiL) \right)_{ij}\right]  
\ee
with $\square =g_{MN}D^MD^N $ the Laplacian built from background-covariant derivatives. 
The covariant derivatives give rise to a background-dependent field strength $\Omega_{MN}=[D_M,D_N]$, encoding both gauge and curvature connections. It takes the general form
\be
\Omega_{MN} = -i F^a_{MN}t_a-\frac{i}{2}R_{MN}^{~~~~PQ}J_{PQ}
\ee
where $t_a$ and $J_{PQ}$ are the generators of the gauge and spin representation of the quantum fluctuation. 
$X$ is the ``field-dependent mass matrix'' of the quantum fluctuations, it is a local background-dependent quantity. 
The effective field strength $\Omega_{MN}$ and the effective mass $X$ are, together with the curvature tensor, the building blocks of the heat kernel coefficients. 
Using the heat kernel method reviewed in App.~\ref{app:HK}, $\Gamma_{1-{\rm loop}} $ takes the form
\begin{align}
& \Gamma_{1-{\rm loop}}  =  (-)^F \frac{1}{2}
\frac{1}{(4\pi )^{\frac{d+1}{2}}}
\int_{\cal M} d^{d+1}x \sqrt{\g} \sum^\infty_{r=0} \frac{ \Gamma(r-\frac{d+1}{2})   }{m^{2r-d-1}}{\rm tr}\,b_{2r}(x) 
 \nn \\
& ~~~~ + (-)^F \frac{1}{2} 
\int_{\partial \cal M} d^{d}x \sqrt{\bar \g} \sum^\infty_{r=0} \left( \frac{ \Gamma(r-\frac{d+1}{2})   }{(4\pi )^{\frac{d+1}{2}} m^{2r-d-1}}{\rm tr}
\,b^\partial_{2r}(x)+
\frac{ \Gamma(r-\frac{d}{2})   }{(4\pi )^{\frac{d}{2}} m^{2r-d}}{\rm tr}
\,b^\partial_{2r+1}(x)
\right)
\label{eq:Gam1_b}
\end{align}
with $\rm tr$ the trace over internal (non-spacetime) indexes. Analytical continuation in $d$ has been used, the expression is valid for any dimension. The local quantities $b_{2r}$ and $b^\partial_{2r}$ are referred to as the bulk and boundary heat kernel coefficients.

For odd bulk dimension    all the bulk terms in Eq.\,\eqref{eq:Gam1_b} are finite. There are log divergences from the $b^\partial_{2r+1}$ terms for $r\leq \frac{d}{2}$, which renormalize the boundary-localized fundamental action. 
For even bulk dimension there are log-divergences from both bulk terms and from the $b^\partial_{2r}$ terms for $r\leq \frac{d+1}{2}$. These log divergences renormalize both the bulk and boundary fundamental actions. 

The terms with negative powers of masses in Eq.\,\eqref{eq:Gam1_b}  are finite. They amount to an expansion for large $m$ and give rise to a long-distance effective action, $S_{\rm eff}=\int d^{d+1}x \sqrt{|\g|} {\cal L}_{\rm eff}+\int d^{d}x \sqrt{|\bar\g|} {\cal L}^{\partial}_{\rm eff}$ 
with
\be
 {\cal L}_{\rm eff} = (-)^F
\frac{1}{2}
\frac{1}{(4\pi )^{\frac{d+1}{2}}}
\sum^\infty_{r=[(d+3)/2]} \frac{ \Gamma(r-\frac{d+1}{2})   }{m^{2r-d-1}} {\rm tr}\, b_{2r}(x)
\label{eq:Leff}
\ee
\be
 {\cal L}^\partial_{\rm eff} =(-)^F \frac{1}{2} 
\left(  \sum^\infty_{r=[(d+3)/2]}  \frac{ \Gamma(r-\frac{d+1}{2})   }{(4\pi )^{\frac{d+1}{2}} m^{2r-d-1}}{\rm tr}
\,b^\partial_{2r}(x)+
 \sum^\infty_{r=[d/2]+1} \frac{ \Gamma(r-\frac{d}{2})   }{(4\pi )^{\frac{d}{2}} m^{2r-d}}{\rm tr}
\,b^\partial_{2r+1}(x)
\right)
\ee

Only the first heat kernel coefficients are explicitly known, we use up to $b_6$ and $b^\partial_5$ respectively for bulk and boundary coefficients.

\subsubsection{Bulk Contributions}

The general expressions for the bulk coefficients are \cite{Vassilevich:2003xt}
\begin{align}
b_0=&I \nn \\
b_2=&\frac{1 }{6}RI-X \nn \\ \nn 
b_4=&\frac{1}{360}\Big(
12 \square R+5 R^2-2R_{MN}R^{MN}+2R_{MNPQ}R^{MNPQ} \Big)I \label{eq:b4}
\\
& - \frac{1}{6} \square X - \frac{1}{6} R X + \frac{1}{2} X^2 + \frac{1}{12} \Omega_{MN}\Omega^{MN}
\end{align}
\begin{align}
b_6 =  \frac{1}{360}\bigg( &
8 D_{P}\Omega_{MN}D^{P}\Omega^{MN}
+2 D^{M}\Omega_{MN} D_{P}\Omega^{PN}
+12 \Omega_{MN}\square \Omega^{MN}
-12 \Omega_{MN}\Omega^{NP}\Omega^{~~M}_{P}  \nn  \\ \nn 
&-6 R_{MNPQ}\Omega^{MN}\Omega^{PQ}
-4 R_{M}^{~N}\Omega^{MP}\Omega_{NP} + 5 R\Omega_{MN}\Omega^{MN} \\ & \nn
- 6 \square^2 X +60 X\square X+ 30 D_M X D^M X - 60 X^3
\\ & \nn
- 30 X \Omega_{MN}\Omega^{MN} - 10 R \square X - 4 R_{MN} D^N  D^M X - 12 D_M R D^M X + 30 XX R \\ & 
- 12 X \square R  - 5 X R^2 + 2 X R_{MN}R^{MN} - 2 X R_{MNPQ}R^{MNPQ}
\bigg)   +O\Big(R^3\Big) \nn \\ &  \label{eq:b6}
\end{align}
with $I$ the identity matrix for internal indexes. 
Here we give only the part of $b_6$ relevant for our applications, the full $b_6$ coefficient is given in appendix, Eq.\,\eqref{eq:b6full}.

The invariants are built from background fields expressed in holographic variables, including the boundary components such as $\Phi_0\cb K$. The boundary-to-bulk propagator satisfies the bulk EOM, hence the terms involving Laplacians such as those arising from $\square X$ can be evaluated using the  bulk EOMs.

\subsubsection{Boundary Contributions}

More invariants are involved in the boundary contributions to the heat kernel coefficients. 
Using the conventions from Sec.~\ref{se:conv}, we introduce the extrinsic curvature $L=L_{\mu\nu}\bar \g^{\mu\nu}$, with $L_{\mu\nu}=\Gamma^n_{\mu\nu}$ where the normal $n^M$ vector is outward-pointing from the boundary.\,\footnote{In Ref.\,\cite{Branson:1999jz} the $n_M$ vector is inward pointing. The sign of $\partial_\perp$,$L_{ab}$, and consequently $D_\perp$,  is flipped in the boundary heat kernel coefficients  under this change of convention 
} 
One also needs to characterize the boundary conditions of the fluctuation, which can  either be Dirichlet  or  Neumann. This is done in general via the projectors $\Pi_{\pm}$, satisfying $(\Pi_\pm)^2=\Pi_\pm$ and $\Pi_++\Pi_-=I$. The $\Pi_-$ ($\Pi_+$) selects the components of the fluctuation satisfying Dirichlet (Neumann) boundary conditions. One also introduces
\be
\chi=\Pi_+-\Pi_-
\ee
Finally $S$ is the boundary term appearing in  the Neumann BC 
of the fluctuation as $(D_\perp -S)$, \textit{e.g.} $(\partial_\perp -S)\Phi|_{\partial \cal M}^{\rm fl} =0 $.
The boundary heat kernel coefficients are known for arbitrary boundary condition of the fluctuation up to $b^\partial_5$ \cite{Branson:1999jz}. The definitions of the $\Pi_\pm$ are set accordingly.

Focusing on the terms involving $X$ and $\Omega$, the boundary heat kernel coefficients are given by \cite{Branson:1999jz}
\begin{align}
b^\partial_3 &=-\frac{1}{4} \chi X \nn \\ 
b^\partial_4 & = \left(\frac{2}{3}\Pi_+- \frac{1}{3}\Pi_-\right) \partial_\perp X +\frac{1}{3} L X  - 2 SX
\end{align}
\begin{align}
b^\partial_5  = \frac{1}{5760}\bigg(&
-360\chi D_\perp \partial^\perp X +1440 S \partial_\perp X  +720 \chi X^2 \nn \\ & -240 \chi \partial_\mu\partial^\mu X -240 R \chi  X - 2880 S^2 X
  -
(90\Pi_++450\Pi_-)L \partial_\perp X \nn \\
 & +1440 LSX 
- (195\Pi_+-105\Pi_-)L L X
- (30\Pi_++25\Pi_-)L_{\mu\nu}L^{\mu\nu}  X \nn \\
& -180X^2 +180\chi X\chi X-\frac{105}{4}\Omega_{\mu\nu}\Omega^{\mu\nu}
+120\chi\Omega_{\mu\nu}\Omega^{\mu\nu}
+\frac{105}{4}\chi\Omega_{\mu\nu}\chi\Omega^{\mu\nu}
\nn \\
&
-45\Omega_{\mu\perp}\Omega^{\mu\perp}
+180\chi\Omega_{\mu\perp}\Omega^{\mu\perp}
-45\chi\Omega_{\mu\perp}\chi\Omega^{\mu\perp}
\bigg)+\ldots
\end{align}
The ellipses represent pure curvature terms which are irrelevant for our applications. 

The invariants contain the boundary components of the background such as $\Phi_0\cb K$. When a normal derivative acts on such a term, it can be evaluated using the discontinuity equation on the boundary, Eq.\,\eqref{eq:DiscN}, just like in the evaluation of the free part of the holographic action (see Sec.~\ref{se:gen}).

\subsection{Scalar Fluctuation}
\label{se:s0}

The action for a set of  massive scalar fields $\Phi$ with mass $m^2$ is 
\begin{align}
S[\Phi]=-\int_{\cal M} d^{d+1}x\sqrt{\g} & \left( \frac{1}{2} D_M \Phi^a D^M \Phi^a + \frac{1}{2}m^2\Phi^a\Phi^a + V(\Phi,\PhiL) +\frac{1}{2}\xi R \Phi^a\Phi^a  \right) \nn  \\ & + \frac{1}{2}\int_{\partial {\cal M}} d^d y\sqrt{\bar \g}_y  \Phi^a_0 B \Phi^a_0 \,
\label{eq:free_action_scalar}
\end{align}
We have included the $\xi$ coupling to the scalar curvature.  
The potential can depend on other fields denoted collectively as $\PhiL$.
We split the quantum field as $\Phi=\Phi|_{\rm bg}+\Phi|_{\rm fl}$.
The wave operator for the bulk fluctuation $\Phi|_{\rm fl}$ is
\be
{\cal D}^{ab}=-  (D_MD^M)^{ab} +(V'')^{ab} + \xi R\delta^{ab} +m^2 \delta^{ab}
\ee
with $V''=\frac{\delta^2} {\delta \Phi\delta \Phi} V$.

 The canonical invariants needed to evaluate the heat kernel coefficients are thus 
\be
X = (V''(\PhiL|_{\rm bg})^{ab} + \xi R\delta^{ab} \label{eq:Xscal}
\ee
\be
\Omega_{MN}=-iF_{MN}|_{\rm bg} \label{eq:Omegascal}
\ee
Here $F_{MN}=F_{MNa}t_\Phi^a$ where $t_\Phi^a$ 
are the generators of the gauge representation of $\Phi$. 

Regarding boundary conditions of the fluctuation, the $\Pi_\pm$ projectors associated to the BC  are $\Pi_+=1$, $ \Pi_-=$ (Neumann) or $\Pi_+=0$, $ \Pi_-=1$ (Dirichlet).

\subsection{Vector Fluctuation}

\label{se:s1}

The YM action is\,\footnote{
Definitions:  $F^a_{MN}=\partial_M A_N^a- \partial_N A_M^a + f^{abc} A_M^b A_N^c= D_M A_N^a- D_N A_M^a + f^{abc} A_M^b A_N^c$, $D_M=D^{(R)}_M-iA^a_M t_G^a $, $(t_G^b)_{ac}=i f^{abc}$ with  $D^{(R)}_M $ the Riemann covariant derivative. 
}
\be
S[A]=-\frac{1}{4g^2_{YM}}\int_{\cal M} d^{d+1}x\sqrt{\g } F^a_{MN}F^{MNa}
\ee
Splitting the field as $A_M= A_M|_{\rm bg} +{A}_M|_{\rm fl}$, 
the bilinear action for the fluctuation is
\be
-\frac{1}{2g^2_{YM}}\int_{\cal M} d^{d+1}x\sqrt{\g }\bigg( A^a_M|_{\rm fl}\Big[-\square^{ab} \g_{MN} 
+ R_{MN} \delta^{ab}
  +2 f^{acb} F^{ c}_{MN}\Big] A^b_N|_{\rm fl} -  (D^MA^a_M|_{\rm fl} )^2
\bigg) +{\rm bdry~ term} \label{eq:YM_quad}
\ee
where we have used $D^MA^a_N D^NA^a_M = (D^MA^a_M)^2 - f^{abc}F^b_{NM} - R_{NM}\delta^{ab}$. 
The terms in bracket in Eq.\,\eqref{eq:YM_quad} is the wave operator. The subsequent canonical invariants are
\be
(X)^{ab}_{MN}=R_{MN} \delta^{ab}
  +2 f^{acb} F^{ c}_{MN}|_{\rm bg} \label{eq:XYM}
\ee
\be
(\Omega_{MN})^{Pab}_{~~Q}= -R^P_{~~QMN}\delta^{ab} +\delta^P_Q f^{acb} F^c_{MN}|_{\rm bg} \,. \label{eq:OmegaYM}
\ee

\subsubsection{Boundary Conditions}

Near the boundary, the gauge field decomposes as  $A_N=(A_\perp,A_\mu)$. The BC of the  fluctuation are constrained by gauge invariance to either be Neumann for $A_\mu$ and Dirichlet for $A_\perp$ or the converse (see \textit{e.g.} \cite{Vassilevich:2003xt}).
These two  BCs are sometimes respectively referred to as ``absolute'' and ``relative''.
Explicitly, the two possible BCs are 
\be
A_\perp|_{\partial \cal M}=0\,,\quad \partial_\perp A_\mu|_{\partial \cal M} =0 \quad\quad\quad{\rm (absolute)} \label{eq:BCabsolute}
\ee
\be
(D_\perp-L)A_\perp|_{\partial \cal M}=0\,,\quad  A_\mu|_{\partial \cal M} =0 \quad\quad\quad{\rm (relative)} \label{eq:BCrelative}
\ee
 The $\Pi_\pm$ projectors associated to the fluctuation $A_M|_{\rm fl}$ are \cite{Vassilevich:2003xt}
\be
\Pi_+= \delta_{MN} - \delta_{M \perp} \delta_{N \perp}\,\quad \quad\quad 
\Pi_-= \delta_{M\perp} \delta_{N \perp } \,.
\ee



\subsubsection{Gauge Fixing}

We fix the gauge redundancy  of the fluctuation using the Faddeev-Popov procedure. 
As usual in background field  calculations (see \textit{e.g.} \cite{Vassilevich:2003xt,Peskin:257493,Hoover:2005uf,Fichet:2013ola}), we pick the background version of the Feynman-`t\,Hooft gauge 
\be
S[A]_{\rm FP}=-\frac{1}{4g^2_{YM}}\int_{\cal M} d^{d+1}x\sqrt{\g }   (D^MA^a_M|_{\rm fl} )^2
 \label{eq:YM_FP}
\ee
which cancels the corresponding term in Eq.\,\eqref{eq:YM_quad} . The quadratic Lagrangian of the ghost  is simply $\bar c^a \square^{ab} c^b$.
The corresponding canonical invariants are those of a scalar fluctuation, Eqs.\,\eqref{eq:Xscal},\,\eqref{eq:Omegascal}  with $V=0$, $\xi=0$. If the gauge fluctuation satisfies the BC Eq.\,\eqref{eq:BCabsolute} [resp. Eq.\,\eqref{eq:BCrelative}], the ghost has Neumann BC $\partial_\perp c^a|_{\partial \cal M}=0$   [resp.  Dirichlet BC $c^a|_{\partial \cal M}=0$].  
Since ghosts anticommute ($F=$\,odd) the total, gauge-invariant heat kernel  coefficients from the YM fluctuation are 
\be
b^{(\partial)}_{\rm tot}=b^{(\partial)}_{A}- 2b^{(\partial)}_{\rm gh}\,.
\ee

\section{Application 1: Scalar Loops in AdS}

\label{se:AdSLoop}

In this section we apply the formalism laid down in section~\ref{se:HOLEA} to scalar QFTs in AdS$_{d+1}$ spacetime with  boundary truncated at $z=z_0$. 
That is, we will extract one-loop data from the boundary effective action --- and translate it into CFT language --- without the need of calculating AdS loops.

We will  focus on  the pieces of the heat kernel invariants that depend on the boundary components $K\cb \varphi_0(x)$.\,\footnote{This amounts to restricting our interest to  the holographic vertices, \textit{i.e.} the simplest operators  encoded in the boundary effective action (see  Sec.~\ref{se:interactions}). }
Importantly, the boundary component of the holographic basis in \textit{on-shell in the bulk}. 
This fact will help simplifying the heat kernel coefficients, because the derivatives acting on $K$ are evaluated using the bulk equation of motion. 

Having background fields which are on-shell in the bulk while off-shell on the boundary is a condition needed  for the  AdS/CFT correspondence to hold. 
In an example we will show  that a piece of CFT data from an exact AdS loop is correctly reproduced at all available orders by the heat kernel coefficients upon use of the bulk equations of motion.

We   work in the long distance  regime  $(y-y')^2\gg z z_0$ for which $\zeta(x,x')\gg \frac{1}{k^2}$.
The $K|_{\zeta\gg 1}$ approximation satisfies the EOM to $O(\frac{1}{\zeta})$ accuracy. 
The normal derivative on the boundary is $\partial_\perp=-kz_0 \partial_z$. The extrinsic curvature is $L^\perp_{\mu\nu}=-k g_{\mu\nu}$, $L= - kd $.

\subsection{Evaluating the Boundary One-loop Effective Action}

\label{se:AdSLoopScalar}

We consider a massive scalar $\Phi$ whose fluctuations we will integrate out.
The scalar $\Phi$   interacts with light scalars $\varphi_i$ via the  coupling 
\be V = \frac{1}{2}\Phi^2 \prod^\n_{i=1} \varphi_i \,. 
\ee
Each scalar has a distinct mass given by $m^2_i=\Delta_i(\Delta_i-d)k^2$.
We choose $\Delta_i= \Delta_{i,+}$. 
 We set the coupling to curvature $\xi=0$ to zero (the generalization to nonzero $\xi$ is straigthforward). 

We are interested in the $\varphi_i$ background, the term from  $V$ we focus on  is $\frac{1}{2}\Phi \prod^\n_{i-1} \varphi_{i,0} \cb K_i (x) $ with $K_i (x)=K_{\Delta_i} (x)$.  It follows that the canonical background-dependent mass is
\be
X=\prod^\n_{i-1} \varphi_{i,0} \cb K_i (x) 
\ee
The field strength $\Omega_{MN}$ vanishes.

The $\Phi$ field may have a  bilinear boundary-localized action $B$ influencing its propagator. This boundary action may be tuned so that the field contains a light mode. Here we assume $B=0$,  implying that there is no light mode in $\Phi$. As a result both $\Phi_D$ and $\Phi_0$ can be integrated out in a large-mass expansion without further subtleties, which then gives a long-distance EFT for $\varphi$ (see also discussion in section \ref{se:EFT_discussion}).

\begin{figure}
\centering
	\includegraphics[width=0.3\linewidth,trim={0cm 10cm 13cm 0cm},clip]{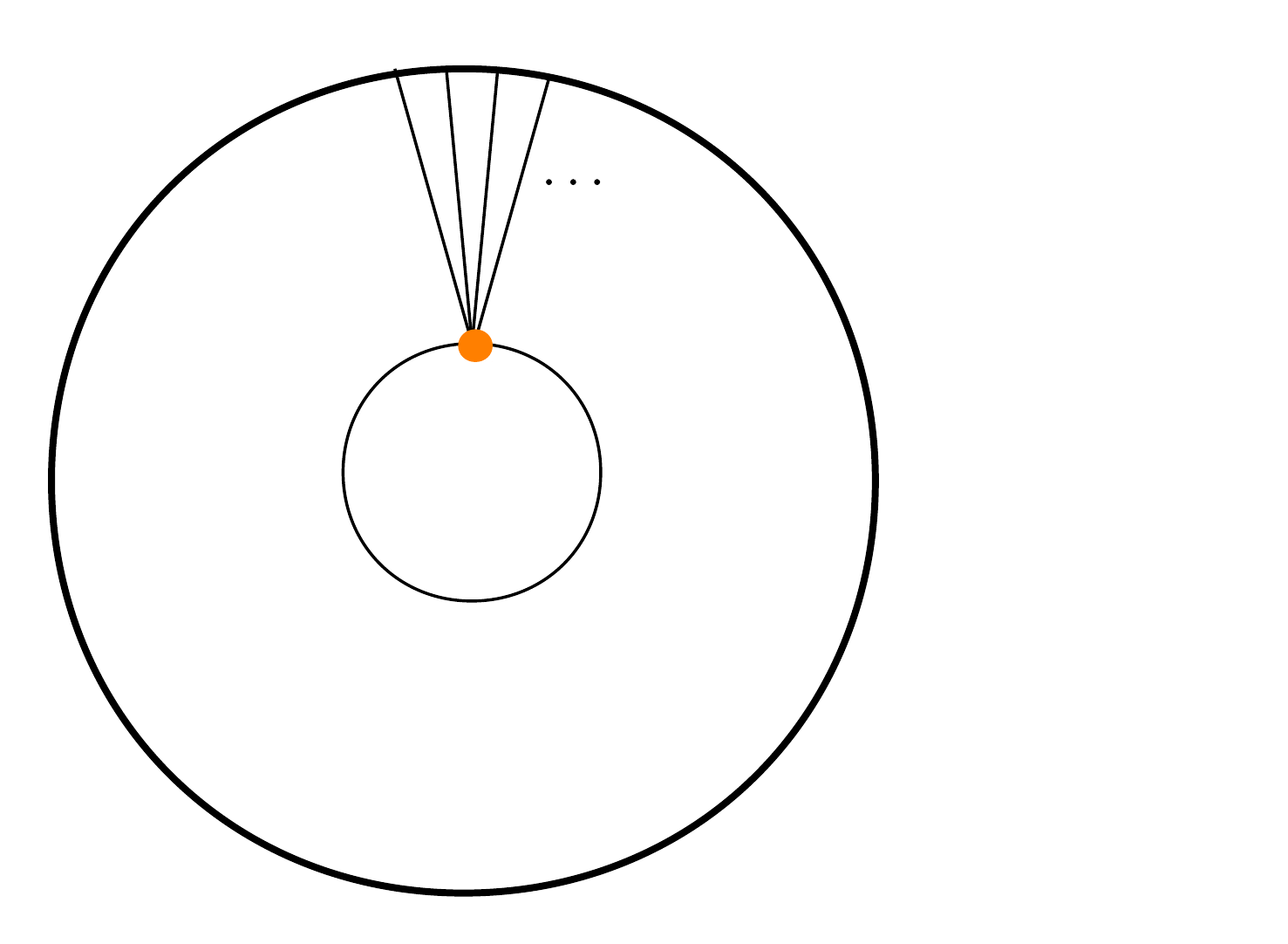}
	\includegraphics[width=0.3\linewidth,trim={0cm 10cm 13cm 0cm},clip]{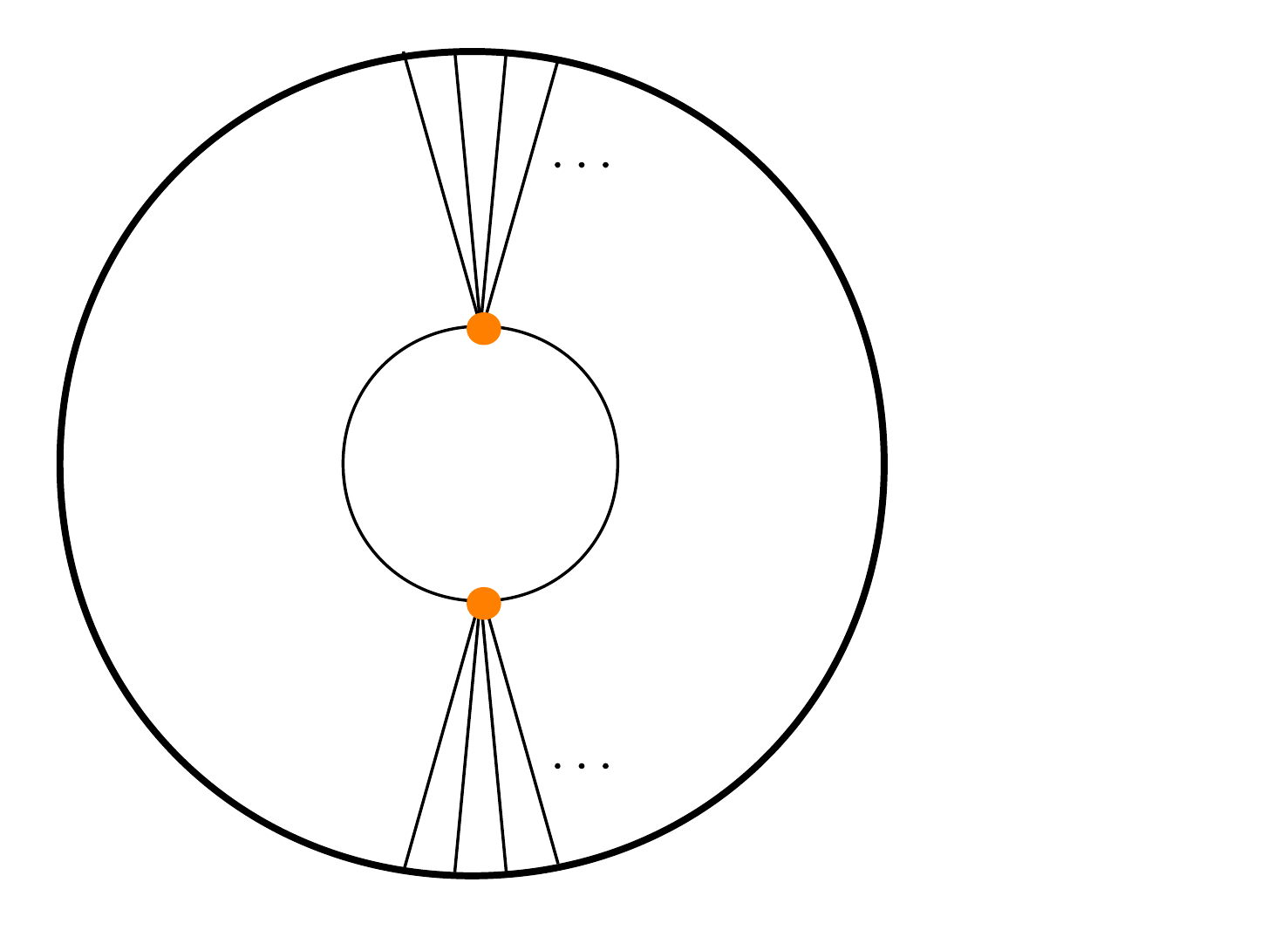}
	\includegraphics[width=0.3\linewidth,trim={0cm 10cm 13cm 0cm},clip]{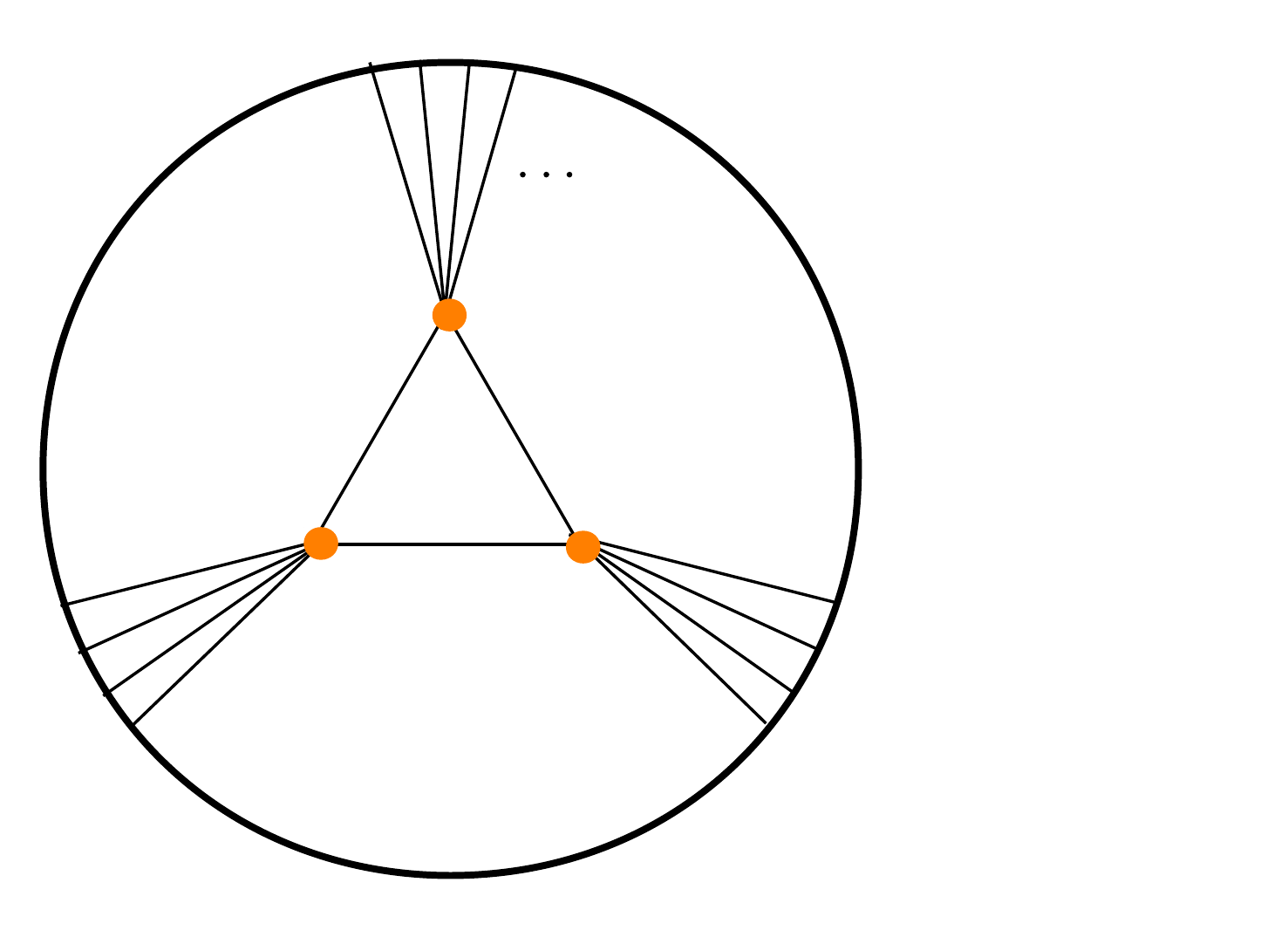}
	\label{eq:AdS_scalar_diags}
    \caption{AdS$_{d+1}$ Witten diagrams  contributing to the first heat kernel coefficients. Internal lines can either have  Dirichlet ($\Delta_+$) or Neumann ($\Delta_-$) boundary condition. In the latter case each internal line can be further decomposed into a Dirichlet line and a boundary line. 
    }
    \label{fig:diags_AdS}
\end{figure}

The  diagrammatic contributions to the  heat kernel  coefficients  computed here are shown in Fig.\,\ref{fig:diags_AdS}. 

\subsubsection*{Bulk Coefficients}

For the bulk  heat kernel coefficients it is convenient to introduce the integrated coefficient
\be
\bb_r = \int_{\rm AdS} d^{d+1}x \sqrt{|\g|} b_r(x) \,.
\ee
 The nonlocal holographic vertices discussed in section~\ref{se:interactions} appear in the bulk coefficients. We introduce the vertices
\be\Lambda_{\n,r}(x_{0,i})=\int_{z_0 }^\infty \frac{dz}{(kz)^{d+1}}  \left(\prod^\n_{i=1} K_i(x_{0,i},x) \right)^r
\label{eq:Lambdanr}
\ee
We define the shortcut notations $X_0(x)=\V(x)  $ and
\be
\int_{z_0 }^\infty \frac{dz}{(kz)^{d+1}} \left( \prod^\n_{i=1} K_i  \cb \varphi_{i,0} (x) \right)^r  = \Lambda_{\n,r} \cb \left[\V\right]^r(x) = \Lambda_\n \cb [X_0]^r (x)
\ee
where each of the endpoints of $\Lambda_n$ is convoluted using the  $\cb$ product with the relevant boundary variable $\varphi_{i,0}$. 
Our bulk results will be expressed in terms of these quantities. We remind that $\Lambda_n$ generates a $n$pt contact Witten diagram upon differentiation of the action in the boundary fields $\varphi_{i,0}$.

To evaluate the bulk heat kernel coefficients we need to evaluate $\square X$. When both derivatives hit the same field inside $X$,  the EOM can be used and gives  a $\sum^\n_{i=1} \Delta_i(\Delta_i-d)k^2 $ contribution. When each derivative hits a different field inside $X$, one needs to evaluate the cross terms $\partial_M K_i\partial^M K_j$. The exact answer is complicated. However, in the $\zeta\gg \frac{1}{k^2}$ limit, these cross terms  simplifies to $\partial_M K_i\partial_M K_j\sim \Delta_i\Delta_j k^2$. 
The cross terms combine with the square term to give
\be
\square X|_{\zeta\gg 1} =\left(\sum^\n_{i=1}\Delta_i\right)\left(\sum^\n_{i=1}\Delta_i-d\right) k^2 X|_{\zeta\gg 1} \label{eq:boxX_AdS}
\ee
We see that the $\Delta_i$ add up. This is the expected result from AdS/CFT: The dimension of the $X$ composite  is $\Delta_X=\sum^\n_{i=1}\Delta_i+O(\frac{1}{N^2})$ at leading order in the large $N$ expansion. The coefficient showing up in Eq.\,\eqref{eq:boxX_AdS} is the quadratic Casimir of $X$. 
Beyond the  $\zeta\gg \frac{1}{k^2}$ regime, \textit{i.e.} for distances shorter that $|y-y'|\sim z_0$, the relation Eq.\,\eqref{eq:boxX_AdS} does not hold. This is a way to see that AdS/CFT breaks down at distances shorter than $z_0$. 

The contributions  to the bulk coefficients from the scalar interactions encoded in $X$ are found to be
\begin{align}
 {\rm tr} \,\bb_2&= - \Lambda_\n \cb X_0   \label{eq:b2_scal} \\
 {\rm tr} \, \bb_4& = \frac{1}{2} \Lambda_{\n,2} \cb [X_0]^2 + \frac{1}{6}\left(- \DeltaX(\DeltaX-d)+d(d+1) \right) k^2 \Lambda_\n \cb X_0 
 \label{eq:b4_scal}
\end{align}
\begin{align}
 {\rm tr} \, \bb_6  = &  - \frac{1}{6} \Lambda_{\n,3} \cb [X_0]^3 (x)+
\frac{1}{12} \left( \DeltaX(\DeltaX-d)- d(d+1)\right) k^2  \Lambda_{\n,2} \cb [X_0]^2 (x) \nn \\ &  \label{eq:b6_scal}
-\frac{1}{360}\Big(
8d^3+5d^4+6 \DeltaX^2(\DeltaX-d)^2 +d^2 (7-22\DeltaX(\DeltaX-d)) \\ \nn  &  \quad\quad\quad\quad\quad\quad
+d (4-26\DeltaX(\DeltaX-d))
\Big) k^4 \Lambda_\n \cb X_0 (x)
\end{align}
In the $\bb_6$  coefficients, we performed  integration by parts on certain terms \textit{i.e.} we used Green's first identity. The generated boundary term contributes to the boundary coefficient $b^\partial_6$. In this work we stopped at $b^\partial_5$ --- the general $b^\partial_6$ has apparently not been computed in the heat kernel literature --- hence the boundary term arising from integration by parts in $b_6$ is neglected.

\subsubsection*{Boundary Coefficients}

We need other identities to evaluate the boundary coefficients. Normal derivatives evaluated on the boundary (such as  $\partial_\perp K$) appear. It is convenient to  display $\partial_\perp K$ in terms of a CFT 2pt function. 
We have (see section \ref{se:AdSB2b})
\begin{align}
\partial_\perp K_i = & -\frac{1}{\sqrt{\bar \g}}\left(\frac{{\cal C}_{\Delta_i}\eta_{\Delta_i}}{(y-y')^{2\Delta_i}}
+\Delta_{i}^- k \delta^d(y-y') 
\right) \nn \\ 
= & - \frac{1}{\sqrt{\bar \g}}\left(
\langle{\cal O}_i(y){\cal O}_i(y')\rangle
+\Delta_{i}^- k \delta^d(y-y') 
\right)
\end{align}
where in the second line we have introduced the scalar primary ${\cal O}_i$ with dimension $\Delta_i$, normalized such that the ${\cal C}_{\Delta_i}\eta_{\Delta_i}$ coefficients are absorbed.  

We then evaluate $\partial_\perp X$. The contact terms combine and make the dimension $\Delta_{X}^-=\sum^\n_{i=1} \Delta_{i}^-= \sum^\n_{i=1} (d-\Delta_{i}) $ appear.
The $\partial_\perp X$ derivative  takes the form
\be
\partial_\perp X =-\Delta_{X}^- k  X_0 - \frac{1}{\sqrt{\bar \g}}
\sum^\n_{i=1}\langle{\cal O}_i{\cal O}_i\rangle \cb X_0
\label{eq:delXAdS}
\ee
where one has introduced a shortcut notation for the $\cb$ convolutions: the  correlator for ${\cal O}_i$ is convoluted with the corresponding $\varphi_{i,0}$ inside $X$. 

We turn to the evaluation of $D^\perp \partial_\perp X$. 
We evaluate $D^\perp \partial_\perp K$ in momentum space and get only a contact term $D^\perp \partial_\perp K=(\Delta(d-\Delta)k^2  - k^2 z_0^2 \partial_\mu\partial^\mu )\delta^d(y-y')  $.  The cross terms $\partial_\perp K_i \partial^\perp K_j$ are evaluated using Eq.\,\eqref{eq:delXAdS}. The contact terms from the cross terms combine with those from the diagonal terms, leading to a coefficient $\Delta_{X}^-(d-\Delta_{X}^-)$ for the total contact term. We see that the $\Delta_{X}^-$ add up, which is consistent with the underlying conformal symmetry.  Again, this combination happens from the $|y-y'|\gg z_0$ regime of the boundary-to-bulk propagators. Beyond this regime no such simplification  occurs, illustrating the breakdown of AdS/CFT.  

The full term reads
\begin{align}
D^\perp \partial_\perp X = &  \Delta^-_{X}(d-\Delta^-_{X}) k^2 X_0-
k^2 z_0^2\sum^\n_{i=1}(\partial_\mu\partial^\mu)_{i} X_0 \\ \nn &
+ 
\frac{2}{\sqrt{\bar \g}} \Delta^-_{X} k 
\sum^\n_{i=1}\langle{\cal O}_i{\cal O}_i\rangle \cb X_0
+ 
\frac{1}{\sqrt{\bar \g}}\frac{1}{\sqrt{\bar \g}}
\sum^\n_{i\neq j}[\langle{\cal O}_i{\cal O}_i\rangle \cb][\langle{\cal O}_j{\cal O}_j\rangle \cb] X_0\,.
\end{align}
Again we use a shortcut notation for the convolutions (correlators with label $i$ contract with the $\varphi_{i,0}$ fields inside $X_0$).

We choose Neumann BC for the fluctuation (\textit{i.e.} we integrate out both $\Phi_D$ and $\Phi_0$). 
Using that $\Pi_+=1$, $\Pi_-=0$, $\chi=1$, $S=0$, the boundary heat kernel coefficients are found to be
\begin{align}
{\rm tr}\, b^\partial_3 &=-\frac{1}{4} X_0 \\
{\rm tr}\, b^\partial_4 & = -\frac{2}{3} \frac{1}{\sqrt{\bar \g}}
\sum^\n_{i=1}\langle{\cal O}_i{\cal O}_i\rangle \cb X_0
-\frac{1}{3}(2\DeltaX^- +  d) k X_0 \,
\end{align}
\begin{align}
{\rm tr}\, b^\partial_5  = \frac{1}{5760}\bigg(& 720 X_0^2  + 
\left(-360  \Delta^-_{X}(d-\Delta^-_{X}) k^2 + 90 d \DeltaX^- k^2 
+15d(14+ 3d)k^2 \right) X_0
\nn \\
&  + 360
k^2 z_0^2\sum^\n_{i=1}(\partial_\mu\partial^\mu)_{i} X_0 
-240 k^2 z_0^2 \partial_\mu\partial^\mu X_0
\nn \\
& + (90 d-720 \DeltaX^- )k  \frac{1}{\sqrt{\bar \g}}
\sum^\n_{i=1}\langle{\cal O}_i{\cal O}_i\rangle \cb X_0
\nn \\
& 
-360\frac{1}{\sqrt{\bar \g}}\frac{1}{\sqrt{\bar \g}}
\sum^\n_{i\neq j}[\langle{\cal O}_i{\cal O}_i\rangle \cb][\langle{\cal O}_j{\cal O}_j\rangle \cb] X_0
\bigg)
\end{align}
The result from Dirichlet BC (\textit{i.e.} from integrating out  $\Phi_D$ and not $\Phi_0$) is shown in App.~\ref{app:b_Dir} for completeness.

Finally we emphasize that all the bulk and boundary coefficients obtained in this section are consistent with the  conformal symmetry of the boundary  in the sense that the conformal dimensions of the components of $X$ appear  only via the total dimensions $\Delta_X$, $\Delta^-_X$, and never in other combinations nor individually.  This feature emerges via nontrivial combinations of coefficients which occur only in the AdS/CFT validity regime $|y-y'|\gg z_0$.

\subsection{Anomalous Dimensions of CFT Operators}

Here we extract information from the one-loop effective action providing the leading non-planar  corrections to the conformal dimensions of  the CFT operators ${\cal O}_i$  associated with  the  light fields  $\varphi_i$. 
Depending on the spacetime dimension, these corrections are either finite and evaluated at large $m_\Phi$ or contain a log divergence. Below we only focus on odd dimensions hence the corrections are finite.

The heavy  field $\Phi$ is in correspondence with a CFT operator with large conformal dimension $\DeltaH \gg \Delta_i$, with 
\be
m^2_\Phi= \DeltaH (\DeltaH -d)k^2 \,. 
\ee
Depending on cases our results are expressed using either $m_\Phi$ or $\DeltaH$.


\subsubsection{Single-trace Operator  }

When the heat kernel coefficients contribute to the bulk mass of a field, this translates as contribution to the anomalous dimension of the corresponding CFT operator. 
We define the anomalous dimension
\be
\Delta_\pm=\Delta^0_\pm\pm\gamma \label{eq:gamma_def}
\ee
where $\Delta^0_\pm$ corresponds to the conformal dimension in the free theory in AdS \textit{i.e.} the CFT with $N\to \infty$.

\paragraph{Anomalous Dimensions from Bubbles }

We consider the interaction Lagrangian 
${\cal L}_{\rm int}=-y\frac{1}{2}\Phi^2\varphi$, giving
$X=y\varphi$. 
The anomalous dimensions arising from bubble diagrams are read from the $\bar b_4$ and $\bar b_6$ coefficients, Eqs.\,\eqref{eq:b4_scal},\,\eqref{eq:b6_scal}.
 We obtain 
\begin{align}
\gamma^{d=2}_{{\cal O}, {\rm bubble}} &=  - \frac{y^2}{16\pi (\Delta-1)} \left(\frac{1}{2\,(\DeltaH (\DeltaH-2))^{\frac{1}{2}}} 
+ \frac{\Delta(\Delta-2)-6}{24\,(\DeltaH (\DeltaH-2))^{\frac{3}{2}}}\right)  +O\left(\frac{1}{\DeltaH^5}\right)
\label{eq:gam2bub}
\\
\gamma^{d=4}_{{\cal O}, {\rm bubble}} &= 
 \frac{y^2}{32\pi^2 (\Delta-2)} \left(\frac{(\DeltaH (\DeltaH-4))^{\frac{1}{2}}}{2} 
- \frac{\Delta(\Delta-4)-20}{24\,(\DeltaH (\DeltaH-4))^{\frac{1}{2}}} \right) +O\left(\frac{1}{\DeltaH^3}\right)
 \label{eq:gam4bub}
\\
\gamma^{d=6}_{{\cal O}, {\rm bubble}} &= 
- \frac{y^2}{192\pi^3 (\Delta-3)} \left(\frac{(\DeltaH (\DeltaH-6))^{\frac{3}{2}}}{2} 
- \frac{\Delta(\Delta-6)-42}{8}(\DeltaH (\DeltaH-6))^{\frac{1}{2}} \right) +O\left(\frac{1}{\DeltaH}\right)
 \label{eq:gam6bub}
\end{align}
In the CFT one has $y\sim \frac{1}{N}$ thus these  are $\sim \frac{1}{N^2} $ corrections to the conformal dimension of $\cal O$.

\paragraph{Comparison with Exact AdS$_3$ Bubble } 

The AdS$_3$ bubble was found in \cite{Carmi:2018qzm} to be
\be
{\cal B}(i\nu, \DeltaH)=iy^2\frac{\psi(\DeltaH-\frac{1+i\nu}{2})- \psi(\DeltaH-\frac{1-i\nu}{2}) }{8\pi \nu}
\ee
in the spectral representation. When evaluated at $\pm i\nu=\Delta-1$ it gives the exact anomalous dimension of the single trace operator via
\begin{align}
& \gamma^{d=2}_{{\cal O}, {\rm bubble}}=\frac{{\cal B}(\Delta-1,\DeltaH)}{2\Delta-2} 
\label{eq:gam2bubexact}
\\\nn 
 & 
\overset{{\rm large}~\DeltaH} {=}  -\frac{y^2}{16\pi(\Delta-1)}\left(
\frac{1}{2\DeltaH}+\frac{1}{2\DeltaH^2}+\frac{12+\Delta(\Delta-2)}{24\DeltaH^3}
+\frac{4+\Delta(\Delta-2)}{8\DeltaH^4}\right)+O\left(\frac{1}{\DeltaH^5}
\right)  
\end{align}
We can then verify that our result Eq.\,\eqref{eq:gam2bub} reproduces exactly Eq.\,\eqref{eq:gam2bubexact} upon expanding in $\DeltaH$ up to $O(\frac{1}{\DeltaH^5})$. 
This matching at fourth order provides a nontrivial sanity check of our results.

In our result Eq.\,\eqref{eq:gam2bub}, the next-to-leading order term originates from the  derivative term  $X\square X$ in  the heat kernel coefficient (Eq.\,\eqref{eq:b6}). This $\square $ was then simplified using the bulk EOM (see Eq.\,\eqref{eq:b6_scal}). 
This illustrates that having background fields on-shell in the bulk is necessary in order to obtain CFT data from the one-loop boundary effective action.

\paragraph{Anomalous Dimensions from Tadpoles }

We consider the interaction Lagrangian 
${\cal L}_{\rm int}=-\frac{1}{4}\lambda\Phi^2\varphi^2$, giving
$X=\frac{1}{2}\lambda\varphi^2$. From the 
$\bar b_{2,4}$ coefficients Eqs.\,\eqref{eq:b2_scal},\,\eqref{eq:b4_scal} we obtain the anomalous dimensions
\be
\gamma^{d=2}_{{\cal O}, {\rm tad}} = 
\frac{\lambda}{16\pi (\Delta-1)} \left(
(\DeltaH (\DeltaH-2))^{\frac{1}{2}}
+
\frac{\Delta(\Delta-2)-6}{12 (\DeltaH (\DeltaH-2))^{\frac{1}{2}} }
\right)+O\left(\frac{1}{\DeltaH^3}\right)
\ee
\be
\gamma^{d=4}_{{\cal O}, {\rm tad}} = 
- \frac{\lambda}{96\pi^2 (\Delta-2)} \left( (\DeltaH (\DeltaH-4))^{\frac{3}{2}}
-
\frac{\Delta(\Delta-4)-20}{4}(\DeltaH (\DeltaH-4))^{\frac{1}{2}}
\right)+O\left(\frac{1}{\DeltaH}\right)
\ee
\be
\gamma^{d=6}_{{\cal O}, {\rm tad}} = 
 \frac{\lambda}{960\pi^3 (\Delta-3)} \left( (\DeltaH (\DeltaH-6))^{\frac{5}{2}}
 + 
 \frac{\Delta(\Delta-6)-42}{16}(\DeltaH (\DeltaH-6))^{\frac{3}{2}}
 \right) +O\left(\DeltaH\right)
\ee
The next-to-leading term can similarly be obtained from the $\bar b_{6}$ coefficient Eq.\,\eqref{eq:b6_scal} and is not shown here for convenience.

\subsubsection{Double-trace Operators  }
\label{se:DT_scalar}

The $-\frac{1}{4}\lambda\Phi^2\varphi^2$ coupling also generates at one-loop  a quartic  coupling  $\frac{1}{4!}\eta \varphi^4$  with $\eta\propto \lambda^2$.  This is encoded in the  $\bar b_4$, $\bar b_6$ coefficients. 
This quartic coupling induces in turn an anomalous correction to the double trace operators $[{\cal O}{\cal O}]_n \equiv {\cal O} \square^n {\cal O}$. Using the perturbation theory approach established in \cite{Fitzpatrick:2010zm} we obtain 
\be
\gamma_{n} = c_n \eta_{\rm 1-loop}
\ee
with 
\be
c^{d=2}_{n} = \frac{1}{8\pi(2\Delta+2n-1)}
\ee
\be
c^{d=4}_{n} = \frac{(n+1)(\Delta+n-1)(2\Delta+n-3)}{8\pi^2(4\Delta+4n-6)}
\ee
\be
c^{d=6}_{n} = \frac{(n+1)(n+2)(\Delta+n-2)(\Delta+n-1)^2(2\Delta+n-5)(2\Delta+n-4)}{64\pi^3(2\Delta+2n-5)(2\Delta+2n-3)}
\ee
computed from \cite{Fitzpatrick:2010zm}
and the loop-generated couplings
\be
\eta^{d=2}_{\rm 1-loop} = -\frac{3\lambda^2}{16\pi(\DeltaH(\DeltaH-2))^{\frac{1}{2}}}
\ee
\be
\eta^{d=4}_{\rm 1-loop} = 
\frac{3\lambda^2(\DeltaH(\DeltaH-2))^{\frac{1}{2}}}{32\pi^2}
\ee
\be
\eta^{d=6}_{\rm 1-loop} = 
-\frac{\lambda^2(\DeltaH(\DeltaH-2))^{\frac{3}{2}}}{64\pi^3}
\ee
obtained from our effective action. 
In the CFT one has  $\lambda\sim \frac{1}{N^2}$ thus these are $\sim\frac{1}{N^4}$ corrections to the conformal dimension of the double trace operators.

\subsection{ Corrections to Boundary-Localized Operators}

The boundary heat kernel coefficients lead to boundary operators which take in general  the form of local operators  with possibly CFT 2pt correlators appended to them. The 2pt correlators are the consequence of normal derivatives acting on $X$ on the boundary. 

In this subsection we focus on the interaction Lagrangian ${\cal L}_{\rm int}=-\frac{1}{4}\lambda\Phi^2\varphi^2$ giving
$X=\frac{1}{2}\lambda\varphi^2$.
For example for AdS$_5$ the boundary coefficients are
\begin{align}
{\rm tr}\, b^\partial_3 &=-\frac{1}{8} \lambda \varphi^2_0 \\
{\rm tr}\, b^\partial_4 & = -\frac{2\lambda}{3} \frac{1}{\sqrt{\bar \g}}
 (\langle{\cal O}{\cal O} \rangle \cb  \varphi_0)\, \varphi_0
+\frac{\lambda}{3}(2\Delta-10) k  \varphi^2_0 \,
\end{align}
\begin{align}
{\rm tr}\, b^\partial_5  = \frac{1}{5760}\bigg(&
 360 \lambda\,
k^2 z_0^2 \, \varphi_0 \partial_\mu\partial^\mu \varphi_0 
 + 
60 \left(12\Delta^2 -78 \Delta + 133\right)
\lambda \, k^2 \varphi^2_0 + 180 \lambda^2 \varphi_0^4 
\nn \\
& 
+ 360(4\Delta-15 )
\lambda\,k  \frac{1}{\sqrt{\bar \g}}
 (\langle{\cal O}{\cal O} \rangle \cb  \varphi_0)\, \varphi_0
-360\lambda\,\frac{1}{\sqrt{\bar \g}}\frac{1}{\sqrt{\bar \g}}
 \left( \langle{\cal O}{\cal O} \rangle \cb  \varphi_0\right)^2
\bigg)
\end{align}
and     enter in the holographic effective action as 
\be
\Gamma^{\partial,{\rm fin}}_{\rm 1-loop}=\frac{1}{2} 
\int_{\partial \rm AdS} d^{4}x \sqrt{\bar \g}  \left( 
   -\frac{m_\Phi}{16\pi^2}
{\rm tr} \,b^\partial_{4}(x)
\right)  \label{eq:Gamfinphi4AdS}
\ee
\be
\Gamma^{\partial,{\rm div}}_{\rm 1-loop}=\frac{1}{2} 
\int_{\partial \rm AdS} d^{4}x \sqrt{\bar \g}  \left(
\frac{m^2_\Phi}{8\pi^2}
{\rm tr}
\,b^\partial_{3}(x) 
-\frac{1}{8\pi^2}
{\rm tr}
\,b^\partial_{5}(x)
\right) \log\frac{m_\Phi}{\mur}\, \label{eq:Gamdivphi4AdS}
\ee
where $\mur$ is the renormalization scale.

\subsubsection{Correction to OPE Coefficients }

\label{se:OPE_corr_AdS}

The effective action encoding the $b^\partial_{4,5}$ coefficients contains terms of the form 
\be
\int d^dy  \langle{\cal O}_i{\cal O}_i\rangle \cb X_0 = 
 \int d^dy_1 d^dy_2 \varphi(y_1)\langle{\cal O}_i(y_1){\cal O}_i(y_2)\rangle \varphi(y_2) \,.
\ee
We can see that this term  induces a one-loop correction to the normalization of the  CFT 2pt function.
For $d=2$ and $d=4$ we find the corrections
\be \left(1 + \frac{\lambda }{12\pi^2m_\Phi} - \frac{8\Delta-31}{128 \pi m^2_\Phi}\lambda k
\right) \langle{\cal O}(y_1){\cal O}(y_2)\rangle
\ee 
\be \left(1 - \frac{\lambda m_\Phi}{24\pi^2} +\frac{4\Delta-15}{128 \pi^2}\lambda k\, 
\log\frac{m_\Phi}{\mur}\right) \langle{\cal O}(y_1){\cal O}(y_2)\rangle
\ee
where the second term is from $b^\partial_4$ and the third from $b^\partial_5$.

In the CFT these corrections amount, upon unit-normalization of the operators, to a correction to the OPE coefficients involving $\cal O$. That is, the  $ (1+\delta_{\cal O} )$ factor turns into a correction to the OPE coefficients by $c_{\cal O \ldots }= c^0_{{\cal O} \ldots }(1 -\frac{1}{2}\delta_{\cal O} )$ where $c^0_{\cal O \ldots }$ is the OPE coefficient value in the planar limit.  
This type of correction amounts to an AdS/CFT version of  ``wavefunction renormalization''.

\subsubsection{Operator Mixing }

The effective action containing  $b^\partial_{5}$  encodes a term
\be
\int d^dy  \left( \langle{\cal O}{\cal O} \rangle \cb  \varphi_0\right)^2 = \int d^dy_1\int d^dy_2\int d^dy_3
\varphi(y_1)\langle{\cal O}(y_1){\cal O}(y_2)\rangle
\langle{\cal O}(y_2){\cal O}(y_3)\rangle
\varphi(y_3) \,.
\ee
Because of conformal symmetry, this term reduces to the 2pt function of another CFT operator, ${\cal O}^2$, which has dimension $2\Delta$ at this order in perturbation theory. Evaluating the $\int d^d y_2$ integral we have
\begin{align}
\int d^{d}y_2  
  \langle{\cal O}(y_1){\cal O}(y_2)\rangle
\langle{\cal O}(y_2){\cal O}(y_3)\rangle
& =  - i \pi^{\frac{d}{2}}\frac{\Gamma^2(\frac{d}{2}-\Delta) \Gamma(2\Delta)}{\Gamma^2(\Delta)\Gamma(d-2\Delta)} 
\frac{{\cal C}^2_\Delta}{{\cal C}_{2\Delta}}
\frac{\eta^2_\Delta}{\eta_{2\Delta}}
\langle{{\cal O}^2}(y_1){\cal O}^2(y_3) \rangle \,\nn \\
& =  -  i \frac{(2\Delta-d)^2}{(4\Delta-d)} kz^{-d}_0 
\langle{{\cal O}^2}(y_1){\cal O}^2(y_3) \rangle \,. 
\label{eq:DT_conv}
\end{align}
The $i$ factor comes from the Lorentzian metric. This  explicitly shows  how a double trace operator  emerges on the AdS boundary as a result of loop corrections. 

The combination arising when taking twice the derivative of the boundary action in $\varphi_0$ ends up being
\be
\langle{{\cal O}}(y_1){\cal O}(y_2)
-   i \frac{(2\Delta-d)^2}{(4\Delta-d)} \frac{\lambda}{64\pi m^2_\Phi} kz^{-d}_0 
\langle{{\cal O}^2}(y_1){\cal O}^2(y_2) \rangle \,.
\label{eq:DT_mix}
\ee
In the CFT language, this correction corresponds to a mixing between ${\cal O}$ and ${\cal O}^2$ induced by non-planar corrections. 

\subsubsection{Corrections to local operators/multitrace deformations}

Local boundary-localized operators are also generated at one-loop.  Defining the tree-level boundary couplings
\be
{\cal L}_0^\partial = -\frac{1}{2}\varphi_0\left(Z_{b,0}z^2_0\partial_\mu\partial^\mu +m^2_{b,0} \right)\varphi_0 -\frac{1}{4!}\lambda_{b,0} \varphi^4_0
\ee
we obtain the following one-loop corrections from the $b^\partial_{3,4,5}$, heat kernel coefficients, 
\be
Z_{b,d=2} = Z^0_b-\frac{\lambda k^2}{64\pi m^2_\Phi} +O(m^{-3}_\Phi)
\ee
\be
m^2_{b,d=2} = m^2_{b,0}+\frac{\lambda}{16\pi}\log\frac{\mur}{m_\Phi}
-\frac{\lambda (2\Delta-5)}{24\pi m_\Phi}+O(m^{-2}_\Phi)
\ee
\be
(\lambda)_{b,d=2} = \lambda_{b,0}-  \frac{3\lambda^2}{32\pi m^2_\Phi} +O(m^{-3}_\Phi)
\ee

\be
Z_{b,d=4} = Z^0_b-\frac{\lambda k^2}{128\pi^2}\log\frac{\mur}{m_\Phi} +O(m^{-1}_\Phi)
\ee
\be
m^2_{b,d=4} = m^2_{b,0} 
-\frac{\lambda m_\Phi^2}{64\pi^2}\log\frac{\mur}{m_\Phi}
+\frac{\lambda (\Delta-5)m_\Phi}{24\pi^2 }+O(m^0_\Phi)
\ee
\be
(\lambda)_{b,d=4} = \lambda_{b,0} - \frac{3\lambda^2}{64\pi^2}\log\frac{\mur}{m_\Phi} +O(m^{-1}_\Phi)
\ee
The radiative generation of such boundary-localized local operators is sometimes   discussed in the extradimension literature at a qualitative level. Here we have obtained their exact coefficients. 

From the viewpoint of AdS/CFT, these local boundary-localized operators may, under certain conditions on the value of $\Delta$, correspond to multitrace deformations in the dual CFT. For example, when $\Delta<\frac{d}{2}+1$ one identifies $\varphi_0$ as the CFT operator itself, and $m^2_b$ then corresponds to a double trace deformation  $\int d^dy{\cal O}^2(y)$ whose effects have been well studied (see \textit{e.g.} \cite{Gubser:2002zh, Hartman:2006dy,Giombi:2018vtc}). 
 Here we have found that such operators are generically generated via loops in AdS. 
 

\section{Application 2: Vector Loops in AdS}
\label{se:YM}

In this section we evaluate pieces of the one-loop effective action  generated by integrating out the spin-1 fluctuation from a nonabelian gauge field. We extract the beta functions of the boundary nonabelian field and anomalous dimensions of scalar CFT operators.

\subsection{Yang-Mills Beta Functions in AdS}

Our focus is on the gauge background field $F_{\mu\nu}|_{\rm bg}$. 
We  use  the set of BCs  Eq.\,\eqref{eq:BCabsolute} for the fluctuation.  
 The BC of the $A_M$ fluctuation enforces the $\Pi_\pm$ projectors to be
$
\Pi_+= \delta_{MN} - \delta_{M \perp} \delta_{N \perp}$, $
\Pi_-= \delta_{M\perp} \delta_{N \perp } 
$.  

We assume the gauge background is on-shell thus  $D_M F^{MN}=0$. Fixing the  gauge of the background  to $D_M A^M=0$, the EOM reduces to $\square A_M=0$. The homogeneous solutions are $z^{\frac{d}{2}-1} I_{\frac{d}{2}-1}(pz) $, $z^{\frac{d}{2}-1} K_{\frac{d}{2}-1}(pz) $.

The gauge background contributes to the bulk coefficients $b_4$, $b_6$ and to the boundary coefficient $b^\partial_5$.
To evaluate the $F_{MN}\square F^{MN}$ terms we use Jacobi's identity and
the commutator $[D_M, D_N]F^a_{PQ}=  [F_{MN},F_{PQ}]^a$. 
The result is
\be
{\rm tr }\left[ F_{MN}\square F^{MN} \right] = {\rm tr } \left[
 4 F_{MN}F^{NP}F_{P}^{~~M} -2 (D_M F^{MN})^2\right]\,. 
\ee
In the $b^\partial_5$ coefficient we also notice that the $F_{\mu\perp}F^{\mu\perp}$ terms vanish by EOM and gauge fixing.

 Taking into account the ghost contributions we find that the total  heat kernel coefficients $b_{\rm tot}=b_A-2b_{\rm gh}$ are 
\begin{align}
{\rm tr}\, b_{4,{\rm tot}}  = & -\frac{25-d}{12} C_2(G) F^a_{MN}F^{a,MN}|_{\rm bg} \label{eq:b4gauge}  \\
{\rm tr}\, b^\partial_{5,{\rm tot}}  = & \frac{27-d}{48} C_2(G) F^a_{\mu\nu}F^{a,\mu\nu}|_{\rm bg}  \label{eq:b5gauge} \\
{\rm tr} \,b_{6,{\rm tot}} = &
 \frac{5d^3-154 d^2 +557 d +12 }{360}  C_2(G) k^2 F^a_{MN}F^{a,MN}|_{\rm bg} \nn
\\ 
 & +\frac{d-1}{90}{\rm tr} F_{MN}F^{NP}F_P^{~~M}|_{\rm bg}
 \label{eq:b6gauge}
\end{align}
where $C_2(G)$ is the quadratic Casimir of the gauge representation, $ f^{acb} f^{adb} = C_2(G) \delta^{cd}$.


\subsubsection{$\beta$ functions}

Both  bulk and boundary gauge couplings can be renormalized by the logarithmic divergences appearing in  the one-loop effective action. Which term of $\Gamma_{\rm 1-loop}$ diverges depends on the dimension of spacetime.
Unlike  flat space where a running gauge coupling can   only happen in $d=4$, we will see that in AdS space this can also occur in higher dimensions as a consequence of the curvature. 
The $\beta$ function of the gauge couplings can be evaluated by putting together the one-loop effective action Eq.\,\eqref{eq:Gam1_b} and the classical part. The classical part is 
\be
S = -\frac{1}{4g^2} \int d^{d+1}x\sqrt{|\g|} (F^a_{MN})^2 
-\frac{1}{4g_b^2} \int  d^{d}y\sqrt{|\bar\g|} (F^a_{\mu\nu})^2\Big|_{z=z_0} \,
\ee
where a boundary-localized kinetic term has been introduced. Bulk and boundary divergences respectively renormalize the $g$ and $g_b$ parameters.  

For AdS$_4$ ($d=3$), the log divergence comes from the ${\rm tr}\,  b_{4,{\rm tot}}$ coefficient. The overall  factor is  $-\frac{11}{6}$.
This reproduces the well-known 4d YM $\beta$ function for the bulk coupling, $\beta_{\frac{1}{g^2}}=\frac{11}{24\pi^2 }  C_2(G)$. 

For AdS$_5$ ($d=4$), the log divergence comes from the ${\rm tr}\,  b^\partial_{5,{\rm tot}}$ coefficient. The $\beta$ function for the boundary gauge coupling is found to be
 \be \beta_{\frac{1}{g_b^2}}\Big|_{{\rm AdS}_5}=  -\frac{23}{192\pi^2} C_2(G) \, .\ee 
 
For AdS$_6$ ($d=5$), the log divergence comes from the ${\rm tr}\,  b_{6,{\rm tot}}$ coefficient. The gauge coupling has canonical dimension $[g]=-1$. 
The  coefficient in Eq.\,\eqref{eq:b6gauge}  reduces to $-\frac{107}{90}k^2$. The $\beta$ function for the bulk coupling is found to be
\be
\beta_{\frac{1}{g^2}}\Big|_{{\rm AdS}_6} =  \frac{107}{1440 \pi^3}k^2 C_2(G) \,. 
\ee
Here we can see that  the logarithmic running of this dimensionful gauge coupling is a consequence of the AdS curvature. This can be alternatively understood as the renormalization of the $RFF$ operators  displayed in section \ref{se:YM_6d_gen}.

\subsubsection{Boundary Yang-Mills Effective Action }

Using the above results, let us  write the bilinear boundary action at one-loop. We work in momentum space.  Irrespective of the gauge,  the $A_z$ background can be eliminated (\textit{i.e.} integrated out at classical level)  using the $D_\mu F^{\mu z}=0$ component of the EOM, giving the identity
\be
F_{MN}F^{MN}=F_{\mu\nu}F^{\mu\nu}+4 \left(\eta_{\mu\nu}-\frac{p_\mu p_\nu}{p^2}\right)(\partial_z A^\mu)(\partial_z A^\nu)
\ee
The second term contributes to the boundary action, generating the 2pt function of a conserved CFT current (see \textit{e.g.} \cite{Witten:1998qj}).  
Integrating by part and using the bulk EOM with $D_M A^M=0$ gauge, the boundary action then reads
\be
S=-\int \frac{d^d p}{(2\pi)^d }\sqrt{\bar \g}
\left(\frac{p^2}{2g^2_b} + \frac{1}{g^2} \partial_\perp K  \right) A^\mu_0
\Pi_{\mu\nu} A^\nu_0 \bigg|_{z=z_0}\,
\ee
with $\Pi_{\mu\nu}=\left(\eta_{\mu\nu}-\frac{p_\mu p_\nu}{p^2}\right)$.

Using the same calculations as in section~\ref{se:AdSB2b}, the exact boundary-to bulk propagatar for $A_\mu$ is 
\be
K=\frac{1}{\sqrt{\bar \g}}\int \frac{d^dp}{(2\pi)^d} \frac{z^{\frac{d}{2}-1}K_{\frac{d}{2}-1}(pz)}{z_0^{\frac{d}{2}-1}K_{\frac{d}{2}-1}(pz_0)} \,. 
\ee
We then take the  $pz_0\ll 1$ limit. 
The $\partial_\perp K$ derivative goes as $\propto\log(p)^{-1} $, $p$, $p^2\log p$ and $\propto p^2$ for $d=2,3,4$ and $d\geq 5$ respectively.  
We also integrate  the $\beta$ functions and identify the renormalization scale with $p$. The reference scale is denoted $p_0$.

The  resulting boundary one-loop effective actions for $d=3,4,5$ are
\begin{align}
\Gamma [A_0]\Big|_{AdS_4} = -\int \frac{d^3 p}{(2\pi)^3 }\sqrt{\bar \g}
\left(\frac{p^2}{2g^2_b}+\left(\frac{1}{g_0^2}  +  \frac{11 C_2(G)}{24\pi^2}  \log\frac{p}{p_0} \right)k p z_0 \right) A^\mu_0
\Pi_{\mu\nu} A^\nu_0 
\end{align}

\begin{align}
\Gamma [A_0]\Big|_{AdS_5} = -\int \frac{d^4 p}{(2\pi)^4 }\sqrt{\bar \g}
\left(\frac{1}{2g^2_{b,0}} - \left(  \frac{k z_0^2}{g^2} +  \frac{23 C_2(G)}{384 \pi^2}   \right) \log\frac{p}{p_0}
 \right) p^2 A^\mu_0
\Pi_{\mu\nu} A^\nu_0 
\end{align}

\begin{align}
\Gamma [A_0] \Big|_{AdS_6} = -\int \frac{d^5 p}{(2\pi)^5 }\sqrt{\bar \g}
\left(\frac{1}{2g^2_b}+\left(\frac{1}{g_0^2}  +  \frac{107 C_2(G) k^2 }{1440 \pi^3}   \log\frac{p}{p_0} \right) k z^2_0\right) p^2 A^\mu_0
\Pi_{\mu\nu} A^\nu_0 
\end{align}
In $\Gamma[A_0]\big|_{AdS_5}$ we have absorbed a constant $\frac{2 k^2 z_0^2}{g^2}(\gamma+\log(\frac{p_0z_0}{2}))$ into the reference value of $\frac{1}{g^2_{b,0}}$. 
We see that for AdS$_{4,6}$ the holographic coupling grows in the IR as a  consequence of  bulk one-loop divergences. For AdS$_5$, both  the contribution from the classical action and the boundary one-loop divergence contribute with negative sign. Thus the holographic coupling grows in the UV in AdS$_5$.

\subsubsection{Aside: Arbitrary Background in 6d}
\label{se:YM_6d_gen}

We can also compute the renormalization of gauge operators for general background, in which case  the renormalized operators take the form ``$RFF$'' in 6d. There are three  irreducible invariants.   Here we give the results in terms of the divergent piece of the one-loop effective action  in general 6d background:
\begin{align}
 \Gamma^{\rm YM}_{\rm 1-loop, div }= &\quad\quad \frac{C_2(G)}{128\pi^3}  \Gamma\left(3-\frac{d+1}{2}\right) \times
 \\ &  \int_{\cal M} d^6x 
\left[
\frac{1}{15}R^{MNPQ}F^a_{MN}F^a_{PQ}
-\frac{88}{45} R^{NM}F_{NP}F_M^{a,P}
+\frac{13}{36} R F^a_{MN}F^{a,MN} 
\right] \nn
\end{align}
where the divergence is encoded into the Gamma function with $d\to5$.  Substituing the AdS$_6$ Riemann tensor reproduces the coefficient in Eq.\eqref{eq:b6gauge}. 

\subsection{ Anomalous Dimensions from Massive  Vector Loops}
\label{se:HYM}

Finally we present results involving massive nonabelian vector fluctuations.  The mass matrix of the gauge field is chosen to be a universal mass $m^2_A \delta^{ab}$. Results with a more complicated mass matrix can similarly be obtained. 
In the holographic CFT the massive vector corresponds to a non-conserved current of the global group $G$ with dimension $\Delta_J\gg d-1$, with $m^2_A=(\Delta_J-1)(\Delta_J-d+1)k^2$. 

We assume the presence of a scalar field $\varphi_i$ transforming as a representation $r$ of the gauge group. In analogy with  computations in section \ref{se:AdSLoop}, we evaluate the effect of the heavy vector on the light scalar sector. We focus on odd  spacetime dimensions. 

The action contains the scalar kinetic term 
\be
\frac{1}{2}\int d^{d+1}x (D_M\varphi)_i(D^M\varphi)_i
\ee
in addition to the YM action. The light scalar is treated as background, it therefore contributes to vector  mass via the invariant $X^{ab} = -g^2(t^a_{r}\varphi)_i(t^b_{r}\varphi)_i$. 

From the bulk contributions to the one-loop effective action we obtain the anomalous dimension of the operator ${\cal O}_i$ associated to $\varphi_i$, 
\be
\gamma^{d=2}_{{\cal O}, {\rm tad}} = 
\frac{g^2C_2(r)}{8\pi (\Delta-1)} \left(
\Delta_J-1
+
\frac{\Delta(\Delta-2)-6}{12 (\Delta_J-1) }
\right)+O\left(\frac{1}{\Delta_J^3}\right)
\ee
\be
\gamma^{d=4}_{{\cal O}, {\rm tad}} = 
- \frac{g^2C_2(r)}{48\pi^2 (\Delta-2)} \left( ((\Delta_J-1)(\Delta_J-3))^{\frac{3}{2}}
-
\frac{\Delta(\Delta-4)-20}{4}((\Delta_J-1)(\Delta_J-3))^{\frac{1}{2}}
\right)+O\left(\frac{1}{\Delta_J}\right)
\ee
\be
\gamma^{d=6}_{{\cal O}, {\rm tad}} = 
 \frac{g^2C_2(r)}{480\pi^3 (\Delta-3)} \left( ((\Delta_J-1)(\Delta_J-5))^{\frac{5}{2}}
 + 
 \frac{\Delta(\Delta-6)-42}{16}((\Delta_J-1)(\Delta_J-5))^{\frac{3}{2}}
 \right) +O\left(\Delta_J\right)
\ee
Here $C_2(r)$ is the quadratic Casimir for the  representation $r$. 
The  next to leading order is also known from our results and is not shown here  merely for convenience. 

We can  also compute the one-loop anomalous dimension of double trace operators. 
We find that the effective quartic coupling $\frac{\eta}{4}(\varphi_i)^2(\varphi_j)^2$ is generated at one-loop with  
\be
\eta^{d=2}_{\rm 1-loop} = -\frac{C^2_2(r) g^4}{8\pi(\Delta_J-1)}
\ee
\be
\eta^{d=4}_{\rm 1-loop} = 
\frac{C^2_2(r)g^4((\Delta_J-1)(\Delta_J-3))^{\frac{1}{2}}}{16\pi^2}
\ee
\be
\eta^{d=6}_{\rm 1-loop} = 
-\frac{C^2_2(r)g^4((\Delta_J-1)(\Delta_J-5))^{\frac{3}{2}}}{96\pi^3}
\ee
The anomalous dimensions of the double trace operators $[{\cal O \cal O}]_{i,n}$ associated to the diagonal term $(\varphi_i)^4$ is then given by $
\gamma_{n} = c_n \eta_{\rm 1-loop} $ where the coefficients $c_n$ are given in section \ref{se:DT_scalar}. 
Since the effective quartic coupling also contains nondiagonal terms $(\varphi_i)^2(\varphi_{j\neq i})^2$,  the anomalous dimension matrix of the $[{\cal O \cal O}]_{i,n}$ operators is nondiagonal. There is therefore a mixing between the double trace operators induced by the loop corrections.

\section{A  Boundary Effective Action in dS
}
\label{se:dS}

We turn to an application in de Sitter space dS$_{d+1}$. Our overall goal here is to extract one-loop contributions to late time cosmological correlators from the heat kernel coefficients.  In this context the heat kernel coefficients should arise from  a suitably defined ``cosmological'' boundary effective action. To compute it we will use the analytical continuation between dS and Euclidian AdS, which relates dS calculations to the AdS calculations made in the previous sections.

We consider the flat slicing 
\be
ds_{\rm dS}^2= g^{\rm dS}_{MN}x^M x^N =\frac{\ell^2}{\eta^2}\left(
-d\eta^2 +  (\vec {d y})^2
\right)\ee
with  $\eta\in [-\infty,\eta_0]$, $\eta_0<0$.  $\eta$ is the conformal time, related to the proper time by $d\eta=e^{-\frac{t}{\ell}}dt $. 
The $\eta_0=0$ slice corresponds to future boundary.
This coordinate patch  describes the expanding de Sitter universe with Hubble radius $\ell$. 
It covers  half of the global dS space, the other half amounts to positive $\eta$. 
We are interested in correlators with endpoints on the $\eta=\eta_0$ time slice.

In this section we deal with various kinds of two-point function (time-ordered, anti-time-ordered, Wightman), hence we  explicitly write the time (anti-)ordering operators $T$($\bar T$). 

\subsection{Cosmological Correlators and In-in Formalism}

The cosmological correlators are conveniently computed in the in-in (or Keldish) formalism \cite{Maldacena:2002vr,Weinberg:2005vy,Seery:2007we,Adshead:2009cb,Senatore:2009cf, Adshead:2009cb, Senatore:2016aui}. A in-in correlator is evaluated by performing a time-ordered integral from the initial time to the time of interest $\eta_0$, and then performing an anti-time-ordered  integral back to the initial time. This can be implemented by sligthly shifting the time integral domain away from the real axis, such that the    $-\eta$ variable takes  values in either  $ \mathbb{R}_+(1+i\epsilon)$  or  $\mathbb{R}_+(1-i\epsilon)$. 
In the interaction picture, for a given interaction Hamiltonian $H_I$, one has \be
\langle\Phi(\eta_0,\vec {y}_1)\ldots \Phi(\eta_0,\vec {y}_n)\rangle = \frac{ \langle 0 | \bar T\left(e^{i\int^{\eta_0}_{-\infty_+ }d\eta H_I}\right) 
\Phi(\eta_0,\vec {y}_1)\ldots \Phi(\eta_0,\vec {y}_n)
 T\left( e^{-i\int^{\eta_0}_{-\infty_- }d\eta H_I}\right) | 0\rangle}
{\langle 0 | \bar T\left(e^{i\int^{\eta_0}_{-\infty_+ }d\eta H_I}\right) T\left( e^{-i\int^{\eta_0}_{-\infty_- }d\eta H_I}\right)| 0\rangle }
\label{eq:cosmo_corr_def}
\ee
with $\infty_\pm=\infty(1\pm i \epsilon)$. 
Here $\langle 0 | $ is the free Bunch-Davis vacuum, for which Minkowski spacetime is recovered at early times ($t,t'\ll \ell $) or short distances $|t-t'|^2+|x-x'|^2\ll \ell^2 \eta^2 $.

The in-in correlators are obtained perturbatively by expanding the left and right exponentials in Eq.\,\eqref{eq:cosmo_corr_def} and performing Wick contractions (see \textit{e.g.} \cite{Weinberg:2005vy}). Contractions between two right vertices or a right vertex and an external field are done via a time-ordered propagator $\langle T
\Phi(x)\Phi(x') \rangle=G_{--}(x,x')$.  Contractions between two left vertices are done via an anti-time-ordered propagator $\langle \bar T
\Phi(x)\Phi(x') \rangle=  G_{++}(x,x')$.  Contractions between a left vertex and a right vertex or an external field are done via Wightman functions
$\langle \Phi(x)\Phi(x') \rangle=G_{+-}(x,x')$, $\langle \Phi(x')\Phi(x) \rangle=G_{-+} (x,x')=G^*_{+-} (x,x')$.\,\footnote{We use the convention $\langle \Phi(x)\Phi(x') \rangle=\langle 0| \Phi(t+i\epsilon,y)\Phi(t'-i\epsilon,y) |0\rangle_{\epsilon\to 0}$. See \textit{e.g.} \cite{Streater:1989vi}.   }

\subsection{dS Propagators }

We derive the dS propagators in Fourier space. We will next see that Fourier space renders  the analytical continuation from dS to EAdS straightforward. 
Fourier space is also usually the preferred space to study cosmological correlators. 

Particles in dS space are classified following the unitary irreps of $SO(1,d+1)$ \cite{Schwarz71,Harmonic_analysis}. Their  masses are conveniently parametrized as
\be
\ell^2 m^2=\Delta(d-\Delta)=\frac{d^2}{4}+\nu^2 
\label{eq:mdS}
\ee
In the nontachyonic representation one distinguishes the series of ``heavy'' states with mass $m>\frac{d}{2}\ell^{-1}$  labelled by $\Delta_{i\nu}=\frac{d}{2}+i\nu$ with $\nu\in \mathbb{R}$, and the series of ``light'' states  with mass $0<m<\frac{d}{2}\ell^{-1}$    labelled by $\Delta_{-\mu}=\frac{d}{2}-\mu$ with $\mu\in [0,\frac{d}{2}]$. More generally the $\Delta_{i\nu}$ function is analytically continued into the whole complex plane.

The (anti-)time-ordered  propagators for a scalar particle satisfy the EOM 
\be 
{\cal D}_{\rm dS}G_{--(++)}(x,x')=\mp i\frac{1}{\sqrt{|g|_{\rm dS}}}\delta^{d+1}(x-x')\ee  with ${\cal D}_{\rm dS}=-\square_{\rm dS}+m^2$ the scalar Laplacian. 

Since the EOM applies independently on each endpoint, the propagators must have a factorized structure. Explicitly it is 
\begin{align} G_{--}(p;\eta,\eta') & = \frac{i}{C}F_<(\eta_<)F_>(\eta_>)\,,\quad G_{++}(p;\eta,\eta')=- \frac{i}{\tilde C}\tilde F_<(\eta_<)\tilde F_>(\eta_>) \nn \\
G_{+-}(p;\eta,\eta') &= \frac{i}{ C} F_<(\eta) F_>(\eta')  = - \frac{i}{ \tilde C} \tilde F_<(\eta') \tilde F_>(\eta)
\end{align}
where we have defined  $-\eta^<=\min(-\eta
,-\eta')$, $-\eta^>=\max(-\eta
,-\eta')$. 
The $C$ constant is determined by the Wronskian $W=F_>'F_<-F_<'F_>$ with $C=W\frac{\ell^{d-1}}{|\eta|^{d-1}}$ as in  Eq.\,\eqref{eq:W_gen}, and similarly for $\tilde C$.
A basis of  solutions to the homogeneous EOM in Fourier space is $z^{\frac{d}{2}}H^{(1,2)}_{i\nu}(-p\eta)$.

The condition implementing Bunch-Davies vacuum in the (anti-)time-ordered propagators $G_{--}$, $G_{++}$ is the following \cite{Gibbons:1977mu,Burges:1984qm,Mottola:1984ar,Allen:1985ux,deBoer:2004nd}.  As a result of the symmetries of dS, apart from the singularity at coincident endpoints,  there can be another singularity 
 when the endpoints are at antipodal positions, $x=x',\eta=-\eta'$.  To implement the Bunch-Davies vacuum we require the propagator to be regular in this antipodal configuration (the remaining singularity ensures flat space behaviour at short distance). 
The antipodal configuration lies outside  the expanding Poincar\'e patch---in fact the two endpoints are separated by a cosmological horizon. The regularity condition is implemented by  extending  the Poincar\'e patch into the $\eta>0$ half of dS. 
In Fourier space the antipodal configuration amounts to taking the  $(p\to\infty, \eta'\to-\eta)$ limit. Therefore the Bunch-Davies vacuum is implemented by 
\be
\lim_{\substack{p \to \infty,\, \eta'\to-\eta}}  G_{--,++}(\eta,\eta'; p) = 0\,. \label{eq:BD}
\ee

Taking into account the $i\epsilon$ shifts, 
we find that the  de Sitter propagators in Fourier space are 
\begin{align}
G^{\rm dS}_{--}(p;\eta,\eta')&= -\frac{\pi}{4 } \ell \left(\frac{\eta\eta'}{\ell^2}\right)^{\frac{d}{2}} H^{(1)}_{i\nu}\left(-p\eta^<\right) H^{(2)}_{i\nu}\left(-p\eta^>\right) \nn
\\ 
G^{\rm dS}_{++}(p;\eta,\eta')&= -\frac{\pi}{4 } \ell  \left(\frac{\eta\eta'}{\ell^2}\right)^{\frac{d}{2}} H^{(2)}_{i\nu}\left(-p\eta^<\right) H^{(1)}_{i\nu}\left(-p\eta^>\right) \nn
\\ 
G_{+-}^{\rm dS}(p;\eta,\eta')&=-\frac{\pi}{4 } \ell \left(\frac{\eta\eta'}{\ell^2}\right)^{\frac{d}{2}} 
H^{(1)}_{i\nu}\left(-p\eta\right) H^{(2)}_{i\nu}\left(-p\eta'\right) 
\label{eq:dS_prop}
\end{align}
One can explicitly verify that $G_{++}=(G_{--})^*$ and that the Wightman function $G_{+-}$ is Hermitian whenever $\nu$ is either purely real or imaginary, which corresponds to the two possible  representations  of states in de Sitter space. 

The amputated boundary-to-bulk propagators  from the (anti-)time-ordered propagators are given by
\be
\sqrt{\bar g}_{\rm dS} K_{-}^{\rm dS}(p; \eta,\eta') =  \left(\frac{\eta}{\eta_0}\right)^{\frac{d}{2}} \frac{ H^{(2)}_{i\nu}\left(-p\eta_{0}\right)}{ H^{(2)}_{i\nu}\left(-p\eta_{0} \right)}  \,
\ee
\be
\sqrt{\bar g}_{\rm dS} K_{+}^{\rm dS}(p; \eta,\eta') =  \left(\frac{\eta}{\eta_{0}}\right)^{\frac{d}{2}} \frac{ H^{(1)}_{i\nu}\left(-p\eta\right)}{ H^{(1)}_{i\nu}\left(-p\eta_{0} \right)}  \,.
\ee
The amputated boundary-to-bulk Wightman function equals either of those depending on which endpoint is put on the boundary.

\subsection{From dS to EAdS }

We use analytical continuation  to express the dS correlators  in Euclidian AdS (EAdS). Similar analytical continuation has been  discussed in \textit{e.g.} \cite{Balasubramanian:2002zh, Maldacena:2002vr, Harlow:2011ke, Anninos:2014lwa} in the (A)dS/CFT context, and has  more recently  been used in \cite{Sleight:2020obc,Sleight:2019mgd, Sleight:2019hfp} to study the dS correlators in the  Mellin-Barnes representation (see also \cite{DiPietro:2021sjt,Sleight:2021plv,Meltzer:2021zin,Premkumar:2021mlz}). 
An advantage of this approach is that much is known about perturbative and spectral techniques in EAdS;  all this knowledge is readily transferred  to  dS via analytical continuation. 
In the context of the present work, there are additional reasons to work in EAdS. The usual heat kernel formalism needs  the fluctuation to have a diagonal Feynman propagator in order to be applicable. In the in-in formalism from the original dS spacetime, the various 2pt functions are simultaneously involved hence we do not know how to straightforwardly proceed. Upon rotating to EAdS one obtains a Euclidian propagator matrix, which is trivially diagonalized.  We can thus evaluate the one-loop effective action directly in the Euclidian space. These steps are realized further below.

The Euclidian AdS metric $g_{\rm EAdS}$ is given by
\be
ds_{\rm EAdS}^2  =\frac{L^2}{z^2}\left(
dz^2 +  (\vec {d y})^2
\right)\ee
with $z\in [z_0, \infty]$. 
The scalar propagator for a particle with mass $ m_{\rm EAdS}^2= (\alpha^2-\frac{d^2}{4})\frac{1}{L^2} $ is 
\be
G_\alpha^{\rm EAdS}(p;z,z')= L \left(\frac{zz'}{L^2}\right)^{\frac{d}{2}}I_\alpha(pz_<)K_\alpha(pz_>)\,. 
\label{eq:GEAdS}
\ee
The boundary-to-bulk propagator is
\be
\sqrt{|\bar g|}_{\rm EAdS} K_\alpha^{\rm EAdS}(p;z)=\left(\frac{z}{z_0}\right)^{\frac{d}{2}}\frac{K_\alpha(pz)}{K_\alpha(pz_0)}
\ee

The transformation from dS to EAdS is implemented by closing the following contours in the $\eta$ plane
\be
	\includegraphics[width=0.6\linewidth,trim={5cm 8.5cm 5cm 5cm},clip]{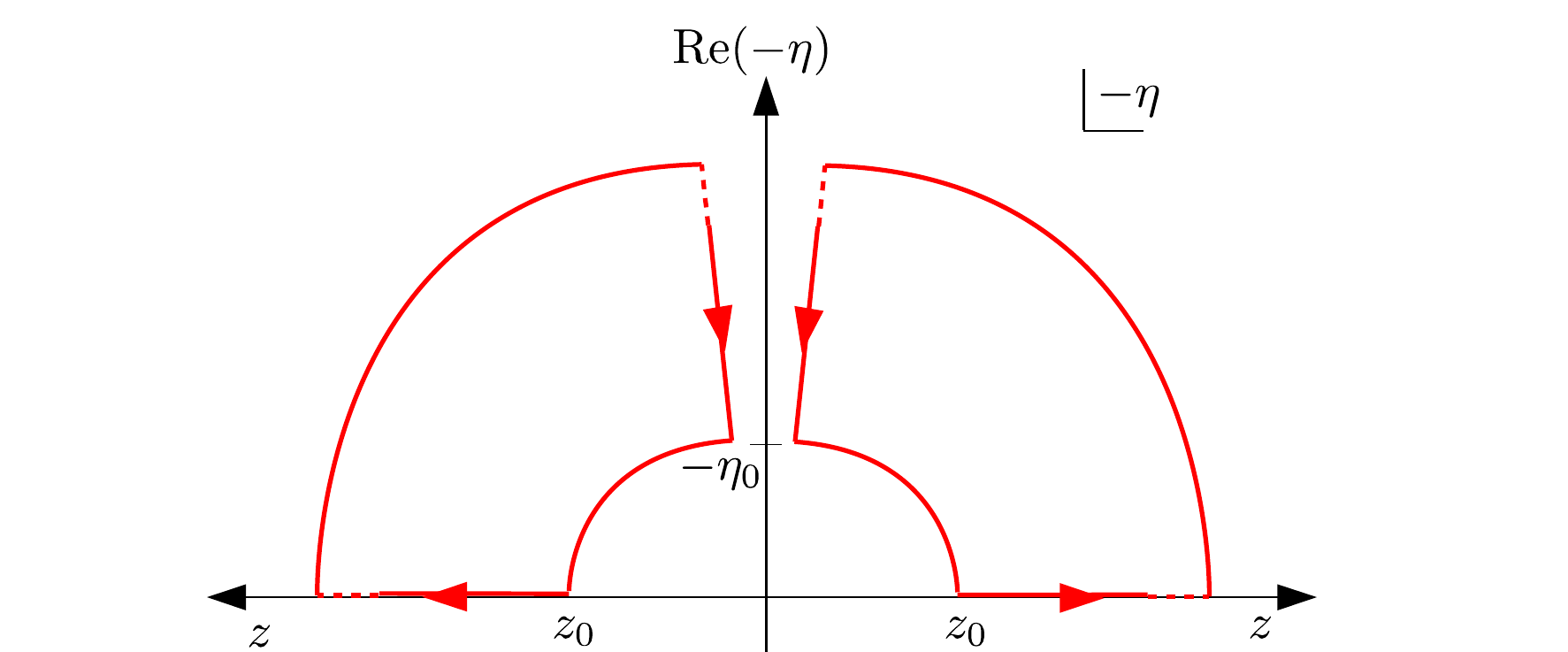} \label{eq:Wick_contour}
\ee
which corresponds to  
the  Wick rotations
\be
(-\eta)_\pm=z e^{\pm i(\frac{\pi}{2}-\epsilon)} \, \label{eq:Wick} \,.
\ee
Equivalent Wick rotations have been used in \cite{Sleight:2020obc,Sleight:2019mgd, Sleight:2019hfp,DiPietro:2021sjt}. One can  readily see that this converts the dS metric to the EAdS one upon  identification of the dS and EAdS radii $ L^2=-\ell^2 $.
For our purposes, however, we will use only the $\ell$ radius. This implies that the EAdS metric resulting from the Wick rotation has only-minus signature,  which will imply the presence of $i^d$, $(-i)^d$ factors in the expressions. 

We denote by an arrow the replacement  Eq.\,\eqref{eq:Wick} in a function, $f(-\eta_-,-\eta_+)\to f(z e^{- i(\frac{\pi}{2}-\epsilon)},z e^{ i(\frac{\pi}{2}-\epsilon)})$. 
We plug the Wick rotation into Eq.\,\eqref{eq:dS_prop}.
The time coordinates of $G_{--}$, $G_{++}$, $G_{+-}$ are rotated as $(\eta_-,\eta_-)$, $(\eta_+,\eta_+)$, $(\eta_+,\eta_-)$ respectively. \footnote{These are in fact the only possibilities for closing the contours with vanishing contribution from the arcs at $|\eta|\to\infty$, as required to perform Wick rotation.}  
For the $G_{+-}$ function, both the $F_<$ and $F_>$ pieces are rotated to a Bessel K function with positive argument. For $G_{--}$, $G_{++}$, the $F_>(\eta_>)$ piece  turns into a Bessel K function $\propto K(pz_>)$ with positive argument. In contrast, the Wick rotation in the $F_<(\eta_<)$ piece gives a Bessel K function with negative argument, which requires careful treatment since $K$ has a branch cut along the negative axis. To evaluate this piece,  the following identity and its complex conjugate are useful:
\be
H^{(1)}_{i\nu} ( p z_< e^{-i(\frac{\pi}{2}-\epsilon)})= \frac{i}{\sin\pi i\nu}\left(
e^{-i\frac{3\pi}{2}\nu} I_{i\nu}( p z_<) - e^{i\frac{\pi}{2}\nu} I_{-i\nu}( p z_<)
\right) \,. 
\ee

We find that  the dS propagators take the form
\begin{align}
 G^{\rm dS}_{--}(p;\eta,\eta')&\to \frac{1}{2\sin (\pi i\nu)}\left( e^{-i\pi \Delta_{i\nu} }G_{i\nu}^{\rm EAdS}(p;z,z') - e^{-i\pi \Delta_{-i\nu} }G_{-i\nu}^{\rm EAdS}(p;z,z')
\right) \label{eq:G--Wick} \\
 G^{\rm dS}_{++}(p;\eta,\eta')&\to \frac{1}{2\sin (\pi i\nu) }\left( e^{i\pi \Delta_{i\nu} }G_{i\nu}^{\rm EAdS}(p;z,z') - e^{i\pi \Delta_{-i\nu}  }G_{-i\nu}^{\rm EAdS}(p;z,z')
\right) \\
 G_{+-}^{\rm dS}(p;\eta,\eta')& \to \frac{1}{2\sin (\pi i\nu) } \left( G_{i\nu}^{\rm EAdS}(p;z,z') - G_{-i\nu}^{\rm EAdS}(p;z,z') \right)
\end{align}
where the EAdS propagators are given by Eq.\,\eqref{eq:GEAdS} with the identification $L\equiv \ell$. We have introduced
\be
\Delta_{\pm i\nu}=\frac{d}{2} \pm i\nu \,. 
\ee
Our result matches exactly the form of the propagators obtained in \cite{DiPietro:2021sjt} up to an overall sign.

For the Wick rotation of the boundary-to bulk propagators the general formula is
\be
\sqrt{\bar g}_{\rm dS} K_\pm^{\rm dS}(p;\eta) \to N_\pm(p)  \sqrt{\bar g}_{\rm EAdS} K^{\rm EAdS}(p;z) \label{eq:K_DS_EAdS}
\ee
with 
\be
N_{\pm }(p) = \mp \frac{2i }{\pi}  e^{\pm i\frac{\pi}{2}\Delta_{-i\nu}}  \frac{K_{i\nu}(-p\eta _0)}{H^{(2)}_{i\nu}(-p\eta_0)}
\ee
The same is true for the boundary-to-bulk Wightman function. 
Let us consider the case of a light mode, which is typically the relevant case for cosmological correlators. We take the small $p z_0$  limit and assume real $i\nu$ with $|i\nu|=\mu$.  
For any sign of $i\nu$ we find 
$N_{\pm }(p) \approx e^{\pm i\frac{\pi}{2}\Delta_{-\mu}} $.
Thus \be
\sqrt{\bar g}_{\rm dS} K_\pm^{\rm dS}(p;\eta)\Big|_{{\rm light}, pz_0\ll 1} \to e^{\pm i\frac{\pi}{2}\Delta_{-\mu}}   \sqrt{\bar g}_{\rm EAdS} K_\mu^{\rm EAdS}(p;z)
\label{eq:KdSEAdSlight}
\ee
We have used $K^{\rm EAdS}_\mu=K^{\rm EAdS}_{-\mu}$. We can see that the dimension of the physical operator $\Delta_{-\mu}=\frac{d}{2}-\mu$ always appear.
The phases in Eq.\,\eqref{eq:KdSEAdSlight} are consistent with the ones obtained in \cite{DiPietro:2021sjt}.

Finally, the integral measures are Wick rotated as
\begin{align}
&i\int^{\eta_0}_{-\infty_+} \frac{d\eta_+ }{ (-\eta_+ \ell^{-1})^{d+1}}\to (-i)^{d-1} \int_{z_0}^{\infty } \frac{dz }{ (z \ell^{-1})^{d+1}} \\
-&i\int^{\eta_0}_{-\infty_-} \frac{d\eta_- }{ (-\eta_- \ell^{-1})^{d+1}} \to   (i)^{d-1}\int_{z_0}^{\infty } \frac{dz }{ (z \ell^{-1})^{d+1}} 
\end{align}

\subsection{The Generator of Cosmological Correlators}

We follow, revisit and expand a proposal from Ref.\,\cite{DiPietro:2021sjt}.  The analytical continuation from dS to EAdS is used to define a functional $ Z^{\rm pert}_{\rm dS}[J]$, expressed in EAdS,  that  generates the cosmological correlators at any order in perturbation theory.

We first realize that, upon the two types of Wick rotation in Eq.\,\eqref{eq:Wick_contour}, the dS field with values on $\mathbb{R}(1\pm i\epsilon)$ is analytically continued as either  a holomorphic or an antiholomorphic function of $\eta$, $\Phi^{\rm dS}(-\eta_\pm)$. This implies that  there are two distinct fields in $Z^{\rm pert}_{\rm dS}$, related to each other by complex conjugation. We define the two fields $\Phi_\pm$ as a function of $z$, with $\Phi_\pm(z)=\Phi_\pm(-\eta_\pm e^{\mp i\frac{\pi}{2}})$. They satisfy $\Phi_+^*=\Phi_-$. 
 In this subsection it is enough to focus on a single pair of EAdS fields  $\Phi_\pm$, \textit{i.e.} a single dS field only.
The $\Phi_-$ field comes from the time-ordered contour, the $\Phi_+$ field comes from the anti-time-ordered contour. The $\Phi_-$ is used for  external legs.

We introduce a boundary source $J$   coupled to the $\Phi_-$ fields. The source is localized on the boundary of EAdS. 
The source can  either be  understood  as a  function of the original dS coordinates or of the EAdS coordinates.

Given these definitions and conventions,  we introduce the generating functional 
\begin{align}
 Z^{\rm pert}_{\rm dS}[J] = 
\nn
 \int {\cal D} \Phi_\pm  \exp\bigg[ & - \frac{1}{2}(\Phi_-,\Phi_+) \cB \hat G^{-1}_{\rm EAdS} \cB (\Phi_-,\Phi_+)^t -   J \cb \Phi_{-,0}  
\\
&
\quad\quad\quad\quad\quad
-\int_{\rm EAdS} \sqrt{|\g|} dz d^dx \left(
i^{d+1}{\cal L}_I[\Phi_-] + (-i)^{d+1} {\cal L}_I[\Phi_+]\right)  
\bigg] \label{eq:gen_cosmo_1}
\end{align}
where $J$ is boundary localized. 
We remind the definition of the convolution products, here in EAdS
\be A \cB B  = \int_{  {\rm EAdS}} \sqrt{|\g|} d^{d+1}x A(x)B(x)
\,,\quad A \cb B = \int_{ \partial {\rm EAdS}} \sqrt{|\bar\g|} d^dy 
A(y)A(y)  \ee with the EAdS radius set to $\ell$. 
We have introduced the propagator matrix
\be
\hat G_{\rm EAdS}=\frac{1}{2\sin (\pi i\nu) }\begin{pmatrix} e^{-i\pi \Delta_{i\nu} }G_{i\nu}^{\rm EAdS} - e^{-i\pi \Delta_{-i\nu} } G_{-i\nu}^{\rm EAdS} &  G_{i\nu}^{\rm EAdS} - G_{-i\nu}^{\rm EAdS}
\\
G_{i\nu}^{\rm EAdS} - G_{-i\nu}^{\rm EAdS}
 & e^{i\pi \Delta_{i\nu} }  G_{i\nu}^{\rm EAdS} - e^{i\pi \Delta_{-i\nu} } G_{-i\nu}^{\rm EAdS} 
\end{pmatrix} \label{eq:hatGEAdS}
\ee
Using $\Phi_+^*=\Phi_-$ one can verify that the fundamental action (\textit{i.e.} the kinetic term and the interaction term) is real.

The connected perturbative cosmological correlators originally defined by Eq.\,\eqref{eq:cosmo_corr_def} are then given by   derivatives of the generating functional \be W^{\rm pert}_{\rm dS}[J]=\log( Z^{\rm pert}_{\rm dS}[J]) \,. \ee
Moreover, as discussed in section \ref{se:interactions}, taking the Legendre transform  $\Gamma[\Phi_{\pm,0}]=-W[J]+J\cb \Phi_{-,0}$ gives the effective action in the boundary variables $\Phi_{\pm,0}$. This effective action $\Gamma[\Phi_{\pm,0}]$ generates the 1P-irreducible boundary diagrams, where irreducible means with respect to boundary-to-boundary lines. These  diagrams have boundary-to-bulk propagators as external legs --- they are  the dS version of Witten diagrams.

The reason why the $\Gamma^{\rm pert}_{\rm dS}[\Phi_{\pm,0}]$ effective action depends on both $\Phi_{-,0}$ and $\Phi_{+,0}$ is that the propagators amputated by the Legendre transform are non-diagonal --- both $\Phi_+$ and $\Phi_-$ show up  when using the chain rule on  $W^{\rm pert}_{\rm dS}[J]$. 
Thus  the operation of amputation requires some care.
In order to evaluate the boundary action that generates the amputated cosmological correlators, we take an alternative route by first performing  a field redefinition.

\subsubsection*{Canonical Normalization}

We can see that the propagator matrix $\hat G_{\rm EAdS}$ contains linear combinations of $G_{i\nu}$ and $G_{-i\nu}$. Thus we introduce a pair of EAdS fields $\Phi$, $\tPhi $ satisfying $\langle\Phi \Phi\rangle\propto G_{-i\nu}$, $\langle\tilde \Phi \tilde \Phi\rangle\propto G_{i\nu}$, $\langle\Phi \tilde \Phi\rangle=0$. 
We introduce the complex coefficients
\be
\c^\pm_{\pm i\nu}=\frac{e^{\pm i\frac{\pi}{2}\Delta_{\pm i\nu}}}{\sqrt{2\sin \pi i\nu}} \,. 
\ee
We find that the transformation 
\be
\begin{pmatrix}
\Phi_- \\ \Phi_+
\end{pmatrix} =U \begin{pmatrix}
\tPhi \\ \Phi
\end{pmatrix}\,,\quad \quad
U=
\begin{pmatrix}
c^-_{i\nu}& i c^-_{-i\nu} \\
c^+_{i\nu}& i c^+_{-i\nu}
\end{pmatrix}\,
\ee
diagonalizes and canonically normalizes 
the matrix of propagators. 

Plugging it into Eq.\,\eqref{eq:gen_cosmo_1} gives canonically normalized kinetic terms.
The generating functional reads
\begin{align}
& Z^{\rm pert}_{\rm dS}[J] = 
\nn \\ \nn
& \int {\cal D} \Phi{\cal D} \tilde\Phi \exp\bigg[- \frac{1}{2}(\Phi\cB  G_{-i\nu}^{-1} \cB  \Phi + \tPhi\cB G_{i\nu}^{-1} \cB \tPhi  )
-
{\cal O}\big[\Phi_0,\tPhi_0 \big]\cb J 
\\
&
\quad\quad\quad\quad
-\int_{\rm EAdS} \sqrt{|\g|} d^{d+1} x \Big(
 i^{d-1}{\cal L}_I[i\c^-_{-i\nu}\Phi + \c^-_{i\nu}\tPhi ]
+(-i)^{d-1} {\cal L}_I[
i\c^+_{-i\nu}\Phi + \c^+_{i\nu}\tPhi] \Big) 
\bigg] \label{eq:gen_cosmo_2}
\end{align}
For a heavy dS mass ($\nu\in \mathbb{R}$), the fields satisfy $\Phi^*=-\tilde\Phi$.   The propagators satisfy $G_{-i\nu}=G^*_{i\nu}$, and therefore   the action is real.  
For a light dS mass ($i\nu \in [-\frac{d}{2},\frac{d}{2}]$), the fields either satisfy  $ \Phi^* = -\Phi$, $ \tilde \Phi^* = \tilde \Phi$ or $ \Phi^* = \Phi$, $ \tilde \Phi^* =- \tilde \Phi$ depending on the value of $i\nu$. The $G_{\pm i\nu}$ propagators are real, and it follows that again the action is real. 

Let us specify the source for a light field. We assume\,\footnote{We can instead assume  $-i\nu=\mu\in[0,\frac{d}{2}]$. In that case $\tPhi$  has Neumann and $\Phi$ has Dirichlet BC.  The source term  is then
$
{\cal O}\big[\Phi_0,\tPhi_0 \big] =  \frac{1}{\sqrt{2\sin\pi\mu }}\,\tPhi_0 \,. 
$
} $i\nu=\mu\in[0,\frac{d}{2}]$. The kinetic term in Eq.\,\eqref{eq:gen_cosmo_2} indicates that the $\Phi$ field has Neumann and $\tPhi$  has Dirichlet BC. 
 By direct comparison with a known cosmological correlator we obtain that the source acts on the Neumann field, with $\Phi=K_{\rm EAdS}\cb \Phi_0+\Phi_D$. This also fixes the normalization, giving
\be
{\cal O}\big[\Phi_0,\tPhi_0 \big]_{\rm light} = \frac{1}{\sqrt{2\sin\pi\mu }}\,\Phi_0 \,. 
\label{eq:OPhi0}
\ee

Putting the pieces together, the amputated cosmological correlators with light fields in external legs are generated by derivatives of the effective action $\Gamma^{\rm pert}_{\rm dS}[\Phi_0]=-W^{\rm pert}_{\rm dS}+J\cb O[\Phi_0] $ as follows,
\be
-{(2\sin \pi i\nu )^\frac{n}{2}}\frac{\delta^n  \Gamma^{\rm pert}_{\rm dS}[\Phi_0] }{\delta \Phi_0(\vec y_1)\ldots \delta \Phi_0(\vec y_n)}
= \langle \Phi_0(\vec y_1) \ldots \Phi_0(\vec y_1) \rangle_{1PI}\,
\label{eq:gen_con_cor_1PI}
\ee
for $i\nu>0$. Equivalently, for $i\nu<0$ the derivatives are taken in $\tPhi$. 
We emphasize that, despite the complex factors present in the action, the resulting cosmological correlators  are real. 

We  remind that, since $\Gamma^{\rm pert}_{\rm dS}[\Phi_0]$ is defined by Legendre transform in the boundary source, 1P-irreducibility is here meant with respect to boundary-to-boundary lines (see also Sec.~\ref{se:interactions}). The 4pt exchange diagram with Dirichlet bulk field exchange, for example, is generated by Eq.\,\,\eqref{eq:gen_con_cor_1PI} as a 1PI diagram.  
As a sanity check one can verify that the dS exchange diagram with cubic vertices and with light states in external legs is correctly reproduced by Eq.\,\eqref{eq:gen_con_cor_1PI}.

\subsection{Towards the Cosmological One-loop Effective Action }

Integrating out a dS bulk fluctuation at tree-level in dS is a fairly simple task. One can use, in particular, the covariant large mass expansion of the propagator given in Eq.\,\eqref{eq:AdS_prop_series}. Integrating  out a dS bulk fluctuation at loop-level in dS is more challenging. 
However, thanks to the analytical continuation to EAdS, we can readily use the AdS heat kernel coefficients evaluated in the previous section to get the heat kernel expansion of the cosmological one-loop effective action, \textit{i.e.} the one-loop piece of $\Gamma_{\rm dS}^{\rm pert}=\Gamma^{\rm dS, pert}_{\rm cl}+\Gamma^{\rm dS, pert}_{\rm 1-loop}+\ldots$.

We focus on integrating out bulk fluctuations of  a scalar field $\Phi$. The field can either be heavy or light in the dS sense. Depending on cases, either Dirichlet or Neumann BC  can be chosen. 
The interaction Lagrangian in terms of dS fields is written  as
\be
{\cal L}^{\rm dS}_I = -\frac{1}{2}\Phi^2 V''(\varphi^{\rm dS}_{i}) 
\ee
where the $\varphi^{\rm dS}_{i}$ are light background fields with masses parametrized by $\mu_i\in[0,\frac{d}{2}]$.
Upon analytical continuation to EAdS, the interactions are $V''_\pm=V''( e^{\pm i\frac{\pi}{2}\Delta_{- \mu_i}} K_{\mu_i} \cb \varphi_{i,0}) $ where the background boundary fields $\varphi_{i,0}$ are  real.

We consider the quadratic piece of the  fundamental action  in $\Phi$,
\begin{align}
& S^{\rm quad}_{\rm EAdS} =     \label{eq:S_cosmo}
\frac{1}{2}(\Phi\cB  G_{-i\nu}^{-1} \cB  \Phi + \tPhi\cB G_{i\nu}^{-1} \cB \tPhi  )
\\ \nn
&
\quad\quad\quad
+ \int_{\rm EAdS} \sqrt{|\g|} d^{d+1}x\Big(\frac{i^{d-1}}{2}\left(i\c^-_{-i\nu}\Phi + \c^-_{i\nu}\tPhi\right)^2 V''_-
+\frac{(-i)^{d-1}}{2}\left(i \c^+_{-i\nu}\Phi + \c^+_{i\nu}\tPhi\right)^2 V''_+\Big)
\end{align}
with the shortcut notation
\be
V''_\pm = V''( e^{\pm i\frac{\pi}{2} \Delta_{- \mu_i}}  K_{\mu_i}\cb\varphi_{i,0}) \,.
\ee

The background-field-dependent mass matrix of the fluctuations can be read from Eq.\,\eqref{eq:S_cosmo}, 
\begin{align}
X &= i^{d} \begin{pmatrix}
i (c^-_{-i\nu})^2 & c^-_{i\nu}c^-_{-i\nu} \\
 c^-_{i\nu}c^-_{-i\nu} & - i (c^-_{i\nu})^2 
\end{pmatrix} V''_- - 
 (-i)^{d} \begin{pmatrix}
i (c^+_{-i\nu})^2 & c^+_{i\nu}c^+_{-i\nu} \\
c^+_{i\nu}c^+_{-i\nu}  & -i c^+_{i\nu})^2 
\end{pmatrix} V''_+ \\
& = 
\frac{i}{2\sin (\pi i\nu)}\begin{pmatrix}
e^{i\pi i\nu} V_-''- e^{-i\pi i\nu} V_+''  & i(V''_+-V''_-) \\
i(V''_+-V''_-)  &    e^{i\pi i\nu} V_+'' -e^{-i\pi i\nu} V_-'' 
\end{pmatrix}
\label{eq:X_dS}
\end{align}
For a light fluctuation, $X$ is self-conjugate. For heavy fields, $X$  is not self-conjugate because conjugation swaps the $\Phi$ and $\tPhi$ components and thus the entries of $X$, however the physical quantities  are the traces $\tr( X^r)$, which  are self-conjugate. 

Finally, we evaluate the trace of arbitrary powers of $X$ and find they take the simple form
\begin{align}
\tr( X^r)= (-1)^r\left( (V''_-)^r + (V''_+)^r \right) \,. \label{eq:trX_dS}
\end{align}
This canonical invariant is the only ingredient needed to evaluate the heat kernel coefficients in the case of scalar interactions considered here.

Evaluating the finite, mass-suppressed parts of the one-loop action, \textit{i.e.} the long-distance EFT, could be interesting. However it may require some more developments due to the fact that one deals with  masses which are tachyonic from the EAdS viewpoint. A careful analytic continuation from nontachyonic to tachyonic EAdS masses might be needed. This is left  to future work. 
Here below we work out the divergent part of the one-loop cosmological action in a simple case.

 \subsection{Renormalization from Scalar Loops in dS$_4$ }
 \label{se:dS4_ren}
 
\begin{figure}
\centering
		\includegraphics[width =0.8\linewidth,trim={0cm 8cm 0cm 0cm},clip]{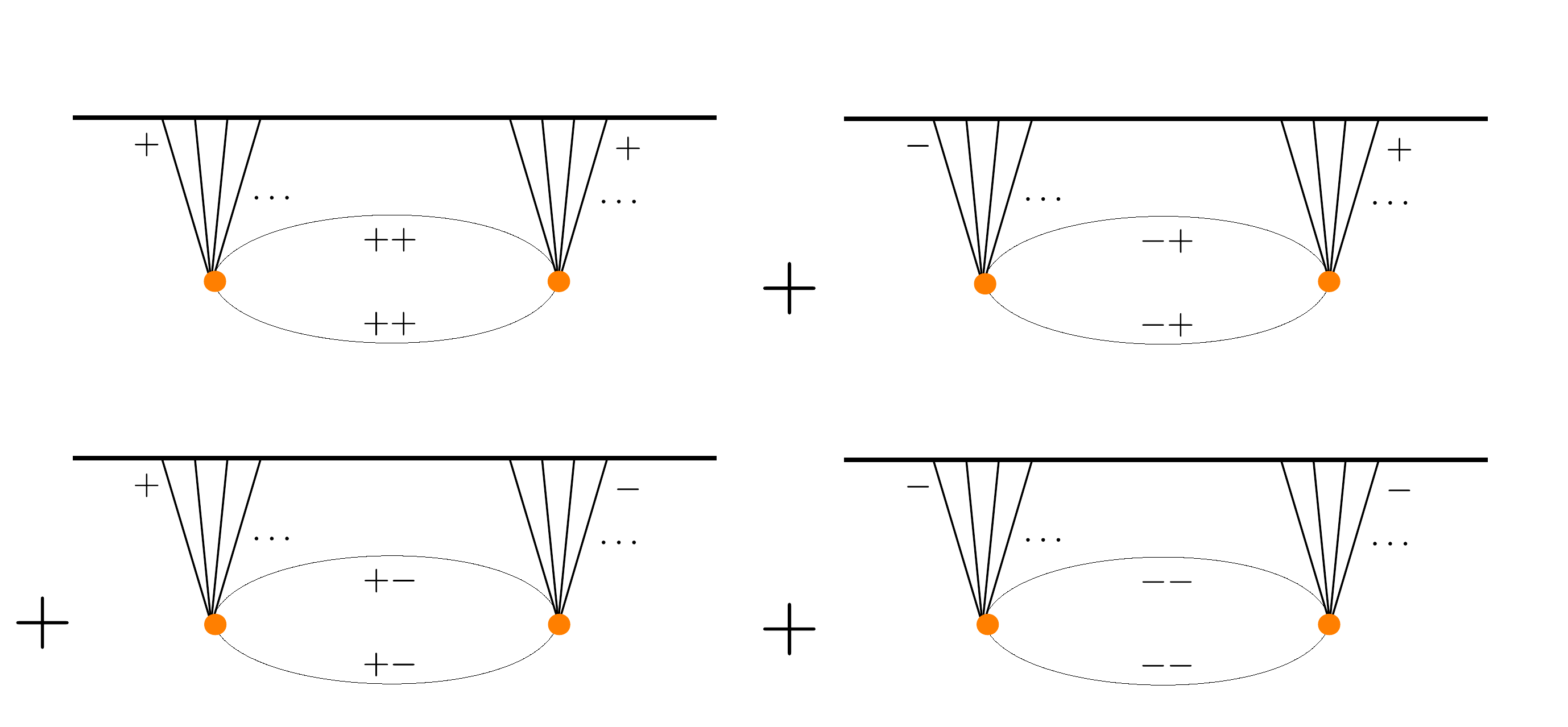}
		\centering
	\includegraphics[width=0.45\linewidth,trim={0cm 13cm 10cm 0cm},clip]{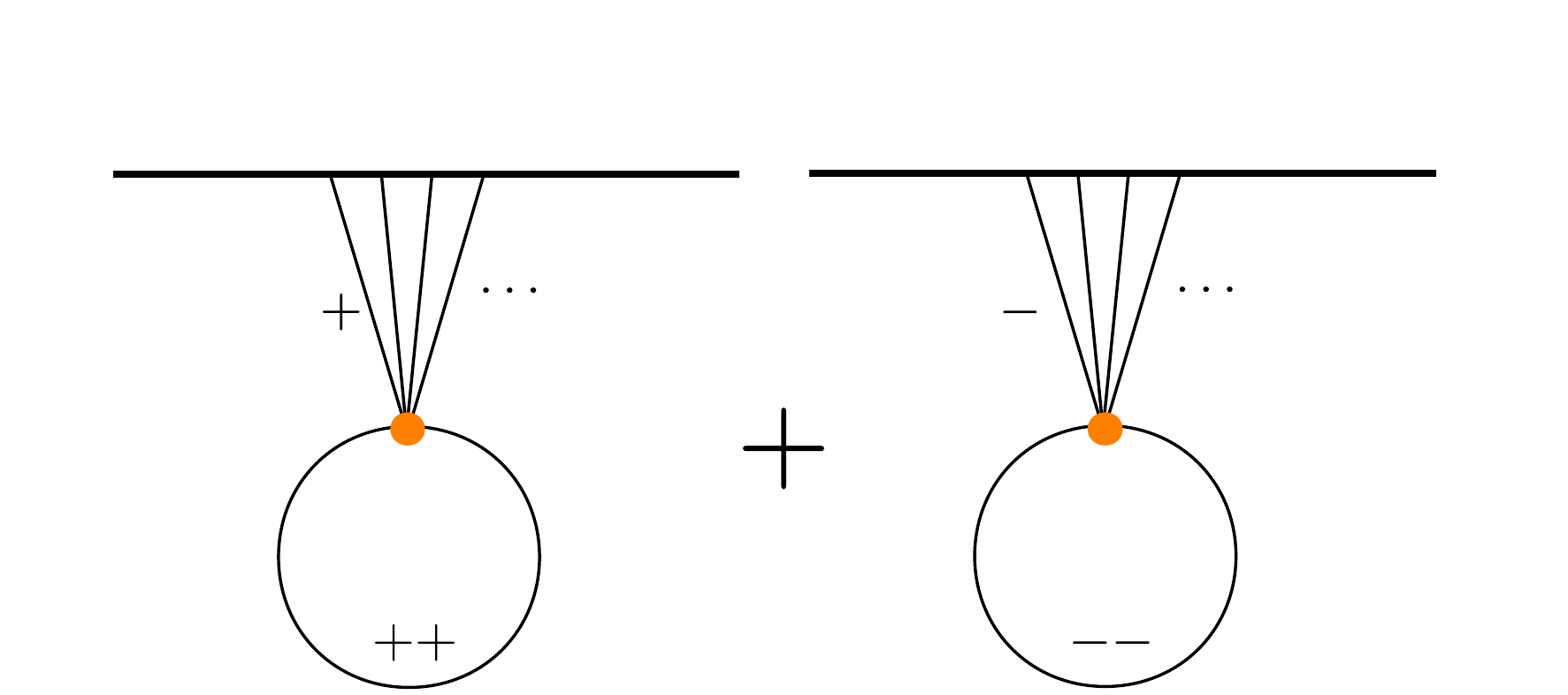} 
    \caption{Analytically-continued dS$_4$ diagrams  contributing to  bulk and boundary divergences in the one-loop effective action. Internal lines are elements of the $\hat G_{{\rm EAdS}_4}$ propagator matrix. 
    }
    \label{fig:diags_dS4}
\end{figure}

 In this section we extract some concrete results in dS$_4$. 
In dS$_4$  one-loop divergences appear from the $b_4$ and $b^\partial_4$ heat kernel coefficients.  The assumed scalar interactions give $V(\varphi_i)=\frac{\Phi^2}{2}\prod^n_{i=1}\varphi_i$. 
The  diagrammatic contributions to the  $b_4,b^\partial_4$   coefficients  are shown in Fig.\,\ref{fig:diags_dS4}. 
The EAdS scalar curvature is $R=-\frac{12}{\ell^2}$. The $\square \prod^n_{i=1}\varphi_i $ and  $\partial_\perp \prod^n_{i=1}\varphi_i $ are evaluated like in 
  the AdS case, see Eqs.\,\eqref{eq:boxX_AdS},\,\eqref{eq:delXAdS}. 
 In the late time limit, the $\Delta_i$ combine to appear only in the
 $\Delta_{X}=\sum^\n_{i=1} \Delta_{i}$,
 $\Delta_{X}^-=\sum^\n_{i=1} (3-\Delta_{i}) $ combinations, in accordance with the underlying conformal symmetry.

 We  take a Neumann BC for the heavy field, accordingly to discussion in Sec.\,\eqref{se:HOLEA}. Using Eq.\,\eqref{eq:trX_dS} with $r=1,2$, we find the heat kernel coefficients
 \begin{align}
 {\rm tr}\, \bb_4 & =   \cos \left(\pi \Delta_X\right)\Lambda_{\n,2} \cb \left[ \prod^n_{i=1}\varphi_{i,0}\right]^2 - \frac{\cos \left(\frac{\pi}{2}\Delta_X\right)}{3\ell^2}\left(- \DeltaX(\DeltaX-3)+12 \right)  \Lambda_\n \cb \prod^n_{i=1}\varphi_{i,0}
\label{eq:b4_dS_light}
\\
{\rm tr}\, b^\partial_4 & =
\frac{4}{3}\cos \left(\frac{\pi}{2}\Delta_X\right)\left( \frac{1}{\sqrt{\bar \g}}
\sum^\n_{i=1}\langle{\cal O}_i{\cal O}_i\rangle \cb  \prod^n_{i=1}\varphi_{i,0}+
\frac{1}{\ell} \left(\DeltaX^- +\frac{d}{2}\right)   \prod^n_{i=1}\varphi_{i,0} \right)\,
 \label{eq:b4bound_dS_light}
 \end{align}
  These enter  in the cosmological one-loop effective action as\,\footnote{
  The calculation is here in Euclidian signature.    There is a relative $-$ sign compared to the equivalent expression in Lorentzian signature, due to the usual convention for the actions. See App.~\ref{app:HK} for details. }
 \begin{align}
  \Gamma_{{\rm dS},\rm div}^{\rm pert}[\varphi_{i,0}] = 
- \frac{ 1 }{16\pi^2 }\left(
  {\rm tr}\,\bb_{4}
+ 
\int_{\partial \cal M} d^{3}x \sqrt{\bar \g}   {\rm tr}
\,b^\partial_{4}(x) \right) \log\left(\frac{\mur}{m_\Phi}\right) \label{eq:GammadS}
 \end{align}

The bulk term contains holographic vertices with $n$ and $2n$ legs. The boundary term contains a $n$pt contact operator. It also contains a nonlocal contribution which amounts to attaching a 2pt CFT correlator  to that local operator.  
The logarithms obtained here via dimensional regularization are similar to those discussed in  \cite{Senatore:2009cf}.

 \subsubsection{Inflationary $\Phi^2\varphi^2$ Example}

 For inflationary correlators the light fields  typically have vanishing mass, which corresponds to $\mu=\frac{d}{2}$ and $\Delta_{-\mu}=0$. 
 We assume a quartic interaction  ${\cal L}^{\rm dS}_I=-\frac{\lambda}{4}\Phi^2\varphi^2$. 
 
 The total dimensions are $\Delta_X=0$, $\Delta^-_X=6$.  We use Neumann BC.
 We get 
 \begin{align}
 {\rm tr }\,\bar b_4&= \frac{\lambda^2}{4}\Lambda_{4} \cb \left[ \varphi_0 \right]^4 - \frac{2 \lambda}{\ell^2} \varphi \cB \varphi
\\
 {\rm tr }\, b^\partial_4&=
\frac{4 \lambda}{3}  \frac{1}{\sqrt{\bar \g}}
(\langle{\cal O}{\cal O}\rangle \cb  \varphi_{0}) \varphi_{0}+
\frac{5\lambda}{\ell}  \varphi_0^2\,
 \,. 
 \end{align}
 
First we make some simple comments. The last contribution to $\bar b_4$ is a  bulk mass term. 
 The first contribution  to $\bar b_4$ amounts to the generation of a bulk $\varphi^4$ operator. One can check that its beta function is the same as in ${\cal M}_4$. 
 The $\Lambda_4$ function (that we called ``holographic vertex'' in the previous sections) is here a 4pt cosmological correlator whose explicit expression is given in App.~\ref{app:dS_corr}.  

 Below we  extract further information from each term and discuss some implications.

  \subsubsection*{Expansion-induced Running Bulk Mass}

 We focus on the bulk mass term. In cosmological correlators,  bulk masses appear via the scaling dimensions $\Delta$ (with $m^2=\Delta(d-\Delta)\ell^2$) arising in the squeezed  limit of 4pt functions.  Here we assumed zero bulk mass for the $\varphi$ field.  
 
 However we can see that in dS$_4$ the bulk mass has a logarithmic running depending on the Hubble radius. A contribution  $\propto m^2_\Phi$ should also most likely appear from the $\bar b_2$ term, however our focus here is only on mass-independent divergences.   We find that the corresponding beta functions for $m_\varphi$ (or $\nu$) is 
 \be
 \beta_{m^2} = \beta_{\nu^2} =\frac{1}{4\pi}\frac{\lambda}{\ell^2} 
 \label{eq:betam2}
 \ee
 Let us compare with 4d Minkowski.  In both ${\cal M}_4$ and dS$_4$, the bulk mass receives logarithmic corrections of the form $\delta m^2\propto m_\Phi^2 \log\mur $ due to  nonderivative interactions if the particle $m_\Phi$ in the loop is massive. However these corrections vanish if $m_\Phi=0$. 
In contrast, Eq.\,\eqref{eq:betam2} dictates that the bulk mass runs as a result of the dS curvature. This effect is independent of $m_\Phi$, which could be zero. Such an effect does \textit{not} exist in ${\cal M}_4$,  \textit{i.e.} the beta function vanishes for $\ell\to 0$.

The upshot is that, in the presence of nonderivative interactions, even if all fields are exactly massless at a given scale, bulk masses of $O(\ell^{-2})$ times loop factor are radiatively generated.

 \subsubsection*{ Correction to OPE Coefficients  }

A CFT 2pt function on the late time boundary arises from the first term in $b_4^\partial$.  At vanishing scaling dimension, the CFT 2pt function in momentum space is simply given by
 \be
 \langle{\cal O}{\cal O}\rangle  = \frac{p^2 \eta_0^2}{1+p\eta_0} \ell^{-1} \,.
 \label{eq:2pt_0mass}
 \ee
 
The same 2pt function arises already at classical level from the kinetic term of $\phi$, thus this one-loop term is a correction to the normalization of $\cal O$. We find the correction
\be 
\left(1 + \frac{\lambda}{12\pi^2}\log\frac{\mur}{m_\Phi} \right) \langle{\cal O}(y_1){\cal O}(y_2)\rangle\,.
\ee
  As already discussed in section \ref{se:OPE_corr_AdS}, upon unit normalization of $\cal O$, this correction amounts  to a CFT version of ``wavefunction renormalization'', which  corrects the OPE coefficients involving $\cal O$.

 \subsubsection*{Boundary-localized Operator}
 
One of the divergences on the late time boundary is a mass term.  
We define the renormalized boundary mass term of the  dS Lagrangian as $b=b_{\rm dS}=-b_{\rm EAdS}$ such that $b_{\rm EAdS}$ enters as 
\be S_{\rm EAdS} \supset \frac{1}{2} \int_{\partial {\rm EAdS}}  d^3x \sqrt{\bar g}  b_{\rm EAdS} (  \varphi^2 +  \tilde \varphi^2) \ee in the EAdS action. We then obtain
the beta function
 \be
 \beta_b = \frac{5\lambda}{8\pi^2 \ell}\,. \label{eq:dS_betab_nonder}
 \ee
 
 This is our first example of the radiative generation of an  operator localized on the boundary of dS.  This term is nontrivial as it results from the finite Hubble radius and would not happen in flat spacetime. 
 The fact that this term is generated radiatively signals that it should be generically included in the fundamental dS action to start with.  
We will see in section \ref{se:propdS_dressed} below that such a term can have significant impact on the propagators.


 \subsubsection{Inflationary $\Phi^2(\partial_M \varphi)^2 $ Example}

We consider a derivative interaction 
 ${\cal L}_I^{\rm dS}=- \frac{\kappa}{4}\Phi^2\partial_M\varphi\partial^M\varphi$ 
and take $\Delta_{-\mu}=0$ such that $\varphi$ has shift symmetry. Here we will 
refer to $\varphi$  as the inflaton. 
The metric factor from the covariant contraction in ${\cal L}_I^{\rm dS}$ gives an extra $i^2$ factor when rotating from dS to EAdS.

We get the heat kernel coefficients
 \begin{align}
 {\rm tr }\,\bar b_4&=  \frac{\kappa^2}{4}\Lambda^\partial_{4} \cb \left[ \varphi_0 \right]^4 + \frac{2 \kappa}{\ell^2} \partial_M \varphi\cB \partial^M \varphi
\\
 {\rm tr }\, b^\partial_4&= - \frac{4 \kappa}{3}  \frac{1}{\sqrt{\bar \g}}
(\langle{\vec \partial\cal O}\vec \partial{\cal O}\rangle \cb  \varphi_{0}) \varphi_{0} - 
 \frac{ 2\kappa z_0^2}{3\ell^3} (\vec\partial \varphi_0)^2
 \,
 \end{align}
where we introduced the 4pt correlator with four derivatives
\be \Lambda^\partial_{4}(x_{0,1\ldots 4}) = \int_{\rm EAdS} d^4x \sqrt{g} \,\partial_M K(x,x_{0,1}) \partial^M K(x,x_{0,2})\partial_N K(x,x_{0,3}) \partial^N K(x,x_{0,4}) \,. \ee
This 4pt correlator  is explicitly given in App.~\ref{app:dS_corr}.

The second term in $\bar b_4$ contributes to the bulk kinetic term, for this term we have written the full bulk field. 
We can see that the 2pt correlator of descendant operators $\vec {\cal O}$ appears  in the boundary coefficient $\bar b^\partial_4$. In momentum space the combination simply amounts to $-p^2\langle{\cal O O} \rangle$.

\subsubsection*{Expansion-induced Running Wavefunction Normalization}

Introducing the renormalized parameters $Z(\mur)$, $\xi(\mur)$ in the dS Lagrangian
\be
{\cal L}^{\rm dS}= -\frac{  Z}{2} (\partial_M\varphi)^2 -\frac{\xi Z^2}{4!} (\partial_M\varphi\partial^M\varphi)^2
\ee
with $Z=1+\delta Z$,  we find the perturbative anomalous dimension for $\varphi$
\be
\gamma_\varphi = -\frac{1}{2}\frac{d\log Z}{d\log\mur} = \frac{\kappa^2}{8\pi^2\ell^2}
\ee
and the beta function
\be
\beta_{\xi} =  \frac{3\kappa^2}{8\pi^2}+2  \gamma_\varphi \xi
\ee

We can  compare these results to renormalization of the NLSM in 4d flat space. In the flat space NLSM, the $\xi$ coupling  would run similarly, but not the bulk kinetic term. That is, the anomalous dimension $\gamma_\varphi$ is a consequence of the nonzero curvature of spacetime;  if $\ell\to\infty$ (keeping constant $z_0/\ell$), then $\gamma_\varphi\to 0$. 
   Comparing $\gamma_\varphi$ and $\beta_{\xi}$ we can see that  the effect of this expansion-induced wavefunction renormalization becomes sizeable when $\xi$ is of order $\ell^2$.

\subsubsection*{Boundary-localized Kinetic Term }

One of the divergences on the late time boundary corresponds to a kinetic term.  
We define the renormalized boundary kinetic term in the dS Lagrangian $c=c_{\rm dS}=c_{\rm EAdS}$ such that $b_{\rm EAdS}$ enters as 
\be S_{\rm EAdS} \supset \frac{1}{2} \int_{\partial {\rm EAdS}}  d^3x \sqrt{\bar g}  c_{\rm EAdS} \frac{z_0^2}{\ell^2}(  (\vec\partial \varphi)^2 +  (\vec\partial \tilde\varphi)^2) \ee in the EAdS action. We then obtain
the beta function
 \be
 \beta_c = \frac{\lambda}{12\pi^2 \ell}\,. \label{eq:dS_betac_der}
 \ee

Again, this RG running is nontrivial as it results from the Hubble radius and would not happen in flat spacetime.  The fact that this term is generated radiatively signals that it should be generically included in the fundamental dS action to start with.

\subsubsection{Effect of Boundary Operators on Propagation }
\label{se:propdS_dressed}

\begin{figure}[t]
\centering
	\includegraphics[width=1.0\linewidth,trim={0cm 12cm 0cm 0cm},clip]{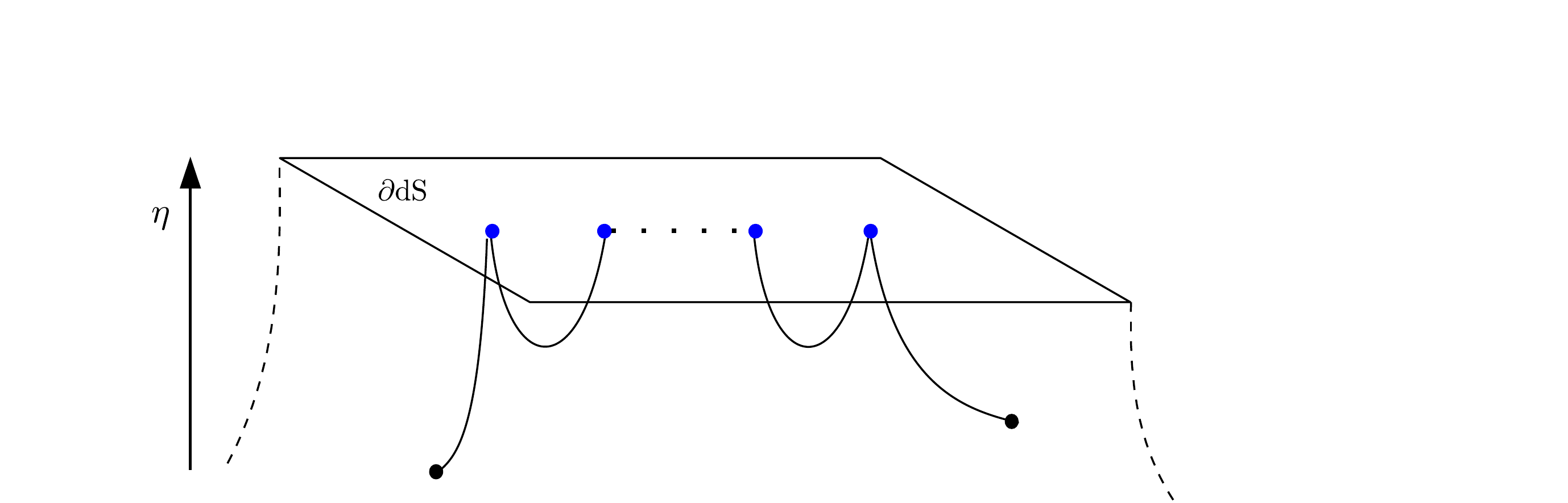}
\caption{ A propagator dressed by insertions of a bilinear operator localized on the late time boundary of de Sitter space. 
\label{fig:dS_dressed}}
\end{figure}

Throughout the above examples we have shown that radiative corrections generate
operators localized on the late time boundary,   in the presence of either derivative and nonderivative couplings of $\varphi$.  The general lesson is that such boundary operators should  be included in the fundamental Lagrangian in a first place.

The bilinear boundary-localized operators influence the dS propagators with Neumann BC. The structure of the propagator in the presence of such operators is the one established in Eqs.\,\eqref{eq:prop_hol_gen}, \eqref{eq:prop_holo_AdS}. 
As explained there, the effect of the boundary-localized operators  can be understood as a dressing of the propagators by boundary-localized insertions.\,\footnote{Renormalization of boundary operators dS$_4$ and dressing of the wavefunction have also been discussed in \cite{Cespedes:2020xqq}  in a different formalism. }

A contribution to this dressing is shown in Fig.\,\ref{fig:dS_dressed}. 
Denoting by $B[-\partial^2]$ the bilinear insertion localized on the late time boundary, and focussing on the massless case for concreteness, we obtain the boundary-to-boundary propagator
\be
G^{\rm dS}_{--,{\Delta=0}}(\vec p;\eta_0,\eta_0) = i
\frac{\eta^3_0(1-i p\eta_0)}{\ell^2\left(p^2\eta^2_0+B \ell(1-i  p\eta_0) \right)}
\label{eq:GbtbdS_dressed}
\ee
and the bulk-to-bulk propagator 
\begin{align}
& G^{\rm dS}_{--,{\Delta=0}}(\vec p;\eta,\eta') =  \label{eq:GdS_dressed}
 \\ \nn &
 \frac{\ell}{2p^3} \left(\frac{\eta\eta'}{\ell^2}\right)^{\frac{3}{2}}
 \frac{1- i p\eta_>}{1-ip\eta_0}
 \left(
- e^{-ip |\eta-\eta'| }(1+i p\eta_<)(1-i p\eta_0)+e^{-ip (2\eta_0-\eta-\eta') }(1-i p\eta_<)(1+i p\eta_0)
 \right)
 \\ \nn & ~~~~~~~~~~~~~ + G^{\rm dS}_{--,{\Delta=0}}(p;\eta_0,\eta_0) \frac{e^{-ip (\eta_0-\eta)  }(1-i p\eta)}{(1-i p\eta_0)}\frac{e^{-ip (\eta_0-\eta') } (1-i p\eta')}{(1-i p\eta_0)}
\end{align}
The first line in Eq.\,\eqref{eq:GdS_dressed} corresponds to the Dirichlet propagator, the second line is the boundary term which encodes all the effects of the boundary operators.
This is a Neumann/Dirichlet relation analogous to the one found in Eqs.\,\eqref{eq:prop_hol_gen}, \eqref{eq:prop_holo_AdS}. This is expected since  dS$\leftrightarrow $EAdS Wick rotations do not affect the structure of boundary conditions.
That is, the dressed dS propagator Eq.\,\eqref{eq:GdS_dressed} can for instance be obtained by starting in EAdS, using the results from section \ref{se:warped_formalism} and rotating to dS using Eq.\,\eqref{eq:G--Wick}.

We can see that whenever the boundary term $B$ is sizeable in units of $\ell$, it can have a significant impact on the propagator. This effect is well-known from the AdS$_5$ literature --- what typically happens is that the propagator tends to be deformed and repelled from the AdS boundary. We conclude that the same kind of effect can  happen in dS.  

In the case of a nonderivative interaction, a boundary-localized mass term $B=$\,cst is present --- at least generated at one-loop as dictated by Eq.\,\eqref{eq:dS_betab_nonder}.  This boundary  mass term is expected to be of order one in units of $\ell$, or at best suppressed by a 4d loop factor. This  term can thus have a strong impact on Eq.\,\eqref{eq:GdS_dressed} when $p\eta_0\ll 1$. 
Note however that the nonderivative case considered here is  tuned in the sense that a nonzero bulk mass \textit{i.e.} a nonzero scaling dimension is also expected, see Eq.\,\eqref{eq:betam2}.  

In the case of a derivative interaction, mass terms cannot be generated since they are protected by shift symmetry. There we have found that $B$ contains a boundary-localized kinetic term  of order $B \propto {\ell} p^2$ --- at least generated at one-loop as dictated by Eq.\,\eqref{eq:dS_betac_der}. This boundary-localized kinetic term  can have significant impact on the propagator if the $\kappa$ coupling is large enough.

It would be certainly interesting to further investigate the impact of boundary-localized operators on the cosmological correlators. This is left for future work.

\section*{Acknowledgments}

I thank Brian Batell, Tony Gherghetta, David Meltzer, Manuel Perez-Victoria, Eduardo Ponton, Allic Sivaramakrishnan and  Flip Tanedo for  useful discussions and comments.
This work has been supported by the S\~ao Paulo Research Foundation (FAPESP) under grants \#2011/11973, \#2014/21477-2 and \#2018/11721-4, by CAPES under grant \#88887.194785, and by the University of California, Riverside.

\appendix

\section{The Discontinuity Equation}
\label{app:der}

Here we give a proof of Eq.\,\eqref{eq:disc}.

We denote by $\partial{ \cal M}^-$ the surface at an infinitesimal normal distance of the actual boundary  $\cal M$ in the  inward direction.    
We define an infinitesimal shell $\sigma_{x_0}$ containing the boundary point $x_0$. The outward part of the boundary of  $\sigma_{x_0}$ is in  $\partial \cal M$, and the  inward part of the boundary of  $\sigma_{x_0}$ is in  $\partial {\cal M}^-$. 
A point belonging to $\partial {\cal M}^-$ is denoted $x_0^-$.

We integrate the bulk equation of motion Eq.\,\eqref{eq:EOM_G} over  the $\sigma_{x_0}$ volume and apply the divergence theorem.
The delta function of the EOM is picked by the integral if the other endpoint of the propagator is inside  the $\sigma_{x_0}$ shell. Taking the infinitesimal limit for $ \sigma_{x_0}$
and using that $\cal M$ is locally Euclidian, we get the discontinuity of the normal derivative evaluated at the boundary. This is Eq.\,\eqref{eq:disc}.

\section{Details and Checks of Propagators in Warped Background}

\label{app:prop}

\subsection{Generic Background  }

\label{app:warped_prop}

Here we show that the propagators given in Sec.~\ref{se:warped_holography} satisfy the general properties derived in Sec.~\ref{se:gen}.

We first define the boundary inner product. In Fourier coordinates it is given by a simple multiplication,
\be
\Phi_0\cb \Psi_0= \rho_0^d e^{\varphi_0} \Phi_0\Psi_0 \,.
\ee
Since the inverses are defined using this product,  inverses involve a $\rho_0^d e^{\varphi_0}$ factor: $[a]^{-1}= \rho_0^d e^{\varphi_0} \frac{1}{a}$. One has $[[a]^{-1}]^{-1}=a$. 

We will check the general relation $G_N=K\cb [G^{-1}_0+{\cal B}]^{-1}\cb K+G_D$.
We evaluate $G_N(p;z,z')-G_D(p;z,z')$ by direct computation:
\begin{align}
G_{N}-G_{D} & = \frac{i\f(z_<)\f(z_>)}{C \f_0 } \frac{C\rho^{d}_0e^{\varphi_0}}{ \rho_0\f'_0+{B}\f_0}
= i \frac{\f(z_<)\f(z_>)}{\f_0 \f_0} G^B_0
\\ & \nn
=K(z_<) \cb G^B_0 \cb K(z_>)
\end{align}
where in the intermediate step one has used the Wronskian in the numerator and we identified the boundary-to-boundary propagator, Eq.\,\eqref{eq:propa_b2b}.
Since $ G_B^N = [(G_B^N)^{-1}]^{-1} $, this verifies the main formula Eq.\,\eqref{eq:prop_hol_gen}.

We can also verify that 
\be
[G^B_0]^{-1}= 
-i\rho_0^{-d}e^{-\varphi_0}\left( \rho_0 \frac{ \f'_0}{\f_0}+{B} \right) = [G_0]^{-1} -i\rho_0^{-d}e^{-\varphi_0} B
\ee
This reproduces the dressing of the boundary-to-boundary propagator by the boundary-localized bilinear insertion $B$.  

It is also instructive to explicitly  verify  the discontinuity formula. We get
\be
(\rho_0 \partial_{z} +B \f_0 ) G^B_N(p;z,z^-_0)|_{z\to z_0} =  i\rho^d_0e^\varphi_0=\frac{i}{\sqrt{|\bar g|}}\,
\ee
where again one has used the Wronskian at $z_0$. 
This  result verifies Eq.\,\eqref{eq:DiscN} upon translation from Lorentzian to Euclidian  conventions.

\subsection{AdS  Propagator in Position Space}

\label{app:AdS_prop}

In Euclidian AdS the propagator satisfying the equation of motion \be \left(-\square_{\rm AdS} +\Delta(\Delta-d)k^2\right)G_\Delta(x,x')=\frac{1}{\sqrt{ \g}}\delta^{d+1}(x-x')\ee is \cite{Burges:1985qq}
\be
G_\Delta(x,x')= \frac{\Gamma(\Delta)}{2\pi^{d/2}\Gamma(\Delta-\frac{d}{2}+1)} \frac{k^{d-1-2\Delta}}{\zeta(x,x')^\Delta}  \,_2F_1\left(
 \Delta,\Delta-\frac{d}{2}+\frac{1}{2}, 2\Delta-d+1 ,-\frac{4}{k^2\zeta(x,x')}
 \right)
\ee
with
\be
\zeta(x,x') = \frac{1}{k^2}\frac{(z-z')^2+(x-x')^2}{zz'}
\ee
the chordal distance in Poincar\'e coordinates. 
Taking one endpoint towards the boundary, the propagator is asymptotically
\be
G_\Delta(x,x')|_{z\to 0}=  
\frac{\Gamma(\Delta)}{2\pi^{d/2}\Gamma(\Delta-\frac{d}{2}+1)} \frac{k^{d-1-2\Delta}}{\zeta(x,x')^\Delta} + O\left(\frac{1}{\zeta^{\Delta+1}}\right) \label{eq:B2b_AdS_Pos}
\ee
for any $\Delta$. 

\section{Holography with Two Boundaries}
\label{app:double}

In this appendix we work out the holographic formalism for a $z_0<z<z_1$ slice of the warped background introduced in Eq.\,\eqref{eq:metric:general}, with \textit{both} boundaries treated holographically. 

Similarly to Sec.~\ref{se:gen}, our starting point is to write the holographic basis in the presence of two boundaries,
\be
\Phi=\Phi_0 \cb L_0+ \Phi_1 \cb L_1 + \Phi_D
\ee
where $\Phi_{0,1}$ are the field values on each boundary, which will be the variables of the holographic action.  The $L_0$, $L_1$ boundary-to-bulk propagators are determined below. 
We work in Fourier space along the transverse Poincar\'e slice, hence the coordinates are $(p^\mu,z)$. 
We consider the classical value $\langle\Phi\rangle$ which satisfies the bulk EOM ${\cal D}\langle\Phi\rangle=0$. It takes in general the form
\be
\langle\Phi\rangle(z)=  a \f(z) + b \h(z)   \,.
\label{eq:Phi_cl_double}
\ee
The associated Wronskian is
\be
W(z) = \f(z) \h'(z) - \f'(z) \h(z) = C \rho^{d-1} e^\varphi \,. 
\ee
The $\f$, $\h$ solutions  and the $a$, $b$  constants  depend  on the  $d$-momentum. We also define 
\be
\f(z_0)=\f_0\,, \quad
\f(z_1)=\f_1\,, \quad
\h(z_0)=\h_0\,, \quad
\h(z_1)=\h_1\, . \label{eq:BC_def_fg}
\ee
Using the definitions Eqs.~\eqref{eq:Phi_cl_double}, \eqref{eq:BC_def_fg}, the $a,b$ constants can be translated into the holographic variables,
\be
a= \frac{\Phi_0 \h_1 - \Phi_1 \h_0 }{\f_0 \h_1-\f_1 \h_0} \, , \quad 
b= - \frac{\Phi_0 \f_1 - \Phi_1 \f_0 }{\f_0 \h_1-\f_1 \h_0} \,.
\label{eq:AB}
\ee
The holographic basis in Fourier space is therefore
\be
\Phi(p,z)=\Phi_0\frac{ \h_1 \f(z) - \f_1 \h(z)  }{\f_0 \h_1-\f_1 \h_0}+
\Phi_1\frac{ \f_0 \h(z) - \h_0 \f(z)  }{\f_0 \h_1-\f_1 \h_0}+\Phi_D\,. \label{eq:Phi_hol_double}
\ee

We plug the solution Eq.~\eqref{eq:Phi_hol_double} into  the quadratic action, giving
\be
S[\Phi_0, \Phi_1] =  \int \frac{d^dp}{(2\pi)^4} \sqrt{|\bar \g|}e^{-\varphi} \frac{1}{2}\left(
\Phi_0 (\rho_0 \partial_z -  b_0 )\Phi|_{z_0}-
\Phi_1 (\rho_1 \partial_z -  b_1 ) \Phi|_{z_1}
\right)  +S_D
\ee
where $b_0$, $b_1$ encode the boundary actions. 
We introduce  \be \hat f_i = \rho_i f'(z_i)-  b_i f(z_i)\ee
and similarly for $\hat g_i$.
Evaluating $\partial_z\Phi_{z_0}$ and $\partial_z\Phi_{z_1}$
using  Eq.~\eqref{eq:Phi_hol_double} we obtain the holographic self-energies
\be
S[\Phi_0, \Phi_1]=
\frac{1}{2}\int \frac{d^4p}{(2\pi)^4} \left(
\Phi_0
\Pi_{0}
 \Phi_0
 +2 \Phi_0
\Pi_{01}
 \Phi_1
 +\Phi_1
\Pi_{1}
 \Phi_1
\right)\,,
\ee
\be
\Pi_{0}=\rho_0^{-d}e^{-\varphi_0}\frac{\hat \f_0 \h_1 - \hat \h_0 \f_1 }{\f_0 \h_1- \h_0 \f_1}\,, \quad 
\Pi_{1}=\rho_1^{-d}e^{-\varphi_1}\frac{\hat \f_1 \h_0 - \hat \h_1 \f_0 }{\f_0 \h_1-\h_0\f_1}\,, \quad
\Pi_{01}=\frac{C }{\f_0 \h_1-\h_0 \f_1}
\ee
where  we have used the Wronskian at $z_0$ and $z_1$ to get the mixed self-energy $\Pi_{01}$.  

We can re-express these self-energies using propagators. 
 The propagator in the presence of two boundaries is
 \be
 G^{\pm \pm}(y,y')=\frac{i}{C}\frac{(\tilde \f_0  \h(z_<)- \tilde \h_0 \f(z_<))
(\tilde \f_1  \h(z_>)- \tilde \h_1 \f(z_>))  }{\tilde \f_0 \tilde \h_1- \tilde \h_0 \tilde \f_1}
\label{eq:Green_gen}
 \ee
where $ \tilde f_i = f_i $
 for a Dirichlet boundary condition ($-$) and 
$ \tilde f_i = \hat f_i$  for a Neumann boundary condition ($+$). 
For the $\Pi_{0}$, $\Pi_{1}$ self-energies we find 
 \be \label{eq:Pi_1}
 \Pi_{0}=\frac{i}{G_p^{+-}\left(
z_0,z_0
 \right)} \, , \quad 
  \Pi_{1}=\frac{i}{G_p^{-+}\left(
z_1,z_1
 \right)}\,. 
 \ee
 The $\Pi_{01}$ self-energy can be expressed as
\begin{align} \label{eq:Pi_01}
\Pi_{01}=\frac{C }{\f_0 \h_1-\h_0 \f_1}&=C
\frac{1}{\tilde \f_0 \tilde \h_1-\tilde \h_0 \tilde \f_1}
\frac{\tilde \f_0 \tilde \h_1-\tilde \h_0 \tilde \f_1}{\tilde \f_0 \h_1-\tilde \h_0 \f_1}
\frac{\tilde \f_0 \h_1-\tilde \h_0 \f_1}{\f_0 \h_1-\h_0 \f_1} \nn
\\
&=
i \frac{G_p^{++}\left(
z_0,z_1
 \right)}{G_p^{++}\left(
z_1,z_1
 \right)G_p^{+-}\left(
z_0,z_0
 \right)}
=
i\frac{G_p^{++}\left(
z_0,z_1
 \right)}{G_p^{++}\left(
z_0,z_0
 \right)G_p^{-+}\left(
z_1,z_1
 \right)}
\,.
\end{align}
We can  recognize (amputated) boundary-to-bulk propagators in the last expressions. We have 
\be
\Pi_{01}=K_0(z_1)\Pi_1=K_1(z_1)\Pi_0\, \label{eq:Pi_01_K}
\ee
where $K_0(z)$, $K_1(z)$ denote the amputated boundary-to-bulk propagators from the $z_0$ and $z_1$ boundaries respectively.

If we let the $\Phi_1$ variable be dynamical \textit{i.e.}  integrate it out in the path integral, we should recover our standard one-boundary holographic action. 
 For a Dirichlet  boundary condition on $z_1$, we have $\Phi_1=0$. It follows trivially that
\be
S[\Phi_0]=
\frac{1}{2}\int \frac{d^4p}{(2\pi)^4} 
\frac{i}{G^{+-}_p\left(z_0,z_0\right)}\Phi_0^2+S_D
\ee
which is the expected result for the holographic action on boundary $0$ with Dirichlet boundary condition on boundary $1$. 
For a Neumann  boundary condition on $z_1$, we have $(\rho_1\partial_z -  b_1 ) \Phi|_{z_1}
=0$, which implies
\be
\Phi_1=\Phi_0\frac{W}{\h_0 \tilde \f_1 -\f_0 \tilde \h_1}\,.
\ee
Substituting $\Phi_1$ with this relation in the two-boundary holographic action gives
\be
S[\Phi_0]=\frac{1}{2}
\int \frac{d^4p}{(2\pi)^4} 
\frac{i}{G^{++}_p\left(z_0,z_0\right)}\Phi_0^2+S_D
\ee
which is again the expected result for the holographic action on boundary $0$ with Neumann boundary condition on boundary $1$.

\section{Elements of the Heat Kernel Formalism}

\label{app:HK}

We work in Euclidian metric and convert to Lorentzian conventions at the end of the calculation. The Euclidian effective action generates the 1PI Euclidian correlators following
\be
-\frac{\delta^n \Gamma_E}{\delta \Phi(x_1)\ldots \delta \Phi(x_n)}=\langle\Phi(x_1)\ldots\Phi(x_n)\rangle\,.
\label{eq:GammaE}
\ee
The one-loop effective action is put in the form of the heat kernel integral
\be
\Gamma^E_{1-{\rm loop}} = -\frac{1}{2}\int \frac{dt}{t} {\rm Tr}\, e^{-t{\cal D}_E}\,. 
\label{eq:Gam1}
\ee
with ${\cal D}_E=-\square_E+m^2+X $. Tr is the trace over all indexes including spacetime coordinates.
The heat kernel trace is expanded as 
\be
{\rm Tr} e^{-t(-\square_E+m^2+X )} =(4\pi t)^{D/2}e^{-tm^2} \int d^Dx_E \sqrt{\g} \sum^\infty_{r=0} {\rm tr}\, b^E_{2r}(x) t^r\,
\ee
where the $b^E_{2r}$ are the Euclidian heat kernel coefficients. 
Using this expansion in Eq.\,\eqref{eq:Gam1}, integrating in $t$ and analytically continuing in $D$ gives
\be
\Gamma^E_{1-{\rm loop}} = -\frac{1}{2}
\frac{\Gamma(r-D/2)}{(4\pi )^{D/2}}
\int d^Dx_E \sqrt{\g} \sum^\infty_{r=0}  \frac{ {\rm tr}\, b^E_{2r}(x) }{m^{2r-D}} \label{eq:GamE}
\ee
for any spacetime dimension. 
Converting to Lorentzian metric gives\,\footnote{Use $x_E^0=i x^0$,  $\Gamma_E^{1-{\rm loop}}=-i \Gamma_{1-{\rm loop}}$. }
\be
\Gamma_{1-{\rm loop}} = \frac{1}{2}
\frac{\Gamma(r-D/2) }{(4\pi )^{D/2}}
\int d^Dx \sqrt{\g} \sum^\infty_{r=0} \frac{ {\rm tr}\,b_{2r}(x)  }{m^{2r-D}}
\label{eq:GamL}
\ee
with $\rm tr$ the trace over internal (non-spacetime) indexes.

The exact $b_0$, $b_2$, $b_4$ coefficients are given in Eq.\,\eqref{eq:b4}.
The exact  $b_6$ coefficient is
\begin{align}
b_6 = & \frac{1}{360}\bigg(
8 D_{P}\Omega_{MN}D^{P}\Omega^{MN}
+2 D^{M}\Omega_{MN} D_{P}\Omega^{PN}
+12 \Omega_{MN}\square \Omega^{MN}
-12 \Omega_{MN}\Omega^{NP}\Omega^{~~M}_{P}  \nn  \\ \nn 
&-6 R_{MNPQ}\Omega^{MN}\Omega^{PQ}
-4 R_{M}^{~N}\Omega^{MP}\Omega_{NP} + 5 R\Omega_{MN}\Omega^{MN} \\ & \nn
- 6 \square^2 X +60 X\square X+ 30 D_M X D^M X - 60 X^3
\\ & \nn
- 30 X \Omega_{MN}\Omega^{MN} - 10 R \square X - 4 R_{MN} D^N  D^M X - 12 D_M R D^M X + 30 XX R \\ & \nn 
- 12 X \square R  - 5 X R^2 + 2 X R_{MN}R^{MN} - 2 X R_{MNPQ}R^{MNPQ}
\bigg)   
\\ & \nn
+\frac{1}{7!}\bigg(
18 \square^2  R + 17 D_M R D^M R  - 2D_PR_{MN} D^PR^{MN}
- 4 D_PR_{MN} D^M R^{PN}  \\ & \nn
+9 D_PR_{MNQL} D^PR^{MNQL}  + 28 R \square R - 8 R_{MN} \square R^{MN} 
 \\ & \nn
+ 24 R_{MN} D_P D^N  R^{MP} + 12 R_{MNQL} \square R^{MNQL} +35/9 R^3 
\\ & \nn
-14/3 R R_{MN} R^{MN} 
+ 14/3 R R_{MNPQ}R^{MNPQ} -208/9 R_{MN} R^{MP} R_{~~P}^{N} 
\\ & \nn
-64/3 R_{MN}R_{PQ} R^{MPNQ}  
-16/3 R^{M}_{~N } R_{MPQL}R^{NPQL}
\\ & 
- 44/9 R^{MN}_{~~~~AB} R_{MNPQ} R^{PQ AB}  -80/9 R_{M~~P~~}^{~~N~~Q} R^{MAP B} R_{NA Q B  } 
\bigg) I 
\label{eq:b6full}
\end{align}

\subsection{Dirichlet $b^\partial_i$ for Scalar Interaction }

\label{app:b_Dir}

In Sec.~\ref{se:AdSLoopScalar} we have used Neumann BC for the fluctuation. If instead one takes Dirichlet BC, using that $\Pi_+=0$, $\Pi_-=1$, $\chi=-1$, $S=0$, the boundary heat kernel coefficients are found to be
\begin{align}
{\rm tr}\, b^\partial_3 &=\frac{1}{4} X_0 \\
{\rm tr}\, b^\partial_4 & = \frac{1}{3} \frac{1}{\sqrt{\bar \g}}
\sum^\n_{i=1}\langle{\cal O}_i{\cal O}_i\rangle \cb X_0
+\frac{1}{3}(\DeltaX^- -d) k X_0 \,
\end{align}
\begin{align}
{\rm tr}\, b^\partial_5  = \frac{1}{5760}\bigg(& -720 X_0^2  + 
\left(360  \Delta^-_{X}(d-\Delta^-_{X}) k^2 - 450 d \DeltaX^- k^2 
-5d(53+ 27d)k^2 \right) X_0
\nn \\
&  - 360
k^2 z_0^2\sum^\n_{i=1}(\partial_\mu\partial^\mu)_{i} X_0 
+240 k^2 z_0^2 \partial_\mu\partial^\mu X_0
\nn \\
& + (720 \DeltaX^--450 d)k  \frac{1}{\sqrt{\bar \g}}
\sum^\n_{i=1}\langle{\cal O}_i{\cal O}_i\rangle \cb X_0
\nn \\
& + 
360\frac{1}{\sqrt{\bar \g}}\frac{1}{\sqrt{\bar \g}}
\sum^\n_{i,j=1}[\langle{\cal O}_i{\cal O}_i\rangle \cb][\langle{\cal O}_j{\cal O}_j\rangle \cb] X_0
\bigg)
\end{align}

\subsection{Elementary Checks from ${\cal M}_3$ and  EAdS$_3$  Bubble}

\subsection*{${\cal M}_3$ Bubble}

We check a  contribution to the Lorentzian  heat kernel coefficient $b_4$ from the scalar bubble diagram in ${\cal M}_3$. 
In ${\cal M}_3$, working in momentum space as in particle physics-style calculations,  the amputated bubble at large $m$ is given by
\be
B_{{\cal M}_3}\Big|_{m\to \infty}= \frac{1}{2}\int \frac{d^3 k}{(2\pi)^3} \frac{1}{-p^2-m^2}\frac{1}{-(p+k)^2-m^2}\Big|_{m\to \infty}=\frac{i}{16\pi m} \ee
Consider the fundamental interaction  $\frac{1}{2}\Phi^2{\cal O}$. 
The bubble diagram would be generated by two  ${\cal O}$ derivatives of the Lorentzian  effective action $\frac{1}{32\pi m}{\cal O}^2$. 
This matches  exactly the $X^2$ term coming from the $b_4$  coefficient (obtained from Eq.\,\eqref{eq:b4}) in  Eq.\,\eqref{eq:GamL}.

\subsubsection*{EAdS$_3$ Bubble } 

We check a  contribution to the Euclidian  heat kernel coefficient $b^E_4$ from the scalar bubble diagram in EAdS. 

\label{app:bubble_check}
Consider the  bubble diagram of a real scalar in EAdS$_3$,
\be
B(x,x')=\frac{1}{2}G(X,Y) ^2 = k^2 \int_{\mathbb{R}} d\nu B(\nu,\Delta)\Omega_\nu(x,x')
\ee
In AdS$_3$, Ref.\,\cite{Carmi:2018qzm} finds 
\be
B(\nu,\Delta)=i\frac{\psi\left(\Delta-\frac{1+i\nu}{2}\right)-\psi\left(\Delta-\frac{1-i\nu}{2}\right)}{16\pi\nu}
\ee
where $\Psi$ is the Digamma function. We expand the functions for large $|\Delta|$ at fixed $\nu$. The leading order  result is
\be
B(\nu,\Delta)|_{|\Delta|\to \infty}= \frac{1}{16\pi \Delta}+O\left(\frac{1}{|\Delta|^2}\right)\,. 
\ee
In position space, using that $\int \Omega_\nu(x,x') = \frac{1}{k^{d+1}\,\sqrt{|\g|}} \delta^{d+1}(x,x')$, we obtain the local operator
\be
B(x,x')|_{|\Delta|\to \infty} =\frac{1}{\sqrt{|\g|}} \frac{1}{16\pi\Delta k } \delta^{3}(x,x')\,. \label{eq:LargeB}
\ee

Consider the fundamental interaction  $\frac{1}{2}\Phi^2{\cal O}$. 
Using Eq.\,\eqref{eq:GammaE}, the bubble diagram would be generated by two  ${\cal O}$ derivatives of the Euclidian  effective action
\be
-\frac{1}{2}\int dx^{d+1} \int dx'^{d+1}\sqrt{|\g|_x} \sqrt{|\g|_{x'}} {\cal O}(x) B(x,x') {\cal O}(x') \,.
\label{eq:EFTB}
\ee
Plugging the limit of Eq.\,\eqref{eq:LargeB} into Eq.\,\eqref{eq:EFTB} gives the local operator 
\be
-\frac{1}{32\pi\Delta k}\int dx^{d+1} \sqrt{|\g|}  {\cal O}^2(x)  \,
\ee
in the Euclidian effective action. 
This matches exactly the $X^2$ term coming from the $b^E_4$  coefficient (obtained from Eq.\,\eqref{eq:b4}) in  Eq.\,\eqref{eq:GamE}.

\section{ Some Elementary 4pt Cosmological Correlators}
\label{app:dS_corr}

Here we give the momentum space expressions for the $\Lambda_4$ and $\Lambda^\partial_4$ holographic vertices with massless fields \textit{i.e.} $\Delta=0$. These quantities in the boundary effective action computed in section \ref{se:dS4_ren}, and are thus contribution to the 4pt cosmological correlators of $\varphi$.  

Defining $p_t=\sum_{i=1}^4p_i$ 
and taking $p_i z_0 \ll 1$, we get
\begin{align}
\Lambda_4(\vec p_1, \vec p_2, \vec p_3, \vec p_4)=& \frac{\ell^4}{3z_0^3}-\frac{\ell^4}{3z_0}(p_1^2+p_2^2+p_3^2+p_4^2)  +\ell^4\frac{p_1p_2p_3p_4}{p_t} \\
& \nn 
- \frac{\ell^4}{3 } \log ({p_t} {z_0})  \left({p_1}^3+{p_2}^3+{p_3}^3+{p_4}^3\right)
\\
& \nn 
+\frac{\ell^4}{3 } 
\left({p_t}
   \sum_{i,j}p_ip_j
   -\sum_i{p_i}\sum_{j\neq i}p_j -  4\sum_{i<j<k } p_ip_jp_k
   \right)
\end{align}

For the $\Lambda^\partial_4$ correlator, using $p_i z_0 \ll 1$ we find
\begin{align}
& \Lambda^\partial_4(\vec p_1, \vec p_2, \vec p_3, \vec p_4)= 
\vec p_1 \cdot \vec p_2\, \vec p_3 \cdot \vec p_4 \, \Lambda(\vec p_1, \vec p_2, \vec p_3, \vec p_4) +\frac{40320}{\ell^4} \frac{p^2_1p^2_2p^2_3p^2_4}{p_t^9} \\
& \nn 
-\frac{24}{\ell^2}\vec p_3\cdot \vec p_4 p_1^2p_2^2 
\frac{
\left(p_1^2+p_2^2 +6p_3^2+6p_4^2 +2p_1p_2 +7(p_1+p_2)(p_3+p_4) +42 p_3p_4\right)+(1,2)\leftrightarrow (3,4)
}{p^7_t}
\end{align}

\bibliographystyle{JHEP}
\bibliography{biblio}

\providecommand{\href}[2]{#2}\begingroup\raggedright\begin{thebibliography}{100}

\bibitem{Aharony:1999ti}
O.~Aharony, S.~S. Gubser, J.~M. Maldacena, H.~Ooguri, and Y.~Oz, {\it {Large N
  field theories, string theory and gravity}},  {\em Phys. Rept.} {\bf 323}
  (2000) 183--386, [\href{http://arxiv.org/abs/hep-th/9905111}{{\tt
  hep-th/9905111}}].

\bibitem{Zaffaroni:2000vh}
A.~Zaffaroni, {\it {Introduction to the AdS-CFT correspondence}},  {\em Class.
  Quant. Grav.} {\bf 17} (2000) 3571--3597.

\bibitem{Nastase:2007kj}
H.~Nastase, {\it {Introduction to AdS-CFT}},
  \href{http://arxiv.org/abs/0712.0689}{{\tt arXiv:0712.0689}}.

\bibitem{Kap:lecture}
J.~Kaplan, ``{Lectures on AdS/CFT from the Bottom Up}.''

\bibitem{Witten:1998qj}
E.~Witten, {\it {Anti-de Sitter space and holography}},  {\em Adv. Theor. Math.
  Phys.} {\bf 2} (1998) 253--291,
  [\href{http://arxiv.org/abs/hep-th/9802150}{{\tt hep-th/9802150}}].

\bibitem{Gubser:1998bc}
S.~S. Gubser, I.~R. Klebanov, and A.~M. Polyakov, {\it {Gauge theory
  correlators from noncritical string theory}},  {\em Phys. Lett.} {\bf B428}
  (1998) 105--114, [\href{http://arxiv.org/abs/hep-th/9802109}{{\tt
  hep-th/9802109}}].

\bibitem{tHooft:1993dmi}
G.~'t~Hooft, {\it {Dimensional reduction in quantum gravity}},  {\em Conf.
  Proc. C} {\bf 930308} (1993) 284--296,
  [\href{http://arxiv.org/abs/gr-qc/9310026}{{\tt gr-qc/9310026}}].

\bibitem{Susskind:1994vu}
L.~Susskind, {\it {The World as a hologram}},  {\em J. Math. Phys.} {\bf 36}
  (1995) 6377--6396, [\href{http://arxiv.org/abs/hep-th/9409089}{{\tt
  hep-th/9409089}}].

\bibitem{Bousso:2002ju}
R.~Bousso, {\it {The Holographic principle}},  {\em Rev. Mod. Phys.} {\bf 74}
  (2002) 825--874, [\href{http://arxiv.org/abs/hep-th/0203101}{{\tt
  hep-th/0203101}}].

\bibitem{CS_WZW}
E.~Witten, {\it {Quantum field theory and the Jones polynomial}},  {\em
  Communications in Mathematical Physics} {\bf 121} (1989), no.~3 351 -- 399.

\bibitem{Ryu:2006bv}
S.~Ryu and T.~Takayanagi, {\it {Holographic derivation of entanglement entropy
  from AdS/CFT}},  {\em Phys. Rev. Lett.} {\bf 96} (2006) 181602,
  [\href{http://arxiv.org/abs/hep-th/0603001}{{\tt hep-th/0603001}}].

\bibitem{DeWitt_original}
B.~S. DeWitt, {\it {Dynamical theory of groups and fields}},  {\em Conf. Proc.
  C} {\bf 630701} (1964) 585--820.

\bibitem{DeWitt_original2}
B.~S. DeWitt, {\it Quantum field theory in curved spacetime},  {\em Physics
  Reports} {\bf 19} (1975), no.~6 295--357.

\bibitem{Gilkey_original}
P.~B. Gilkey, {\it {The spectral geometry of a Riemannian manifold}},  {\em
  Journal of Differential Geometry} {\bf 10} (1975), no.~4 601 -- 618.

\bibitem{McAvity:1990we}
D.~M. McAvity and H.~Osborn, {\it {A DeWitt expansion of the heat kernel for
  manifolds with a boundary}},  {\em Class. Quant. Grav.} {\bf 8} (1991)
  603--638.

\bibitem{Vassilevich:2003xt}
D.~V. Vassilevich, {\it {Heat kernel expansion: User's manual}},  {\em Phys.
  Rept.} {\bf 388} (2003) 279--360,
  [\href{http://arxiv.org/abs/hep-th/0306138}{{\tt hep-th/0306138}}].

\bibitem{Beccaria:2014qea}
M.~Beccaria, G.~Macorini, and A.~A. Tseytlin, {\it {Supergravity one-loop
  corrections on AdS$_7$ and AdS$_3$, higher spins and AdS/CFT}},  {\em Nucl.
  Phys. B} {\bf 892} (2015) 211--238,
  [\href{http://arxiv.org/abs/1412.0489}{{\tt arXiv:1412.0489}}].

\bibitem{Cornalba:2007zb}
L.~Cornalba, M.~S. Costa, and J.~Penedones, {\it {Eikonal approximation in
  AdS/CFT: Resumming the gravitational loop expansion}},  {\em JHEP} {\bf 09}
  (2007) 037, [\href{http://arxiv.org/abs/0707.0120}{{\tt arXiv:0707.0120}}].

\bibitem{Penedones:2010ue}
J.~Penedones, {\it {Writing CFT correlation functions as AdS scattering
  amplitudes}},  {\em JHEP} {\bf 03} (2011) 025,
  [\href{http://arxiv.org/abs/1011.1485}{{\tt arXiv:1011.1485}}].

\bibitem{Fitzpatrick:2011hu}
A.~Fitzpatrick and J.~Kaplan, {\it {Analyticity and the Holographic S-Matrix}},
   {\em JHEP} {\bf 10} (2012) 127, [\href{http://arxiv.org/abs/1111.6972}{{\tt
  arXiv:1111.6972}}].

\bibitem{Alday:2017xua}
L.~F. Alday and A.~Bissi, {\it {Loop Corrections to Supergravity on $AdS_5
  \times S^5$}},  {\em Phys. Rev. Lett.} {\bf 119} (2017), no.~17 171601,
  [\href{http://arxiv.org/abs/1706.02388}{{\tt arXiv:1706.02388}}].

\bibitem{Alday:2017vkk}
L.~F. Alday and S.~Caron-Huot, {\it {Gravitational S-matrix from CFT dispersion
  relations}},  {\em JHEP} {\bf 12} (2018) 017,
  [\href{http://arxiv.org/abs/1711.02031}{{\tt arXiv:1711.02031}}].

\bibitem{Alday:2018pdi}
L.~F. Alday, A.~Bissi, and E.~Perlmutter, {\it {Genus-One String Amplitudes
  from Conformal Field Theory}},  {\em JHEP} {\bf 06} (2019) 010,
  [\href{http://arxiv.org/abs/1809.10670}{{\tt arXiv:1809.10670}}].

\bibitem{Alday:2018kkw}
L.~F. Alday, {\it {On Genus-one String Amplitudes on $AdS_5 \times S^5$}},
  \href{http://arxiv.org/abs/1812.11783}{{\tt arXiv:1812.11783}}.

\bibitem{Meltzer:2018tnm}
D.~Meltzer, {\it {Higher Spin ANEC and the Space of CFTs}},  {\em JHEP} {\bf
  07} (2019) 001, [\href{http://arxiv.org/abs/1811.01913}{{\tt
  arXiv:1811.01913}}].

\bibitem{Ponomarev:2019ltz}
D.~Ponomarev, E.~Sezgin, and E.~Skvortsov, {\it {On one loop corrections in
  higher spin gravity}},  {\em JHEP} {\bf 11} (2019) 138,
  [\href{http://arxiv.org/abs/1904.01042}{{\tt arXiv:1904.01042}}].

\bibitem{Shyani:2019wed}
M.~Shyani, {\it {Lorentzian inversion and anomalous dimensions in Mellin
  space}},  \href{http://arxiv.org/abs/1908.00015}{{\tt arXiv:1908.00015}}.

\bibitem{Alday:2019qrf}
L.~F. Alday and E.~Perlmutter, {\it {Growing Extra Dimensions in AdS/CFT}},
  {\em JHEP} {\bf 08} (2019) 084, [\href{http://arxiv.org/abs/1906.01477}{{\tt
  arXiv:1906.01477}}].

\bibitem{Alday:2019nin}
L.~F. Alday and X.~Zhou, {\it {Simplicity of AdS Supergravity at One Loop}},
  \href{http://arxiv.org/abs/1912.02663}{{\tt arXiv:1912.02663}}.

\bibitem{Meltzer:2019pyl}
D.~Meltzer, {\it {AdS/CFT Unitarity at Higher Loops: High-Energy String
  Scattering}},  {\em JHEP} {\bf 05} (2020) 133,
  [\href{http://arxiv.org/abs/1912.05580}{{\tt arXiv:1912.05580}}].

\bibitem{Aprile:2017bgs}
F.~Aprile, J.~Drummond, P.~Heslop, and H.~Paul, {\it {Quantum Gravity from
  Conformal Field Theory}},  {\em JHEP} {\bf 01} (2018) 035,
  [\href{http://arxiv.org/abs/1706.02822}{{\tt arXiv:1706.02822}}].

\bibitem{Aprile:2017xsp}
F.~Aprile, J.~Drummond, P.~Heslop, and H.~Paul, {\it {Unmixing Supergravity}},
  {\em JHEP} {\bf 02} (2018) 133, [\href{http://arxiv.org/abs/1706.08456}{{\tt
  arXiv:1706.08456}}].

\bibitem{Aprile:2017qoy}
F.~Aprile, J.~Drummond, P.~Heslop, and H.~Paul, {\it {Loop corrections for
  Kaluza-Klein AdS amplitudes}},  {\em JHEP} {\bf 05} (2018) 056,
  [\href{http://arxiv.org/abs/1711.03903}{{\tt arXiv:1711.03903}}].

\bibitem{Giombi:2017hpr}
S.~Giombi, C.~Sleight, and M.~Taronna, {\it {Spinning AdS Loop Diagrams: Two
  Point Functions}},  {\em JHEP} {\bf 06} (2018) 030,
  [\href{http://arxiv.org/abs/1708.08404}{{\tt arXiv:1708.08404}}].

\bibitem{Cardona:2017tsw}
C.~Cardona, {\it {Mellin-(Schwinger) representation of One-loop Witten diagrams
  in AdS}},  \href{http://arxiv.org/abs/1708.06339}{{\tt arXiv:1708.06339}}.

\bibitem{Aharony:2016dwx}
O.~Aharony, L.~F. Alday, A.~Bissi, and E.~Perlmutter, {\it {Loops in AdS from
  Conformal Field Theory}},  {\em JHEP} {\bf 07} (2017) 036,
  [\href{http://arxiv.org/abs/1612.03891}{{\tt arXiv:1612.03891}}].

\bibitem{Yuan:2017vgp}
E.~Y. Yuan, {\it {Loops in the Bulk}},
  \href{http://arxiv.org/abs/1710.01361}{{\tt arXiv:1710.01361}}.

\bibitem{Yuan:2018qva}
E.~Y. Yuan, {\it {Simplicity in AdS Perturbative Dynamics}},
  \href{http://arxiv.org/abs/1801.07283}{{\tt arXiv:1801.07283}}.

\bibitem{Bertan:2018afl}
I.~Bertan, I.~Sachs, and E.~D. Skvortsov, {\it {Quantum $\phi^4$ Theory in
  AdS${}_4$ and its CFT Dual}},  {\em JHEP} {\bf 02} (2019) 099,
  [\href{http://arxiv.org/abs/1810.00907}{{\tt arXiv:1810.00907}}].

\bibitem{Bertan:2018khc}
I.~Bertan and I.~Sachs, {\it {Loops in Anti--de Sitter Space}},  {\em Phys.
  Rev. Lett.} {\bf 121} (2018), no.~10 101601,
  [\href{http://arxiv.org/abs/1804.01880}{{\tt arXiv:1804.01880}}].

\bibitem{Liu:2018jhs}
J.~Liu, E.~Perlmutter, V.~Rosenhaus, and D.~Simmons-Duffin, {\it
  {$d$-dimensional SYK, AdS Loops, and $6j$ Symbols}},  {\em JHEP} {\bf 03}
  (2019) 052, [\href{http://arxiv.org/abs/1808.00612}{{\tt arXiv:1808.00612}}].

\bibitem{Carmi:2018qzm}
D.~Carmi, L.~Di~Pietro, and S.~Komatsu, {\it {A Study of Quantum Field Theories
  in AdS at Finite Coupling}},  {\em JHEP} {\bf 01} (2019) 200,
  [\href{http://arxiv.org/abs/1810.04185}{{\tt arXiv:1810.04185}}].

\bibitem{Aprile:2018efk}
F.~Aprile, J.~Drummond, P.~Heslop, and H.~Paul, {\it {Double-trace spectrum of
  $N=4$ supersymmetric Yang-Mills theory at strong coupling}},  {\em Phys. Rev.
  D} {\bf 98} (2018), no.~12 126008,
  [\href{http://arxiv.org/abs/1802.06889}{{\tt arXiv:1802.06889}}].

\bibitem{Ghosh:2018bgd}
K.~Ghosh, {\it {Polyakov-Mellin Bootstrap for AdS loops}},  {\em JHEP} {\bf 02}
  (2020) 006, [\href{http://arxiv.org/abs/1811.00504}{{\tt arXiv:1811.00504}}].

\bibitem{Mazac:2018ycv}
D.~Mazac and M.~F. Paulos, {\it {The analytic functional bootstrap. Part II.
  Natural bases for the crossing equation}},  {\em JHEP} {\bf 02} (2019) 163,
  [\href{http://arxiv.org/abs/1811.10646}{{\tt arXiv:1811.10646}}].

\bibitem{Beccaria:2019stp}
M.~Beccaria and A.~A. Tseytlin, {\it {On boundary correlators in Liouville
  theory on AdS$_{2}$}},  {\em JHEP} {\bf 07} (2019) 008,
  [\href{http://arxiv.org/abs/1904.12753}{{\tt arXiv:1904.12753}}].

\bibitem{Chester:2019pvm}
S.~M. Chester, {\it {Genus-2 holographic correlator on AdS$_{5}\times $ S$^{5}$
  from localization}},  {\em JHEP} {\bf 04} (2020) 193,
  [\href{http://arxiv.org/abs/1908.05247}{{\tt arXiv:1908.05247}}].

\bibitem{Beccaria:2019dju}
M.~Beccaria, H.~Jiang, and A.~A. Tseytlin, {\it {Supersymmetric Liouville
  theory in AdS$_{2}$ and AdS/CFT}},  {\em JHEP} {\bf 11} (2019) 051,
  [\href{http://arxiv.org/abs/1909.10255}{{\tt arXiv:1909.10255}}].

\bibitem{Carmi:2019ocp}
D.~Carmi, {\it {Loops in AdS: From the Spectral Representation to Position
  Space}},  \href{http://arxiv.org/abs/1910.14340}{{\tt arXiv:1910.14340}}.

\bibitem{Aprile:2019rep}
F.~Aprile, J.~Drummond, P.~Heslop, and H.~Paul, {\it {One-loop amplitudes in
  AdS$_{5}\times$ S$^{5}$ supergravity from $ \mathcal{N} $ = 4 SYM at strong
  coupling}},  {\em JHEP} {\bf 03} (2020) 190,
  [\href{http://arxiv.org/abs/1912.01047}{{\tt arXiv:1912.01047}}].

\bibitem{Fichet:2019hkg}
S.~Fichet, {\it {Opacity and effective field theory in anti–de Sitter
  backgrounds}},  {\em Phys. Rev.} {\bf D100} (2019), no.~9 095002,
  [\href{http://arxiv.org/abs/1905.05779}{{\tt arXiv:1905.05779}}].

\bibitem{Meltzer:2019nbs}
D.~Meltzer, E.~Perlmutter, and A.~Sivaramakrishnan, {\it {Unitarity Methods in
  AdS/CFT}},  {\em JHEP} {\bf 03} (2020) 061,
  [\href{http://arxiv.org/abs/1912.09521}{{\tt arXiv:1912.09521}}].

\bibitem{Drummond:2019hel}
J.~Drummond and H.~Paul, {\it {One-loop string corrections to AdS amplitudes
  from CFT}},  \href{http://arxiv.org/abs/1912.07632}{{\tt arXiv:1912.07632}}.

\bibitem{Albayrak:2020isk}
S.~Albayrak, C.~Chowdhury, and S.~Kharel, {\it {An étude of momentum space
  scalar amplitudes in AdS}},  \href{http://arxiv.org/abs/2001.06777}{{\tt
  arXiv:2001.06777}}.

\bibitem{Albayrak:2020bso}
S.~Albayrak and S.~Kharel, {\it {On spinning loop amplitudes in Anti-de Sitter
  space}},  \href{http://arxiv.org/abs/2006.12540}{{\tt arXiv:2006.12540}}.

\bibitem{Meltzer:2020qbr}
D.~Meltzer and A.~Sivaramakrishnan, {\it {CFT Unitarity and the AdS Cutkosky
  Rules}},  \href{http://arxiv.org/abs/2008.11730}{{\tt arXiv:2008.11730}}.

\bibitem{Costantino:2020vdu}
A.~Costantino and S.~Fichet, {\it {Opacity from Loops in AdS}},  {\em JHEP}
  {\bf 02} (2021) 089, [\href{http://arxiv.org/abs/2011.06603}{{\tt
  arXiv:2011.06603}}].

\bibitem{Carmi:2021dsn}
D.~Carmi, {\it {Loops in AdS: From the Spectral Representation to Position
  Space II}},  \href{http://arxiv.org/abs/2104.10500}{{\tt arXiv:2104.10500}}.

\bibitem{Fitzpatrick:2011dm}
A.~Fitzpatrick and J.~Kaplan, {\it {Unitarity and the Holographic S-Matrix}},
  {\em JHEP} {\bf 10} (2012) 032, [\href{http://arxiv.org/abs/1112.4845}{{\tt
  arXiv:1112.4845}}].

\bibitem{Ponomarev:2019ofr}
D.~Ponomarev, {\it {From bulk loops to boundary large-N expansion}},  {\em
  JHEP} {\bf 01} (2020) 154, [\href{http://arxiv.org/abs/1908.03974}{{\tt
  arXiv:1908.03974}}].

\bibitem{Antunes:2020pof}
A.~Antunes, M.~S. Costa, T.~Hansen, A.~Salgarkar, and S.~Sarkar, {\it {The
  perturbative CFT optical theorem and high-energy string scattering in AdS at
  one loop}},  {\em JHEP} {\bf 04} (2021) 088,
  [\href{http://arxiv.org/abs/2012.01515}{{\tt arXiv:2012.01515}}].

\bibitem{Fichet:2021pbn}
S.~Fichet, {\it {Dressing in AdS and a Conformal Bethe-Salpeter Equation}},
  \href{http://arxiv.org/abs/2106.04604}{{\tt arXiv:2106.04604}}.

\bibitem{Burgess:1984ti}
C.~P. Burgess and C.~A. Lutken, {\it {Propagators and Effective Potentials in
  Anti-de Sitter Space}},  {\em Phys. Lett. B} {\bf 153} (1985) 137--141.

\bibitem{Inami:1985wu}
T.~Inami and H.~Ooguri, {\it {One Loop Effective Potential in Anti-de Sitter
  Space}},  {\em Prog. Theor. Phys.} {\bf 73} (1985) 1051.

\bibitem{Camporesi:1993mz}
R.~Camporesi and A.~Higuchi, {\it {Arbitrary spin effective potentials in
  anti-de Sitter space-time}},  {\em Phys. Rev. D} {\bf 47} (1993) 3339--3344.

\bibitem{Gubser:2002zh}
S.~S. Gubser and I.~Mitra, {\it {Double trace operators and one loop vacuum
  energy in AdS / CFT}},  {\em Phys. Rev. D} {\bf 67} (2003) 064018,
  [\href{http://arxiv.org/abs/hep-th/0210093}{{\tt hep-th/0210093}}].

\bibitem{Hartman:2006dy}
T.~Hartman and L.~Rastelli, {\it {Double-trace deformations, mixed boundary
  conditions and functional determinants in AdS/CFT}},  {\em JHEP} {\bf 01}
  (2008) 019, [\href{http://arxiv.org/abs/hep-th/0602106}{{\tt
  hep-th/0602106}}].

\bibitem{Giombi:2013fka}
S.~Giombi and I.~R. Klebanov, {\it {One Loop Tests of Higher Spin AdS/CFT}},
  {\em JHEP} {\bf 12} (2013) 068, [\href{http://arxiv.org/abs/1308.2337}{{\tt
  arXiv:1308.2337}}].

\bibitem{Antoniadis:2011ib}
I.~Antoniadis, P.~O. Mazur, and E.~Mottola, {\it {Conformal Invariance, Dark
  Energy, and CMB Non-Gaussianity}},  {\em JCAP} {\bf 09} (2012) 024,
  [\href{http://arxiv.org/abs/1103.4164}{{\tt arXiv:1103.4164}}].

\bibitem{Creminelli:2011mw}
P.~Creminelli, {\it {Conformal invariance of scalar perturbations in
  inflation}},  {\em Phys. Rev. D} {\bf 85} (2012) 041302,
  [\href{http://arxiv.org/abs/1108.0874}{{\tt arXiv:1108.0874}}].

\bibitem{Maldacena:2011nz}
J.~M. Maldacena and G.~L. Pimentel, {\it {On graviton non-Gaussianities during
  inflation}},  {\em JHEP} {\bf 09} (2011) 045,
  [\href{http://arxiv.org/abs/1104.2846}{{\tt arXiv:1104.2846}}].

\bibitem{Bzowski:2011ab}
A.~Bzowski, P.~McFadden, and K.~Skenderis, {\it {Holographic predictions for
  cosmological 3-point functions}},  {\em JHEP} {\bf 03} (2012) 091,
  [\href{http://arxiv.org/abs/1112.1967}{{\tt arXiv:1112.1967}}].

\bibitem{Kehagias:2012pd}
A.~Kehagias and A.~Riotto, {\it {Operator Product Expansion of Inflationary
  Correlators and Conformal Symmetry of de Sitter}},  {\em Nucl. Phys. B} {\bf
  864} (2012) 492--529, [\href{http://arxiv.org/abs/1205.1523}{{\tt
  arXiv:1205.1523}}].

\bibitem{Mata:2012bx}
I.~Mata, S.~Raju, and S.~Trivedi, {\it {CMB from CFT}},  {\em JHEP} {\bf 07}
  (2013) 015, [\href{http://arxiv.org/abs/1211.5482}{{\tt arXiv:1211.5482}}].

\bibitem{Kundu:2014gxa}
N.~Kundu, A.~Shukla, and S.~P. Trivedi, {\it {Constraints from Conformal
  Symmetry on the Three Point Scalar Correlator in Inflation}},  {\em JHEP}
  {\bf 04} (2015) 061, [\href{http://arxiv.org/abs/1410.2606}{{\tt
  arXiv:1410.2606}}].

\bibitem{Kundu:2015xta}
N.~Kundu, A.~Shukla, and S.~P. Trivedi, {\it {Ward Identities for Scale and
  Special Conformal Transformations in Inflation}},  {\em JHEP} {\bf 01} (2016)
  046, [\href{http://arxiv.org/abs/1507.06017}{{\tt arXiv:1507.06017}}].

\bibitem{Ghosh:2014kba}
A.~Ghosh, N.~Kundu, S.~Raju, and S.~P. Trivedi, {\it {Conformal Invariance and
  the Four Point Scalar Correlator in Slow-Roll Inflation}},  {\em JHEP} {\bf
  07} (2014) 011, [\href{http://arxiv.org/abs/1401.1426}{{\tt
  arXiv:1401.1426}}].

\bibitem{Pajer:2016ieg}
E.~Pajer, G.~L. Pimentel, and J.~V.~S. Van~Wijck, {\it {The Conformal Limit of
  Inflation in the Era of CMB Polarimetry}},  {\em JCAP} {\bf 06} (2017) 009,
  [\href{http://arxiv.org/abs/1609.06993}{{\tt arXiv:1609.06993}}].

\bibitem{Arkani-Hamed:2015bza}
N.~Arkani-Hamed and J.~Maldacena, {\it {Cosmological Collider Physics}},
  \href{http://arxiv.org/abs/1503.08043}{{\tt arXiv:1503.08043}}.

\bibitem{Arkani-Hamed:2018kmz}
N.~Arkani-Hamed, D.~Baumann, H.~Lee, and G.~L. Pimentel, {\it {The Cosmological
  Bootstrap: Inflationary Correlators from Symmetries and Singularities}},
  {\em JHEP} {\bf 04} (2020) 105, [\href{http://arxiv.org/abs/1811.00024}{{\tt
  arXiv:1811.00024}}].

\bibitem{Farrow:2018yni}
J.~A. Farrow, A.~E. Lipstein, and P.~McFadden, {\it {Double copy structure of
  CFT correlators}},  {\em JHEP} {\bf 02} (2019) 130,
  [\href{http://arxiv.org/abs/1812.11129}{{\tt arXiv:1812.11129}}].

\bibitem{Baumann:2019oyu}
D.~Baumann, C.~Duaso~Pueyo, A.~Joyce, H.~Lee, and G.~L. Pimentel, {\it {The
  Cosmological Bootstrap: Weight-Shifting Operators and Scalar Seeds}},
  \href{http://arxiv.org/abs/1910.14051}{{\tt arXiv:1910.14051}}.

\bibitem{Green:2020ebl}
D.~Green and E.~Pajer, {\it {On the Symmetries of Cosmological Perturbations}},
   {\em JCAP} {\bf 09} (2020) 032, [\href{http://arxiv.org/abs/2004.09587}{{\tt
  arXiv:2004.09587}}].

\bibitem{Sengor:2021zlc}
G.~Sengor and C.~Skordis, {\it {Scalar two-point functions at the late-time
  boundary of de Sitter}},  \href{http://arxiv.org/abs/2110.01635}{{\tt
  arXiv:2110.01635}}.

\bibitem{Wang:2021qez}
L.-T. Wang, Z.-Z. Xianyu, and Y.-M. Zhong, {\it {Precision Calculation of
  Inflation Correlators at One Loop}},
  \href{http://arxiv.org/abs/2109.14635}{{\tt arXiv:2109.14635}}.

\bibitem{Goodhew:2020hob}
H.~Goodhew, S.~Jazayeri, and E.~Pajer, {\it {The Cosmological Optical
  Theorem}},  {\em JCAP} {\bf 04} (2021) 021,
  [\href{http://arxiv.org/abs/2009.02898}{{\tt arXiv:2009.02898}}].

\bibitem{Pajer:2020wxk}
E.~Pajer, {\it {Building a Boostless Bootstrap for the Bispectrum}},  {\em
  JCAP} {\bf 01} (2021) 023, [\href{http://arxiv.org/abs/2010.12818}{{\tt
  arXiv:2010.12818}}].

\bibitem{Jazayeri:2021fvk}
S.~Jazayeri, E.~Pajer, and D.~Stefanyszyn, {\it {From locality and unitarity to
  cosmological correlators}},  {\em JHEP} {\bf 10} (2021) 065,
  [\href{http://arxiv.org/abs/2103.08649}{{\tt arXiv:2103.08649}}].

\bibitem{Melville:2021lst}
S.~Melville and E.~Pajer, {\it {Cosmological Cutting Rules}},  {\em JHEP} {\bf
  05} (2021) 249, [\href{http://arxiv.org/abs/2103.09832}{{\tt
  arXiv:2103.09832}}].

\bibitem{Goodhew:2021oqg}
H.~Goodhew, S.~Jazayeri, M.~H. Gordon~Lee, and E.~Pajer, {\it {Cutting
  cosmological correlators}},  {\em JCAP} {\bf 08} (2021) 003,
  [\href{http://arxiv.org/abs/2104.06587}{{\tt arXiv:2104.06587}}].

\bibitem{Baumann:2021fxj}
D.~Baumann, W.-M. Chen, C.~Duaso~Pueyo, A.~Joyce, H.~Lee, and G.~L. Pimentel,
  {\it {Linking the Singularities of Cosmological Correlators}},
  \href{http://arxiv.org/abs/2106.05294}{{\tt arXiv:2106.05294}}.

\bibitem{Raju:2012zr}
S.~Raju, {\it {New Recursion Relations and a Flat Space Limit for AdS/CFT
  Correlators}},  {\em Phys. Rev. D} {\bf 85} (2012) 126009,
  [\href{http://arxiv.org/abs/1201.6449}{{\tt arXiv:1201.6449}}].

\bibitem{Arkani-Hamed:2017fdk}
N.~Arkani-Hamed, P.~Benincasa, and A.~Postnikov, {\it {Cosmological Polytopes
  and the Wavefunction of the Universe}},
  \href{http://arxiv.org/abs/1709.02813}{{\tt arXiv:1709.02813}}.

\bibitem{Arkani-Hamed:2018bjr}
N.~Arkani-Hamed and P.~Benincasa, {\it {On the Emergence of Lorentz Invariance
  and Unitarity from the Scattering Facet of Cosmological Polytopes}},
  \href{http://arxiv.org/abs/1811.01125}{{\tt arXiv:1811.01125}}.

\bibitem{Benincasa:2018ssx}
P.~Benincasa, {\it {From the flat-space S-matrix to the Wavefunction of the
  Universe}},  \href{http://arxiv.org/abs/1811.02515}{{\tt arXiv:1811.02515}}.

\bibitem{Cespedes:2020xqq}
S.~C\'espedes, A.-C. Davis, and S.~Melville, {\it {On the time evolution of
  cosmological correlators}},  {\em JHEP} {\bf 02} (2021) 012,
  [\href{http://arxiv.org/abs/2009.07874}{{\tt arXiv:2009.07874}}].

\bibitem{Sleight:2020obc}
C.~Sleight and M.~Taronna, {\it {From AdS to dS Exchanges: Spectral
  Representation, Mellin Amplitudes and Crossing}},
  \href{http://arxiv.org/abs/2007.09993}{{\tt arXiv:2007.09993}}.

\bibitem{Baumann:2020dch}
D.~Baumann, C.~Duaso~Pueyo, A.~Joyce, H.~Lee, and G.~L. Pimentel, {\it {The
  Cosmological Bootstrap: Spinning Correlators from Symmetries and
  Factorization}},  \href{http://arxiv.org/abs/2005.04234}{{\tt
  arXiv:2005.04234}}.

\bibitem{Meltzer:2021bmb}
D.~Meltzer, {\it {Dispersion Formulas in QFTs, CFTs, and Holography}},  {\em
  JHEP} {\bf 05} (2021) 098, [\href{http://arxiv.org/abs/2103.15839}{{\tt
  arXiv:2103.15839}}].

\bibitem{Meltzer:2021zin}
D.~Meltzer, {\it {The Inflationary Wavefunction from Analyticity and
  Factorization}},  \href{http://arxiv.org/abs/2107.10266}{{\tt
  arXiv:2107.10266}}.

\bibitem{Gomez:2021qfd}
H.~Gomez, R.~L. Jusinskas, and A.~Lipstein, {\it {Cosmological Scattering
  Equations}},  \href{http://arxiv.org/abs/2106.11903}{{\tt arXiv:2106.11903}}.

\bibitem{Bonifacio:2021azc}
J.~Bonifacio, E.~Pajer, and D.-G. Wang, {\it {From Amplitudes to Contact
  Cosmological Correlators}},  \href{http://arxiv.org/abs/2106.15468}{{\tt
  arXiv:2106.15468}}.

\bibitem{Sleight:2019mgd}
C.~Sleight, {\it {A Mellin Space Approach to Cosmological Correlators}},  {\em
  JHEP} {\bf 01} (2020) 090, [\href{http://arxiv.org/abs/1906.12302}{{\tt
  arXiv:1906.12302}}].

\bibitem{Sleight:2019hfp}
C.~Sleight and M.~Taronna, {\it {Bootstrapping Inflationary Correlators in
  Mellin Space}},  {\em JHEP} {\bf 02} (2020) 098,
  [\href{http://arxiv.org/abs/1907.01143}{{\tt arXiv:1907.01143}}].

\bibitem{Balasubramanian:2002zh}
V.~Balasubramanian, J.~de~Boer, and D.~Minic, {\it {Notes on de Sitter space
  and holography}},  {\em Class. Quant. Grav.} {\bf 19} (2002) 5655--5700,
  [\href{http://arxiv.org/abs/hep-th/0207245}{{\tt hep-th/0207245}}].

\bibitem{Maldacena:2002vr}
J.~M. Maldacena, {\it {Non-Gaussian features of primordial fluctuations in
  single field inflationary models}},  {\em JHEP} {\bf 05} (2003) 013,
  [\href{http://arxiv.org/abs/astro-ph/0210603}{{\tt astro-ph/0210603}}].

\bibitem{Harlow:2011ke}
D.~Harlow and D.~Stanford, {\it {Operator Dictionaries and Wave Functions in
  AdS/CFT and dS/CFT}},  \href{http://arxiv.org/abs/1104.2621}{{\tt
  arXiv:1104.2621}}.

\bibitem{Anninos:2014lwa}
D.~Anninos, T.~Anous, D.~Z. Freedman, and G.~Konstantinidis, {\it {Late-time
  Structure of the Bunch-Davies De Sitter Wavefunction}},  {\em JCAP} {\bf 11}
  (2015) 048, [\href{http://arxiv.org/abs/1406.5490}{{\tt arXiv:1406.5490}}].

\bibitem{DiPietro:2021sjt}
L.~Di~Pietro, V.~Gorbenko, and S.~Komatsu, {\it {Analyticity and Unitarity for
  Cosmological Correlators}},  \href{http://arxiv.org/abs/2108.01695}{{\tt
  arXiv:2108.01695}}.

\bibitem{Heckelbacher:2022hbq}
T.~Heckelbacher, I.~Sachs, E.~Skvortsov, and P.~Vanhove, {\it {Analytical
  evaluation of cosmological correlation functions}},
  \href{http://arxiv.org/abs/2204.07217}{{\tt arXiv:2204.07217}}.

\bibitem{Petkou:1994ad}
A.~Petkou, {\it {Conserved currents, consistency relations and operator product
  expansions in the conformally invariant O(N) vector model}},  {\em Annals
  Phys.} {\bf 249} (1996) 180--221,
  [\href{http://arxiv.org/abs/hep-th/9410093}{{\tt hep-th/9410093}}].

\bibitem{Branson:1999jz}
T.~P. Branson, P.~B. Gilkey, K.~Kirsten, and D.~V. Vassilevich, {\it {Heat
  kernel asymptotics with mixed boundary conditions}},  {\em Nucl. Phys. B}
  {\bf 563} (1999) 603--626, [\href{http://arxiv.org/abs/hep-th/9906144}{{\tt
  hep-th/9906144}}].

\bibitem{Blau_GR}
M.~Blau, ``{Lecture Notes on General Relativity}.''
  \url{http://www.blau.itp.unibe.ch/GRLecturenotes.html}.

\bibitem{Jackson:100964}
J.~D. Jackson, {\em {Classical electrodynamics; 2nd ed.}}
\newblock Wiley, New York, NY, 1975.

\bibitem{Karch:2006pv}
A.~Karch, E.~Katz, D.~T. Son, and M.~A. Stephanov, {\it {Linear confinement and
  AdS/QCD}},  {\em Phys. Rev.} {\bf D74} (2006) 015005,
  [\href{http://arxiv.org/abs/hep-ph/0602229}{{\tt hep-ph/0602229}}].

\bibitem{Gursoy:2007cb}
U.~Gursoy and E.~Kiritsis, {\it {Exploring improved holographic theories for
  QCD: Part I}},  {\em JHEP} {\bf 02} (2008) 032,
  [\href{http://arxiv.org/abs/0707.1324}{{\tt arXiv:0707.1324}}].

\bibitem{Gursoy:2007er}
U.~Gursoy, E.~Kiritsis, and F.~Nitti, {\it {Exploring improved holographic
  theories for QCD: Part II}},  {\em JHEP} {\bf 02} (2008) 019,
  [\href{http://arxiv.org/abs/0707.1349}{{\tt arXiv:0707.1349}}].

\bibitem{Batell:2008zm}
B.~Batell and T.~Gherghetta, {\it {Dynamical Soft-Wall AdS/QCD}},  {\em Phys.
  Rev.} {\bf D78} (2008) 026002, [\href{http://arxiv.org/abs/0801.4383}{{\tt
  arXiv:0801.4383}}].

\bibitem{Cabrer:2009we}
J.~A. Cabrer, G.~von Gersdorff, and M.~Quiros, {\it {Soft-Wall Stabilization}},
   {\em New J. Phys.} {\bf 12} (2010) 075012,
  [\href{http://arxiv.org/abs/0907.5361}{{\tt arXiv:0907.5361}}].

\bibitem{vonGersdorff:2010ht}
G.~von Gersdorff, {\it {From Soft Walls to Infrared Branes}},  {\em Phys. Rev.}
  {\bf D82} (2010) 086010, [\href{http://arxiv.org/abs/1005.5134}{{\tt
  arXiv:1005.5134}}].

\bibitem{Fichet:2019owx}
S.~Fichet, {\it {Braneworld effective field theories --- holography,
  consistency and conformal effects}},  {\em JHEP} {\bf 04} (2020) 016,
  [\href{http://arxiv.org/abs/1912.12316}{{\tt arXiv:1912.12316}}].

\bibitem{Batell:2007jv}
B.~Batell and T.~Gherghetta, {\it {Holographic mixing quantified}},  {\em Phys.
  Rev.} {\bf D76} (2007) 045017, [\href{http://arxiv.org/abs/0706.0890}{{\tt
  arXiv:0706.0890}}].

\bibitem{Batell:2007ez}
B.~Batell and T.~Gherghetta, {\it {Warped phenomenology in the holographic
  basis}},  {\em Phys. Rev.} {\bf D77} (2008) 045002,
  [\href{http://arxiv.org/abs/0710.1838}{{\tt arXiv:0710.1838}}].

\bibitem{Falkowski:2008yr}
A.~Falkowski and M.~Perez-Victoria, {\it {Holographic Unhiggs}},  {\em Phys.
  Rev. D} {\bf 79} (2009) 035005, [\href{http://arxiv.org/abs/0810.4940}{{\tt
  arXiv:0810.4940}}].

\bibitem{Giombi:2018vtc}
S.~Giombi, V.~Kirilin, and E.~Perlmutter, {\it {Double-Trace Deformations of
  Conformal Correlations}},  {\em JHEP} {\bf 02} (2018) 175,
  [\href{http://arxiv.org/abs/1801.01477}{{\tt arXiv:1801.01477}}].

\bibitem{Leonhardt:2003qu}
T.~Leonhardt, R.~Manvelyan, and W.~Ruhl, {\it {The Group approach to AdS space
  propagators}},  {\em Nucl. Phys. B} {\bf 667} (2003) 413--434,
  [\href{http://arxiv.org/abs/hep-th/0305235}{{\tt hep-th/0305235}}].

\bibitem{Cornalba:2007fs}
L.~Cornalba, {\it {Eikonal methods in AdS/CFT: Regge theory and multi-reggeon
  exchange}},  \href{http://arxiv.org/abs/0710.5480}{{\tt arXiv:0710.5480}}.

\bibitem{Paulos:2011ie}
M.~F. Paulos, {\it {Towards Feynman rules for Mellin amplitudes}},  {\em JHEP}
  {\bf 10} (2011) 074, [\href{http://arxiv.org/abs/1107.1504}{{\tt
  arXiv:1107.1504}}].

\bibitem{Costa:2014kfa}
M.~S. Costa, V.~Gonçalves, and J.~Penedones, {\it {Spinning AdS Propagators}},
   {\em JHEP} {\bf 09} (2014) 064, [\href{http://arxiv.org/abs/1404.5625}{{\tt
  arXiv:1404.5625}}].

\bibitem{Freedman:1998bj}
D.~Z. Freedman, S.~D. Mathur, A.~Matusis, and L.~Rastelli, {\it {Comments on 4
  point functions in the CFT / AdS correspondence}},  {\em Phys. Lett. B} {\bf
  452} (1999) 61--68, [\href{http://arxiv.org/abs/hep-th/9808006}{{\tt
  hep-th/9808006}}].

\bibitem{Klebanov:1999tb}
I.~R. Klebanov and E.~Witten, {\it {AdS / CFT correspondence and symmetry
  breaking}},  {\em Nucl. Phys. B} {\bf 556} (1999) 89--114,
  [\href{http://arxiv.org/abs/hep-th/9905104}{{\tt hep-th/9905104}}].

\bibitem{Kraus:1999it}
P.~Kraus, {\it {Dynamics of anti-de Sitter domain walls}},  {\em JHEP} {\bf 12}
  (1999) 011, [\href{http://arxiv.org/abs/hep-th/9910149}{{\tt
  hep-th/9910149}}].

\bibitem{Hebecker:2001nv}
A.~Hebecker and J.~March-Russell, {\it {Randall-Sundrum II cosmology, AdS /
  CFT, and the bulk black hole}},  {\em Nucl. Phys.} {\bf B608} (2001)
  375--393, [\href{http://arxiv.org/abs/hep-ph/0103214}{{\tt hep-ph/0103214}}].

\bibitem{Langlois:2002ke}
D.~Langlois, L.~Sorbo, and M.~Rodriguez-Martinez, {\it {Cosmology of a brane
  radiating gravitons into the extra dimension}},  {\em Phys. Rev. Lett.} {\bf
  89} (2002) 171301, [\href{http://arxiv.org/abs/hep-th/0206146}{{\tt
  hep-th/0206146}}].

\bibitem{Brax:2018grq}
P.~Brax and S.~Fichet, {\it {Quantum Chameleons}},  {\em Phys. Rev.} {\bf D99}
  (2019), no.~10 104049, [\href{http://arxiv.org/abs/1809.10166}{{\tt
  arXiv:1809.10166}}].

\bibitem{Loop_Quiros}
S.~Fichet, B.~Jain, E.~Ponton, M.~Quiros, and R.~Rosenfeld, ``{Private
  Communication}.''

\bibitem{Milton_Sphere}
C.~M. Bender and K.~A. Milton, {\it Scalar casimir effect for a d-dimensional
  sphere},  {\em Phys. Rev. D} {\bf 50} (Nov, 1994) 6547--6555.

\bibitem{Ambjorn:1981xw}
J.~Ambjorn and S.~Wolfram, {\it {Properties of the Vacuum. 1. Mechanical and
  Thermodynamic}},  {\em Annals Phys.} {\bf 147} (1983) 1.

\bibitem{abramowitz+stegun}
M.~Abramowitz and I.~A. Stegun, {\em Handbook of Mathematical Functions with
  Formulas, Graphs, and Mathematical Tables}.
\newblock Dover, New York, ninth dover printing, tenth gpo printing~ed., 1964.

\bibitem{Sivaramakrishnan:2021srm}
A.~Sivaramakrishnan, {\it {Towards Color-Kinematics Duality in Generic
  Spacetimes}},  \href{http://arxiv.org/abs/2110.15356}{{\tt
  arXiv:2110.15356}}.

\bibitem{Peskin:257493}
M.~E. Peskin and D.~V. Schroeder, {\em {An introduction to quantum field
  theory}}.
\newblock Westview, Boulder, CO, 1995.

\bibitem{Hoover:2005uf}
D.~Hoover and C.~P. Burgess, {\it {Ultraviolet sensitivity in higher
  dimensions}},  {\em JHEP} {\bf 01} (2006) 058,
  [\href{http://arxiv.org/abs/hep-th/0507293}{{\tt hep-th/0507293}}].

\bibitem{Fichet:2013ola}
S.~Fichet and G.~von Gersdorff, {\it {Anomalous gauge couplings from composite
  Higgs and warped extra dimensions}},  {\em JHEP} {\bf 03} (2014) 102,
  [\href{http://arxiv.org/abs/1311.6815}{{\tt arXiv:1311.6815}}].

\bibitem{Fitzpatrick:2010zm}
A.~L. Fitzpatrick, E.~Katz, D.~Poland, and D.~Simmons-Duffin, {\it {Effective
  Conformal Theory and the Flat-Space Limit of AdS}},  {\em JHEP} {\bf 07}
  (2011) 023, [\href{http://arxiv.org/abs/1007.2412}{{\tt arXiv:1007.2412}}].

\bibitem{Weinberg:2005vy}
S.~Weinberg, {\it {Quantum contributions to cosmological correlations}},  {\em
  Phys. Rev. D} {\bf 72} (2005) 043514,
  [\href{http://arxiv.org/abs/hep-th/0506236}{{\tt hep-th/0506236}}].

\bibitem{Seery:2007we}
D.~Seery, {\it {One-loop corrections to a scalar field during inflation}},
  {\em JCAP} {\bf 11} (2007) 025, [\href{http://arxiv.org/abs/0707.3377}{{\tt
  arXiv:0707.3377}}].

\bibitem{Adshead:2009cb}
P.~Adshead, R.~Easther, and E.~A. Lim, {\it {The 'in-in' Formalism and
  Cosmological Perturbations}},  {\em Phys. Rev. D} {\bf 80} (2009) 083521,
  [\href{http://arxiv.org/abs/0904.4207}{{\tt arXiv:0904.4207}}].

\bibitem{Senatore:2009cf}
L.~Senatore and M.~Zaldarriaga, {\it {On Loops in Inflation}},  {\em JHEP} {\bf
  12} (2010) 008, [\href{http://arxiv.org/abs/0912.2734}{{\tt
  arXiv:0912.2734}}].

\bibitem{Senatore:2016aui}
L.~Senatore, {\it {Lectures on Inflation}},  in {\em {Theoretical Advanced
  Study Institute in Elementary Particle Physics}: {New Frontiers in Fields and
  Strings}}, pp.~447--543, 2017.
\newblock \href{http://arxiv.org/abs/1609.00716}{{\tt arXiv:1609.00716}}.

\bibitem{Streater:1989vi}
R.~F. Streater and A.~S. Wightman, {\em {PCT, spin and statistics, and all
  that}}.
\newblock 1989.

\bibitem{Schwarz71}
F.~Schwarz, {\it {Unitary Irreducible Representations of the Groups SO0(n,
  1)}},  {\em {J. Math. Phys.}} {\bf 12} (1971) 131.

\bibitem{Harmonic_analysis}
V.~Dobrev, G.~Mack, V.~Petkova, S.~Petrova, , and I.~Todorov, {\em Harmonic
  analysis on the n-dimensional Lorentz Group and its application to conformal
  quantum field theory}.
\newblock Springer-Verlag Berlin, 1977.

\bibitem{Gibbons:1977mu}
G.~W. Gibbons and S.~W. Hawking, {\it {Cosmological Event Horizons,
  Thermodynamics, and Particle Creation}},  {\em Phys. Rev. D} {\bf 15} (1977)
  2738--2751.

\bibitem{Burges:1984qm}
C.~J.~C. Burges, {\it {The De Sitter Vacuum}},  {\em Nucl. Phys. B} {\bf 247}
  (1984) 533--543.

\bibitem{Mottola:1984ar}
E.~Mottola, {\it {Particle Creation in de Sitter Space}},  {\em Phys. Rev. D}
  {\bf 31} (1985) 754.

\bibitem{Allen:1985ux}
B.~Allen, {\it {Vacuum States in de Sitter Space}},  {\em Phys. Rev. D} {\bf
  32} (1985) 3136.

\bibitem{deBoer:2004nd}
J.~de~Boer, V.~Jejjala, and D.~Minic, {\it {Alpha-states in de Sitter space}},
  {\em Phys. Rev. D} {\bf 71} (2005) 044013,
  [\href{http://arxiv.org/abs/hep-th/0406217}{{\tt hep-th/0406217}}].

\bibitem{Sleight:2021plv}
C.~Sleight and M.~Taronna, {\it {From dS to AdS and back}},
  \href{http://arxiv.org/abs/2109.02725}{{\tt arXiv:2109.02725}}.

\bibitem{Premkumar:2021mlz}
A.~Premkumar, {\it {Regulating Loops in dS}},
  \href{http://arxiv.org/abs/2110.12504}{{\tt arXiv:2110.12504}}.

\bibitem{Burges:1985qq}
C.~J. Burges, D.~Z. Freedman, S.~Davis, and G.~Gibbons, {\it {Supersymmetry in
  Anti-de Sitter Space}},  {\em Annals Phys.} {\bf 167} (1986) 285.

\end{thebibliography}\endgroup

\end{document}